\documentclass[12pt]{iopart}

\renewcommand{\sp}{\epsilon} 
\newcommand{\dd}{\mathrm{d}}
\newcommand{\wmean}{\vec{\mathsf{w}}}

\newcommand{\avg}[1]{\left\langle #1 \right\rangle}

\newcommand{\osc}[1]{\Breve{#1}}
\newcommand{\nit}[1]{\Tilde{#1}}

\expandafter\let\csname equation*\endcsname\relax
\expandafter\let\csname endequation*\endcsname\relax
\usepackage{amsmath}
\usepackage{mathtools}
\usepackage{amssymb}
\usepackage{physics}

\usepackage{cite} 
\usepackage{setspace}
\usepackage{lscape}
\usepackage{import}
\usepackage{wrapfig}
\usepackage{multicol}
\usepackage{textcomp}
\usepackage{accents}
\usepackage{amsfonts}
\usepackage{calc}
\usepackage{mathrsfs}
\usepackage{amssymb}
\usepackage{array}
\usepackage{color}
\usepackage{bm}
\usepackage{graphicx}
\usepackage{hyperref}
\usepackage{cleveref}
\usepackage{booktabs}
\usepackage{float}
\usepackage{caption}
\usepackage{subcaption}
\usepackage{tabularx}
\usepackage[scr=boondox]{mathalfa}
\usepackage[dvipsnames]{xcolor}
\usepackage[normalem]{ulem}
\usepackage{comment}

\usepackage{ulem} 
\usepackage{bm}
\usepackage{dcolumn}
\usepackage{epsfig}
\usepackage{graphics}
\usepackage[latin1]{inputenc}
\usepackage{latexsym}
\usepackage{rotating}
\usepackage{calligra} \DeclareMathAlphabet{\mathcalligra}{T1}{calligra}{m}{n} \DeclareFontShape{T1}{calligra}{m}{n}{<->s*[2.2]callig15}{}
\usepackage{tikz}
\usepackage{etex}
\usetikzlibrary{arrows.meta}
\usetikzlibrary{decorations.pathmorphing} 
\usepackage{booktabs} 
\usepackage{arydshln} 
\usepackage{array} 
\usetikzlibrary{shapes.geometric, arrows, positioning, calc}
\definecolor{beige}{RGB}{245, 245, 220}

\tikzstyle{box} = [rectangle, rounded corners, minimum width=2.5cm, minimum height=1cm, text centered, draw=black, fill=white!80, align=center]
\tikzstyle{box2} = [rectangle,  minimum width=3cm, minimum height=1cm, text centered, draw=black, fill=beige!80, align=center]
\tikzstyle{highlight} = [rectangle, rounded corners, minimum width=2.5cm, minimum height=1cm, text centered, draw=black, fill=gray!90, text=white, font=\bfseries,align=center]
\tikzstyle{arrow} = [thick, -{Stealth}, dotted]
\tikzstyle{thickbox} = [rectangle, draw=black, line width=0.5mm, fill=gray!20, text centered, text width=3cm, minimum height=1.2cm]
\tikzstyle{thickbox2} = [rectangle, draw=black, line width=0.5mm, fill=gray!20, text centered, text width=8cm, minimum height=1.2cm]
\usetikzlibrary{patterns}

\def\mbar{\overline{m}}

\def\rmd{{\rm d}}

\newcommand{\be}{\begin{equation}}
\newcommand{\ee}{\end{equation}}

\newcommand{\RN}[1]{%
  \textup{\uppercase\expandafter{\romannumeral#1}}%
}

\renewcommand{\sp}{\varepsilon}

\newcommand{\stkout}[1]{\ifmmode\text{\sout{\ensuremath{#1}}}\else\sout{#1}\fi}

\begin{document}

\title[Shifted-geodesic approximation for spinning-body gravitational wave fluxes]{Shifted-geodesic approximation for spinning-body gravitational wave fluxes}

\author{Lisa V. Drummond$^{1}$, Scott A. Hughes$^{2}$, Viktor Skoup\'y$^{3}$, Philip Lynch$^{4}$ \& Gabriel Andres Piovano$^{5,6}$}
\address{$^{1}$Department of Physics and TAPIR, California Institute of Technology, Pasadena, CA 91125, USA}
\address{$^{2}$Department of Physics and MIT Kavli Institute, Massachusetts Institute of Technology, Cambridge, MA 02139, USA}
\address{$^{3}$Institute of Theoretical Physics, Faculty of Mathematics and Physics, Charles University, CZ-180 00 Prague, Czech Republic}
\address{$^{4}$Max Planck Institute for Gravitational Physics (Albert Einstein Institute), Am M\"uhlenberg 1, 14476 Potsdam, Germany}
\address{$^{5}$Universit\'e Libre de Bruxelles, BLU-ULB Brussels Laboratory of the Universe, International Solvay Institutes, CP 231, B-1050 Brussels, Belgium}
\address{$^{6}$Physique de l'Univers, Champs et Gravitation, Universit\'e de Mons - UMONS, Place du Parc 20, 7000 Mons, Belgium}

\vspace{10pt}

\begin{abstract}
We present a shifted-geodesic framework for computing gravitational-wave fluxes from spinning test bodies moving on bound orbits of Kerr black holes.  The method provides a simple and efficient means of evaluating energy and angular momentum fluxes incorporating the leading effect of the smaller body's spin. Because post-adiabatic corrections, including secondary spin contributions, are subdominant to the leading adiabatic terms, this approximation is well justified.  In particular, we find that oscillatory spin terms typically contribute very little to fluxes, but their contribution to the description of orbits is computationally expensive, making such terms a natural target for approximation.  In our framework, orbital frequencies and integrals of the motion are perturbed to include spin effects, while the trajectory retains the global structure of geodesic motion.  This simplifies the computation of gravitational radiation. We compute approximate fluxes in the frequency domain using mode-summed Teukolsky amplitudes and validate them against fluxes computed using ``exact'' spinning-body orbit schemes. The shifted-geodesic approximation is most reliable for orbits with lower eccentricity, lower inclination, and larger semi-latus recta.  The approximation becomes less reliable as we approach the separatrix between stable and unstable orbits; fortunately, many inspirals spend less time in this region of parameter space.  A diagnostic inspiral evolution shows very small dephasing due to use of the shifted-geodesic approximation ($\approx10^{-2}$ radians over 1 year), confirming that spin-induced flux corrections can be accurately included using this simple modification to a geodesic trajectory. This approximation provides a rapid and convenient way to compute spinning-body orbits, but is not intended to replace more accurate treatments.  We propose it as a pragmatic alternative when speed and simplicity are prioritized, enabling efficient EMRI/IMRI flux calculations and supporting parameter-space studies for LISA.
\end{abstract}

\maketitle

\section{Introduction and previous work}

Binary systems in which a stellar mass compact object inspirals into a supermassive black hole, known as extreme mass-ratio inspirals (EMRIs), are both astrophysically significant and theoretically rich. These systems are prime targets for low-frequency gravitational wave (GW) observatories such as the planned Laser Interferometer Space Antenna (LISA) \cite{LISARedBook}. EMRIs are expected to produce GWs as a stellar-mass compact object, typically a black hole or neutron star (the \textit{secondary}) with mass $\mu \sim 1-100 M_\odot$, spirals through the strong-field spacetime of a massive black hole (the \textit{primary}) with mass $M \sim 10^5-10^7 M_\odot$. Precise measurements of EMRI signals will provide invaluable tests of the theory of general relativity (GR) and the nature of black holes \cite{Glampedakis2006, Gair2013, Barack:2006pq, Vigeland:2011ji, Barausse2020, Berry:2019wgg, Speri2024}, making it possible to very accurately measure important properties of the massive black hole \cite{Barack:2003fp, Babak2017, Chapman-Bird2025}.

EMRIs are characterized by extreme mass ratios, with $\varepsilon \equiv \mu/M$ expected to range from $10^{-7}$ to $10^{-5}$ \cite{LISARedBook}. On short time scales, neglecting mass-ratio effects, the smaller body follows a geodesic orbit of the large black hole's Kerr spacetime.  At leading order in $\varepsilon$, the smaller body perturbs this spacetime, causing a self-interaction which modifies orbit properties and drives the smaller body to inspiral into the large black hole \cite{Fujita2017, vandeMeent2018, Pound2020, Pound2021, Barack:2018yvs, Wardell2023, Macedo2022, GomesDaSilva2023, Blanco2023_1}.  Extensive study over several decades has led to a now mature program for computing self-forces in the EMRI context \cite{Wardell2024, 2024Macedo, Leather2024, Warburton2024, Long:2024ltn, Albertini2024, Upton2024,Spiers2024, Bourg2024, Burke2024, Upton2025, Cunningham2025, Kuchler2025, Nasipak2025}.

Self forces are typically computed in a limit which treats the smaller body as pointlike.  Any realistic body will of course have some finite size.  This finite size couples to spacetime curvature, which further changes the body's motion: forcing terms enter the equations of motion describing how the body couples to curvature. In EMRI systems, the dominant finite-size effect arises from the non-zero spin angular momentum of the smaller body.  The interaction of the secondary's spin with spacetime curvature is governed by the Mathisson-Papapetrou-Dixon (MPD) equations \cite{Mathisson2010, Papapetrou1951, Dixon1970}.  The spin-curvature coupling perturbs orbital frequencies and integrals of motion, like energy and angular momentum, and introduces corrections to the gravitational-wave fluxes emitted during inspiral \cite{Witzany2019_2, Piovano2020, Zelenka2020, Skoupy2021, Piovano2021, Skoupy2022, Mathews2022, Blanco2023, Albertini2024_2}.

R{\"u}diger demonstrated that spinning test particles admit conserved energy $E$ and axial angular momentum $L_z$ for spinning test particles \cite{Rudiger1981, Rudiger1983_2}. At linear order in secondary spin, a Carter constant analogue ensuring integrability can be constructed \cite{Rudiger1983_2}; recent work extends this to quadratic order if the secondary body has the spin-induced quadrupole moment of a Kerr black hole \cite{2023CompereDruart}. Flux balance laws relate the radiated energy and angular momentum fluxes to the average rates of change in the secondary's conserved quantities along bound orbits \cite{Sago2005,Sago2006,Hughes2005,Isoyama2019}. For a spinning secondary, the balance laws governing the evolution of $E$ and $L_z$ are presented in Refs. \cite{Akcay2020,Mathews2022}, with the first computations of these fluxes along generic orbits conducted in Ref.\ \cite{Skoupy2023}. Recently, Piovano and collaborators used the Hamilton-Jacobi formalism from \cite{Witzany2019} to perform the generic flux calculation \cite{Piovano2024}. A key challenge in this field has been deriving flux balance laws associated with the Carter constant analogue for spinning particles. Grant recently provided expressions for this generalized Carter constant flux balance law \cite{Grant2024}; this can be combined with the closed-form expressions for gauge-invariant actions that are derived in Ref.\ \cite{Witzany2024} by Witzany and collaborators.  
 It was also proven in Ref.\cite{SkoupyWitzany2024_1} that the component of the spin parallel to the orbital angular momentum is conserved. A full scheme for generic 1PA-order waveform generation with a spinning secondary is outlined in Ref.\ \cite{Mathews2025}; quasi-circular waveforms for a slowly spinning primary and precessing secondary are presented in Ref.\ \cite{Mathews2025_2}.

\section{Motivation for this work}

In this paper, we introduce a \textit{shifted-geodesic} framework for modeling gravitational wave fluxes from a spinning secondary in an EMRI system. In this approximation, we shift the orbital frequencies and conserved quantities associated with a geodesic to mimic the effects of the smaller body's spin, offering a computationally efficient alternative to fully modeling the spinning-body dynamics. Adiabatic-order analyses of EMRI waveforms incorporate orbit-averaged, leading-order gravitational self-force effects, and techniques for modeling these effects are now well-established. This approach leverages the mature infrastructure used in the adiabatic-order analysis of gravitational wave fluxes, but extends it to incorporate the leading corrections due to secondary spin in a linear approximation. The approach we develop here fills a gap between adiabatic models (0PA), which neglect finite-size effects, and more sophisticated post-adiabatic models (1PA) that incorporate exact secondary spin corrections.

A major justification for the approach that we propose is that post-adiabatic corrections do not need to be computed to the same level of accuracy as leading-order adiabatic terms. Leading order adiabatic terms need to be known to a relative accuracy of $\sim \varepsilon$  \cite{Khalvati:2025znb}, while post-1-adiabatic corrections only need to be known to $\sim 10^{-2} - 10^{-3}$ \cite{Osburn:2015duj,Burke:2021xrg}. The spin of the secondary enters the gravitational-wave phase at post-adiabatic order through the conservative spin-curvature force $f_\text{SCF}$ and the dissipative dipole contribution to the stress-energy tensor $f_\text{dipole,diss}$ \cite{Warburton2017,Mathews2025,2025LRRWaveforms}:
\begin{equation}
\Phi_\text{GW} = \underset{\text{adiabatic: $f^{\alpha}_{\text{ad}}$}} {\underbrace{\varphi_0 \varepsilon^{-1}}} + \underset{\substack{\text{post-1-adiabatic:}\;  f^\alpha_{\text{SCF}} + \\ f^{\alpha}_{\text{oscil}} + f^{(1)\alpha}_{\text{cons}} +  f^{(2)\alpha}_{\text{diss}} +  f^{\alpha}_{\text{dipole,diss}} }} {\underbrace{\varphi_1 \varepsilon^0}}
\end{equation}
Because spinning-body effects are suppressed by a factor of the mass ratio $\varepsilon$ relative to the adiabatic contributions, they can be computed with correspondingly lower precision, typically by a factor of $\varepsilon\sim10^{-4}-10^{-5}$. The shifted-geodesic approximation captures the leading-order impact of spin-curvature coupling while avoiding the full complexity of self-consistent spinning-body models, enabling more accurate flux computations than purely adiabatic methods with significantly reduced computational cost.

The spin of the secondary introduces two types of effects: (i) dissipative corrections, which on average extract energy and other conserved quantities from the orbit, driving inspiral, and (ii) conservative corrections, which preserve conserved quantities but modify orbital properties such as frequencies. Some components of the spinning-secondary trajectory oscillate and average to zero over an orbit, while others, such as the leading dissipative term, accumulate secularly over many orbits. In the shifted geodesic approximation, we propose excluding the purely oscillatory components of the orbital spinning-secondary corrections in the GW flux calculation, arguing that these contributions are the least significant, while being the most expensive to compute. This expense arises because calculating the full spinning-body trajectory often requires evaluating a large number of Fourier coefficients, which can become a computational bottleneck in the flux calculation. Moreover, while generic inspirals are a key requirement for EMRI data analysis with the LISA mission, fully generic spinning-body orbits are significantly more complicated and slower to generate, requiring many evaluations across parameter space.

The advantages of this approach are its efficiency and simplicity. By modeling the system as an adiabatic inspiral with frequency shifts proportional to the secondary spin, we can leverage well-tested tools for adiabatic-order calculations, while including analytic corrections for the spinning-secondary frequencies and shifts to the integrals of motion. Incorporating these shifts directly to geodesic orbits preserves the essential features of secondary-spin-induced secular evolution while minimizing computational cost and complexity. We emphasize that while the shifted-geodesic approximation does not capture \textit{all} post-adiabatic effects, it captures the most important spinning-secondary effects across most of parameter space. We also emphasize that this calculation is not intended to replace full spinning-secondary analyses, but rather to provide an efficient complementary approximation. This model is well-suited for computationally intensive applications such as LISA data and science studies, where rapid evaluation of gravitational waveforms and efficient computation of large flux datasets are essential \cite{Chapman-Bird2025}.

The paper is organized as follows. In Sec.\ \ref{sec:ComputinggenericGWFluxes}, we review several approaches to computing the trajectory of a spinning body on generic orbits, including the frequency-domain method, the Hamilton-Jacobi formalism and a fully analytic technique based on a world-line shift. In Sec.\ \ref{sec:shiftedgeoframework}, we introduce the shifted-geodesic approximation for computing trajectories, which is the central focus of this paper. In Sec.\ \ref{sec:evalualtingSGfluxes}, we examine the impact of using the shifted-geodesic approximation in spinning-body gravitational-wave flux evaluations. Sec.\ \ref{sec:results} assesses the utility of this approximation and defines its applicable parameter space and Sec.\ \ref{sec:inspiral} presents the dephasing due to the approximation for an inspiral calculation. Finally, Sec.\ \ref{sec:conclusion} discusses the broader implications of our approach and outlines future directions for the model.

By developing this efficient method for computing gravitational wave fluxes in EMRIs with spinning secondaries, we aim to bridge the gap between simple geodesic models and more complex post-adiabatic frameworks, providing a tool that can be immediately used for GW data analysis while also serving as a foundation for more detailed studies of secondary spin effects in EMRIs \cite{Burke2024}.

\section{Computing spinning-body motion along a generic trajectory}
\label{sec:ComputinggenericGWFluxes}

The Mathisson-Papapetrou-Dixon (MPD) equations govern how a spinning body's internal structure couples to spacetime curvature, producing a conservative spin-curvature force that perturbs its trajectory and drives the precession of its spin vector. These equations are closed by a choice of spin supplementary condition (SSC); see \ref{sec:motionspin} for more details. For extreme mass-ratio systems, linearizing in the secondary spin simplifies the MPD system to
\begin{align}
    \frac{Du^\alpha}{d\tau} &\equiv f^\alpha_S= -\frac{1}{2\mu}{R^\alpha}_{\nu\lambda\sigma}u^\nu S^{\lambda\sigma}\;,
    \label{eq:spinforcelin}\\
    \frac{DS^\alpha}{d\tau} &= 0\;,
    \label{eq:spinpreclin}\\
    u_\alpha S^\alpha &= 0\;,
    \label{eq:spinsupplin}
\end{align}
where $u^\alpha$ is the four velocity $dx^\alpha/d\tau$ (with $\tau$ proper time measured by the orbiting body), ${R^\alpha}_{\nu \lambda _\sigma}$ is the Riemann tensor and $S^{\lambda\sigma}$ is the spin tensor for the secondary body, directly related to the spin vector $S^{\mu}$ via Eq.\ \ref{eq:spinvec}.  The operator $D/d\tau \equiv u^\gamma\nabla_\gamma$, where $\nabla_\gamma$ denotes a covariant derivative with respect to the coordinate $x^\gamma$.  When $DA^\alpha/d\tau = 0$, $A^\alpha$ is parallel transported along the trajectory whose tangent is the 4-velocity $u^\gamma$; a non-zero right-hand side describes a forcing term which pushes the evolution of $A^\alpha$ away from parallel transport. The magnitude of the spin vector is denoted $S\equiv \sqrt{S^\alpha S_\alpha}$, and we define two commonly-used dimensionless spin parameters $s = S/\mu^2$ ($0 \le s \le 1$) and $\sigma = S/(\mu M)$ ($0 \le \sigma \le \varepsilon$).  The secondary spin vector evolves according to parallel transport in Kerr spacetime \cite{Marck1983,vandeMeent2020} with explicit expressions for its components along the trajectory given in \ref{sec:motionspin}.

\subsection{Fixed turning-point gauge for a spinning test-body}
\label{sec:lineartrajfreq}

The choice of reference geodesic determines the ``spin gauge" or parameterisation used to describe the linearized  spinning-body trajectory as described by Eqs.\ (\ref{eq:spinforcelin})--(\ref{eq:spinpreclin}) (see Appendix A of Ref.\ \cite{Drummond2022_2} and Sec.\ III.\ D.\ in Ref.\ \cite{Piovano2024}). In this work, we adopt the \textit{fixed turning-point} gauge used in Refs.\ \cite{Drummond2022_1, Drummond2022_2, Skoupy2021, Skoupy2022, Skoupy2023, WitzanyPiovano2024, Piovano2024, Skoupy2025, Piovano:2025aro}, in which the reference geodesic is chosen to have the same averaged radial and polar turning points. This choice admits a quasi-Keplerian form in which the motion closely resembles geodesic motion but includes spin-dependent corrections. Importantly, this parameterization is especially useful in our context because it provides clear intuition for separating which corrections are purely oscillatory and which contribute secular growth.

The radial $r$ and polar $z\equiv\cos\theta$ trajectories take the form
\begin{align}
    r &= \frac{p M}{1+e\cos(\Upsilon_r \lambda + \delta\hat{\chi}_r(\lambda) + \delta\chi_r^S(\lambda))} + \delta\mathcalligra{r}^S(\lambda) \; ,\label{eq:spintrajectoryr} \\
    z &= \sin I \cos( \Upsilon_z \lambda + \delta\hat{\chi}_z(\lambda) + \delta\chi^S_z(\lambda) ) + \delta\mathcalligra{z}^S(\lambda) \; ,
    \label{eq:spintrajectorytheta}
\end{align}
with the frequencies shifted due to the secondary spin such that
\begin{align}
    \Upsilon_r = \hat{\Upsilon}_r + \Upsilon_r^S \; , \ \ \ \ \Upsilon_z = \hat{\Upsilon}_z + \Upsilon_z^S \; , \ \ \ \ \Upsilon_\phi = \hat{\Upsilon}_\phi + \Upsilon_\phi^S\;.
\end{align}
In these quantities and in what follows, a hat accent indicates a geodesic quantity and an ``$S$'' superscript denotes a spinning-secondary correction. Here $\Upsilon_r$, $\Upsilon_z$ and $\Upsilon_\phi$  are the radial, polar and azimuthal frequencies conjugate to Mino-time $\lambda$, which is defined by $d\lambda=d\tau/\Sigma$ and $\Sigma=r^2+a^2\cos^2\theta$. The quantities ($\delta\hat\chi_r, \delta\chi^S_r$) and $(\delta \hat\chi_z, \delta\chi^S_z)$ are purely oscillatory terms which contribute to the Darwin-Hughes anomaly angles, as defined in Refs.\ \cite{Drummond2022_2, Lynch:2024hco}. Finally,  $\delta\mathcalligra{r}^S(\lambda)$ and $\delta\mathcalligra{z}^S(\lambda)$ are purely oscillatory spinning-secondary corrections that modify the turning points/libration region of the trajectory.
 
 The spin-precession frequency conjugate to $\lambda$ is denoted by $\Upsilon_p$. The quasi-Keplerian parameters $p$, $e$ and $I$ denote semi-latus rectum, eccentricity and inclination respectively. We also define $x_I\equiv \cos I$.  For convenience, we also define
\begin{align}
    \Upsilon_r^S = \sigma\delta\Upsilon_r^S \; , \ \ \ \ \Upsilon_z^S = \sigma\delta\Upsilon_z^S \;, \ \ \ \ \Upsilon_\phi^S = \sigma\delta\Upsilon_\phi^S\;.
\end{align}
Note that $(\Omega_r,\Omega_z,\Omega_\phi)$ are the radial, polar and azimuthal frequencies conjugate to Boyer-Lindquist coordinate time $t$; the quantity $\Gamma$ transforms between the coordinate-time frequencies and the Mino-time frequencies via $\Omega_{r,\theta,\phi}=\Upsilon_{r,\theta,\phi}/\Gamma$.

The linearized equations of motion admit conserved integrals: energy $E$, axial angular momentum $L_z$, and a Carter-like constant $Q$ or $K \equiv Q + (L_z - aE)^2$.  They are offset from geodesic values by terms proportional to $\sigma$, such that
\begin{align}
\label{eq:CoM}
E &= \hat E + \sigma\delta E^S,\quad
L_z = \hat L_z + \sigma\delta L_z^S\,\quad
Q = \hat Q + \sigma\delta Q^S\;.
\end{align}
The quantities $\delta\mathcalligra{r}_S$ and $\delta\mathcalligra{z}_S$ encode spin corrections and are periodic in the spin-corrected frequencies.  The anomaly angles are also shifted from their geodesic counterparts by $\delta\chi_r^S$ and $\delta\chi_z^S$.  See Refs.\ \cite{Drummond2022_1, Drummond2022_2} for further notational details.

The conserved quantities and frequencies of bound spinning-body orbits can be defined in \textit{any} spin gauge or parameterization. Whether they exhibit shifts relative to their geodesic values depends on how the reference geodesic is specified. The Hamilton-Jacobi formulation itself is spin-gauge-independent, but different studies adopt different parameterizations. For instance, Refs.\ \cite{Witzany2019, Witzany2024} employed a \textit{fixed constants-of-motion} parameterization, where shifts are measured relative to a geodesic with the same $(E,L_z,Q)$.  By contrast, Refs.\ \cite{Piovano2024,Skoupy2025,Piovano:2025aro} used the \textit{fixed turning-points} gauge. As shown in the Appendices of Ref.\ \cite{Piovano2024}, one can transform trajectories between these gauges straightforwardly. A third useful choice is the \textit{fixed-frequency} gauge, in which the reference geodesic is chosen to match the orbital frequencies of the spinning-body orbit. Finally, Ref.\ \cite{Piovano:2025aro} introduces a \textit{fixed-eccentricity} gauge which is useful for avoiding divergences in certain quantities at the last stable orbit (LSO). In practice, it is straightforward to translate between these parameterizations and any of them may be adopted within a given framework.

We define the phases $w_r = \Upsilon_r \lambda$, $w_z = \Upsilon_z \lambda$ and $ w_p = \Upsilon_p \lambda$. Linearizing the trajectory's $r$ and $z$ coordinate motions in secondary spin yields
\begin{align}
r(w_r,w_z,w_p) &= \hat{r}(w_r) + r^S(w_r,w_z,w_p)\;,
\\
z(w_r,w_z,w_p) &= \hat{z}(w_z) + z^S(w_r,w_z,w_p)\;,
\end{align}
where $\hat r(w_r)$ and $\hat z(w_z)$ denote the functional forms for a geodesic orbit (however, note that they depend on the \textit{spinning-orbit} mean anomalies). The linear-in-spin contributions to these trajectories are expressed as
\begin{align}
    r^S &= \frac{e pM \delta\chi_r^S \sin(w_r + \delta\hat{\chi}_r)}{(1+e\cos(w_r + \delta\hat{\chi}_r))^2} + \delta \mathcalligra{r}^S \; , \\
    z^S &= -\sin I \delta\chi_z^S \sin(w_z + \delta\hat{\chi}_z) + \delta\mathcalligra{z}^S \; .
\end{align}
The terms $r^S$ and $z^S$ encode entirely oscillatory corrections beyond the secular effects arising from frequency shifts.  Expressions for the coordinate time $t$ and azimuthal angle $\phi$ are obtained by integrating the corresponding components of the four-velocity.  Starting with
\begin{equation}
u_t = -\hat{E} + u_t^S(\lambda)\;,\qquad
u_\phi = \hat{L}_z + u_\phi^S(\lambda)\;,
\end{equation}
we raise indices as usual: $u^t = g^{tt}u_t + g^{t\phi}u_\phi$, $u^t = g^{\phi t}u_t + g^{\phi\phi}u_\phi$.  We then change our time parameter from proper time $\tau$ to Mino time $\lambda$:
\begin{equation}
U^\alpha \equiv \frac{\dd x^\alpha}{\dd\lambda} = \Sigma \frac{\dd x^\alpha}{\dd\tau} = \Sigma\,u^\alpha\;.
\end{equation}
The components $U^t$ and $U^\phi$ are then integrated to yield
\begin{align}
t &= \Gamma \lambda + \Delta t(\Upsilon_r\lambda,\Upsilon_z\lambda,\Upsilon_p\lambda)\;, \\
\phi &= \Upsilon_\phi \lambda + \Delta\phi(\Upsilon_r\lambda,\Upsilon_z\lambda,\Upsilon_p\lambda)\;.
\end{align}
The coefficients $\Gamma$ and $\Upsilon_\phi$ describe the secular accumulation of $t$ and $\phi$ as functions of $\lambda$, while $\Delta t$ and $\Delta\phi$ collect all purely oscillatory pieces.  To obtain these oscillatory pieces, we decompose the associated components of $U^\alpha$ into geodesic and linear-in-spin parts,
\begin{equation}
U^t = \hat{U}^t + U^t_S \; , \qquad
U^\phi = \hat{U}^\phi + U^\phi_S \; .
\end{equation}
The spin-dependent contributions $U^t_S$ and $U^\phi_S$ can be written as perturbations of the geodesic potentials with respect to both the orbital coordinates and the conserved quantities. Explicitly, they take the form [see Eq.\ (36) of Ref.\ \cite{Skoupy2023}]
\begin{align}\label{eq:UtS}
U^t_S &= \pdv{V^t}{r} r^S + \pdv{V^t}{z} z^S
        - \pdv{V^t}{E} u_t^S + \pdv{V^t}{L_z} u_\phi^S \; , \\
\label{eq:UphiS} U^\phi_S &= \pdv{V^\phi}{r} r^S + \pdv{V^\phi}{z} z^S
        - \pdv{V^\phi}{E} u_t^S + \pdv{V^\phi}{L_z} u_\phi^S  \; ,
\end{align}
where the geodesic potentials $V^t$ and $V^\phi$ are given by
\begin{align}
V^t &= \frac{r^2+a^2}{\Delta}\!\left[(r^2+a^2)E-aL_z\right]
      - a^2E(1-z^2)+aL_z \; , \\
V^\phi &= \frac{a}{\Delta}\!\left[(r^2+a^2)E-aL_z\right]
      + \frac{L_z}{1-z^2}-aE  \; .
\end{align}
The oscillatory pieces $\Delta t$ and $\Delta\phi$ can then be obtained by Fourier-expanding $U^t$ and $U^\phi$ (in Mino time) and integrating each harmonic mode. This procedure is described in detail in Ref.~\cite{Drummond2022_1}; we summarize the resulting frequency-domain formulation in the following section.

\subsection{Comparing approaches for generic spinning test-body motion}
\label{sec:HamiltonJacobi}

We now compare three approaches to computing generic spinning-body trajectories in Kerr: (i) the frequency-domain method, (ii) the Hamilton-Jacobi formulation, and (iii) the worldline-shift approach. Each method offers a different trade-off between computational efficiency, analytic tractability and numerical precision.

Unlike the bound Kerr geodesic case, the radial and polar motions of a spinning secondary do not decouple to evolve with single frequencies $\Upsilon_r$ and $\Upsilon_z$ respectively. Instead, they receive contributions from linear combinations of the form $n^s \Upsilon_r + k^s \Upsilon_z + j \Upsilon_p$, with integer indices $n^s, k^s,$ and $j$. Hence, the frequency-domain method of Refs.\ \cite{Drummond2022_1,Drummond2022_2} expands all unknowns in Eqs.\ (\ref{eq:spintrajectoryr})--(\ref{eq:spintrajectorytheta}) as Fourier series,
\begin{equation}
f(\lambda) = \sum_{n^s,k^s,j} f_{n k j}, e^{-i n^s \Upsilon_r \lambda - i k^s \Upsilon_z \lambda - i j \Upsilon_p \lambda} \; ,
\end{equation}
which are then substituted into Eq.\ (\ref{eq:spinforcelin}) to yield a second-order system solved in the frequency domain. The summation ranges depend on which function $f$ we are expanding. For $\delta\chi_r^S$, the sum runs only over positive and negative $n^s$; for $\delta\chi_z^S$, it runs over positive and negative $k^s$. In the case of $\mathcalligra{r}^S$, the indices $k^s$ and $j$ cannot both vanish, while for $\mathcalligra{z}^S$, the same restriction applies for  $n^s$ and $j$. Formally, the $n^s$ and $k^s$ sums go to $\pm \infty$; in practice, they are truncated at values $\pm n^s_{\rm max}$ and $\pm k^s_{\rm max}$ which are determined empirically, as discussed in detail in Refs.\ \cite{Drummond2022_1,Drummond2022_2}.  The index $j$ ranges from $-1$ to $1$. 

The first step in the frequency-domain method is to isolate the purely oscillatory contributions to $\Delta t$ and $\Delta\phi$ by Fourier decomposing the Mino-time four-velocity components and integrating mode-by-mode. We write the $t$-component of the Mino-time four-velocity as
\begin{equation}\label{eq:dtdlambda}
\dv{t}{\lambda} \equiv U^t(\lambda)
= \sum_{n^s,k^s,j} U^t_{n^s k^s j}\,
e^{-i\,(n^s\Upsilon_r + k^s\Upsilon_z + j\Upsilon_p)\lambda}\,;
\end{equation}
an analogous form describes $U^\phi \equiv d\phi/d\lambda$.  Separating $t(\lambda)$ into a secularly growing piece and an oscillatory remainder,
\begin{equation}
t(\lambda)=\Gamma\,\lambda+\Delta t(\lambda),
\end{equation}
the oscillatory part inherits the harmonic structure,
\begin{equation}\label{eq:DeltatFourier}
\Delta t(\lambda)
= \sum_{n^s,k^s,j}
\Delta t_{n^s k^s j}\,
e^{-i\,(n^s\Upsilon_r + k^s\Upsilon_z + j\Upsilon_p)\lambda}\, .
\end{equation}
Integrating Eq.~\eqref{eq:dtdlambda} mode-by-mode then gives
\begin{equation}\label{eq:dtnkj}
\Delta t_{n^s k^s j}
= \frac{iU^t_{n^s k^s j}}
{\,(n^s\Upsilon_r + k^s\Upsilon_z + j\Upsilon_p)} \, ,
\end{equation}
where $U^t_{n^s k^s j}$ is the corresponding harmonic mode of $U^t$.  A similar analysis relates $\Delta\phi_{n^s k^s j}$ to $U^\phi_{n^s k^s j}$. Expanding to linear order in the secondary spin yields the spin-dependent corrections $\Delta t^S_{n^s k^s j}$ and $\Delta\phi^S_{n^s k^s j}$ in terms of the Fourier modes $U^{t,S}_{n^s k^s j}$ and $U^{\phi,S}_{n^s k^s j}$, together with the geodesic modes $\hat U^t_{n^s k^s j}$ and $\hat U^\phi_{n^s k^s j}$.

The covariant four-velocity component corrections $u^{S}_{t,n^s k^s j}$ and $u^{S}_{\phi,n^s k^s j}$ are related to the contravariant Mino-time components $U^t_S$ and $U^\phi_S$ through Eqs.\ (\ref{eq:UtS}) and (\ref{eq:UphiS}). Their Fourier coefficients are obtained from the corresponding source terms via
\begin{align}
u^{S}_{t,n^s k^s j}
&= \frac{i\,\mathcal{R}_{t,n^s k^s j}}
{n^s \hat{\Upsilon}_r + k^s \hat{\Upsilon}_\theta + j \Upsilon_p} \; ,
\qquad
u^{S}_{\phi,n^s k^s j}
= \frac{i\,\mathcal{R}_{\phi,n^s k^s j}}
{n^s \hat{\Upsilon}_r + k^s \hat{\Upsilon}_\theta + j \Upsilon_p} \; ,
\label{eq:uSfourier}
\end{align}
where $\mathcal{R}_{t,n^s k^s}$ and $\mathcal{R}_{\phi,n^s k^s}$ are the Fourier coefficients of the functions $\mathcal{R}_t=\Sigma f^S_t$ and $\mathcal{R}_\phi=\Sigma f^S_\phi$ as defined in Appendix~C of Ref.~\cite{Skoupy2023}. 
These expressions break down at resonances where $n^s \hat{\Upsilon}_r + k^s \hat{\Upsilon}_\theta + j \Upsilon_p = 0$. In practice, such resonant configurations form a set of measure zero in parameter space and can be avoided when interpolating over grid points that do not lie on the resonance surface. Nevertheless, resonances represent a general limitation of these perturbative constructions and must be treated separately in fully self-consistent inspiral evolutions.

The remaining Fourier coefficients
$u^S_{t,00}$, $u^S_{\phi,00}$, $\delta \chi^S_{r,n^s}$, $\delta \chi^S_{z,k^s}$,
$r^S_{n^s k^s}$, $z^S_{n^s k^s}$,
as well as the frequency shifts $\Upsilon^S_{r}$ and $\Upsilon^S_{z}$
are obtained by solving a linear system
\begin{equation}
\mathbf{M}\cdot\mathbf{v}+\mathbf{c}=0\; ,
\end{equation}
with $\mathbf{v}$ the unknown coefficients, $\mathbf{c}$ built from Fourier terms independent of them, and $\mathbf{M}$ assembled from the functions listed in Ref.~\cite{Drummond2022_1}.
 These functions depend only on geodesic quantities and are presented in the Supplemental Material of Ref.~\cite{Drummond2022_1}. The spin corrections to the conserved quantities $(E,L_z,Q)$ depend on $u^S_{t,00}$, $u^S_{\phi,00}$, as well as the perturbed trajectory $(r,z)$. This framework enabled the first computation of fluxes from generic spinning-body orbits \cite{Skoupy2023}. However, the need to solve the full second-order system together with large Fourier decompositions renders this approach rather slow.

\begin{table}[htbp]
\centering
\caption{Methods for solving the generic linearized spinning-body trajectory.}
\label{tab:lincompare}

\renewcommand{\arraystretch}{1.7} 
\resizebox{0.95\textwidth}{!}{
\begin{tabular}{m{4cm}!{\vrule width 2pt}m{3.6cm}:m{3.6cm}:m{3.6cm}}

& 
\small{\textbf{Frequency-domain} ($2$nd-order equation)}
& 
\small{\textbf{Hamilton-Jacobi} ($1$st-order equation)}
&
\small{\textbf{Wordline-shift} \hspace{0.3cm}(Fully analytic)}
\\ \hline

\vspace{0.2cm}
\begin{tikzpicture}

\node[draw, rotate=90, font=\footnotesize\itshape, anchor=north, align=center] at (-2.25, 0) {Orbital\\[-0.4em] kinematics};

\fill[black] (0,0) circle (0.38cm);
\draw[dashed] (0,0) ellipse [x radius=1.2cm, y radius=0.57cm];
\fill[gray] (0.85,-0.38) circle (0.15cm);
\draw[-{Stealth}, thick] (0.9,-0.6) .. controls (0.85,-0.7) and (0.1,-0.8) .. (0.1,-0.8);
\end{tikzpicture} & \footnotesize{Drummond \& Hughes, 2022a,b,  Refs.\ \cite{Drummond2022_1,Drummond2022_2}} & \footnotesize{Witzany, 2019, Ref.\ \cite{Witzany2019_2}} & \footnotesize{Skoup\'y  \& Witzany, 2024, Ref.\ \cite{SkoupyWitzany2024_2}} \\ \hdashline
\vspace{0.2cm}
\begin{tikzpicture}

\node[draw, rotate=90, font=\footnotesize\itshape, anchor=north] at (1.8, 0) {GW fluxes};

\fill[black] (4,0) circle (0.38cm);
\draw[dashed] (4,0) ellipse [x radius=1.2cm, y radius=0.57cm];
\fill[gray] (4.85,-0.38) circle (0.15cm);
\begin{scope}[rotate=30,shift={(0.35cm, -1.35cm)}]
\draw[decorate, decoration={snake, amplitude=0.05cm, segment length=0.3cm}] (4.3,-0.8) -- (5.1,-0.8);
\draw[decorate, decoration={snake, amplitude=0.05cm, segment length=0.3cm}] (4.3,-0.9) -- (5.1,-0.9);
\draw[decorate, decoration={snake, amplitude=0.05cm, segment length=0.3cm}] (4.3,-1.0) -- (5.1,-1.0);
\end{scope}
\end{tikzpicture}  & \footnotesize{Skoup\'y et al., 2023, Ref.\ \cite{Skoupy2023}} & \footnotesize{Piovano et al., 2024, Ref.\ \cite{Piovano2024}} & \raisebox{-25pt}{\begin{tikzpicture}
\fill[pattern=north east lines, pattern color=gray!70] (0,0) rectangle (3.6,1.95);
\end{tikzpicture}} \\ \hdashline
\vspace{0.2cm}
\begin{tikzpicture}

\node[draw, rotate=90, font=\small\itshape, anchor=north] at (5.75, 0) {Inspiral};

\fill[black] (8,0) circle (0.38cm);
\draw[dashed] (8,0) ellipse [x radius=1.2cm, y radius=0.57cm];
\draw[dashed,color=gray!60] (8,0) ellipse [x radius=0.85cm, y radius=0.42cm];
\fill[gray] (8.7,-0.2) circle (0.15cm);
\end{tikzpicture} & \raisebox{-25pt}{\begin{tikzpicture}
\fill[pattern=north east lines, pattern color=gray!70] (0,0) rectangle (3.6,1.7);
\end{tikzpicture}} &  \raisebox{-25pt}{\begin{tikzpicture}
\fill[pattern=north east lines, pattern color=gray!70] (0,0) rectangle (3.6,1.7);
\end{tikzpicture}} &\raisebox{-25pt}{\begin{tikzpicture}
\fill[pattern=north east lines, pattern color=gray!70] (0,0) rectangle (3.6,1.7);
\end{tikzpicture}}  \\ \hline \hline

\small{\textit{\textbf{Unknowns to solve:} \ \ \ \ \ \ \ \ \ \ \ \ \ \ \ Fourier coefficients}} & 
$(\delta\mathcalligra{r}^S)_{nkj}$, $(\delta\mathcalligra{z}^S)_{nkj}$,
$(\delta\chi_r^S)_{nkj}$, \ \ $(\delta\chi_z^S)_{nkj}$ & 
\ \ \  $(\xi_r)_{nkj}$, \   $(\xi_z)_{nkj}$ & 
\raisebox{-10pt}{\begin{tikzpicture}
\fill[pattern=north east lines, pattern color=gray!70] (0,0) rectangle (3.6,1.1);
\end{tikzpicture}}\\ \hdashline
\small{\textit{\textbf{Unknowns to solve:} \ \ \ \ \ \ \ \ \ \ \ \ \ \ \ Orbital constants}} & 
\multicolumn{3}{c}{$\delta\Upsilon_i\equiv\{\delta\Upsilon_r^S$, $\delta\Upsilon_z^S$, $\delta\Upsilon_\phi^S$, $\Upsilon_{s}$\}; $\delta C_i\equiv\{\delta E^S$, $\delta L_z^S$, $\delta K^S\}$} \\
\end{tabular}}
\vspace{0.5em} 
\parbox{0.9\textwidth}{\footnotesize
\textit{Note.} By ``generic'' we mean orbits that are both eccentric and inclined. Circular \cite{Piovano2020}, equatorial \cite{Skoupy2022}, and spherical \cite{Skoupy2025} inspirals of spinning bodies have been computed; in this table we focus on fully generic trajectories.
}
\end{table}

The Hamilton-Jacobi formulation of spinning-body motion, introduced by Witzany in Ref.\ \cite{Witzany2019} and expanded on in detail by Piovano et al.\ in Ref.\ \cite{Piovano2024}, uses the separability of parallel transport to recast the equations of motion as a first-order system:
\begin{align} \frac{dr}{d\lambda} &= \pm \Delta \sqrt{w_r^{\prime 2} - e_{0r} e^{\kappa}_{C;r} e_{\kappa D}\tilde{s}^{CD}}\; , \label{eq:drdlambdaWitzany} \\ \frac{d\theta}{d\lambda} &= \pm \sqrt{w_{\theta}^{\prime 2} - e_{0\theta} e^{\kappa}_{C;\theta} e_{\kappa D}\tilde{s}^{CD}}\; , \label{eq:dthetadlambdaWitzany} \\ \frac{d\psi_p}{d\lambda} &= \sqrt{K_c}\left(\frac{(r^2 + a^2)E_c - aL_{c}}{K_c + r^2} + a\frac{L_{c} - a(1-z^2)E_c}{K_c - a^2 z^2}\right)\; , \label{eq:paralleltransport2} 
\end{align}
with $\Delta = r^2 - 2Mr + a^2$. Here $w_r'$ and $w_\theta'$ are given by Eqs.\ (26) and (27) of Ref.\ \cite{Witzany2019}; the tetrad $e_{\mu}^A$ is defined in Eqs.\ (11)--(21) of the same reference.  The set $(E_c, L_c, K_c)$ are the background geodesic congruence constants used to define $e_{\mu}^A$. The spin-precession phase $\psi_p$ is related to $w_p$ by Eq.\ (3.47) of Ref.\ \cite{Piovano2024}. 

In this description, the radial and polar motions remain formally coupled, but only through the spin coefficients $\mathcal{S}_r$ and $\mathcal{S}_z$, which are known in closed form (see Appendix C of Ref.\ \cite{Piovano2024}). This limited coupling makes the system tractable, enabling analytic determination of turning points and quadrature expressions for frequency shifts,
\begin{equation}
\label{eq:HJfreq}
\delta\Upsilon_{y}^S = \frac{\hat\Upsilon_{y}}{(2\pi)^3}\int_0^{2\pi}\int_0^{2\pi}\int_0^{2\pi}
\frac{\delta \mathcal{Y}_y(w_r,w_z,w_p)}{\mathcal{Y}_{yg}(w_y)} \text{d} w_r \text{d} w_z \text{d} w_p\;, \quad  
\end{equation}
where $y=r,z$ and where $\delta \mathcal{Y}_y/\mathcal{Y}_{yg}$ is a function defined in Eqs.\ (3.37)--(3.40) of Ref.\ \cite{Piovano2024}.\footnote{Note we use $\mathcal{Y}$ here as opposed to $Y$ as in Ref.\ \cite{Piovano2024}, in order to distinguish from the $Y$ variable used in Eq.\ \ref{eq:transformation1}.} Relative to the frequency-domain method, the Hamilton-Jacobi formulation requires Fourier expansions of only two quantities, $\xi_r$ and $\xi_z$ (defined in Eq.\ \ref{eq:xirz}), significantly improving computational efficiency due to its semi-analytic tractability. See \ref{sec:altmethoddetails} for more details on the Hamilton-Jacobi approach to spinning-body motion.

Skoup\'y and Witzany introduced a novel coordinate transformation which maps the spinning-body worldline $x^\mu$ to a virtual geodesic $\tilde{x}^\mu = x^\mu + \delta x^\mu$, with $\delta x^\mu = \mathcal{O}(s)$ expressed in terms of basis vectors $\delta x^\mu_a$ ($a = 1, 2, 3$) defined in Eqs.\ (14b)--(14c) of Ref.\ \cite{SkoupyWitzany2024_2}. An auxillary vector $\Tilde{v}^\mu$ is defined in Eqs.\ (22a)--(22d) of Ref.\ \cite{SkoupyWitzany2024_2}. The transformed motion is governed by equations formally identical to geodesics but with shifted constants $(\tilde{E},\tilde{J}_z,\tilde{K})$ and with the four-velocity defined as $\Tilde{v}^\mu$. These equations are given in Eqs.\ (23a)--(23d) of Ref.\ \cite{SkoupyWitzany2024_2}. This method yields a complete analytic description of spinning-body orbits, making it the most efficient developed so far.

Table \ref{tab:lincompare} summarizes and compares the three approaches. The frequency-domain method is the most conceptually straightforward, requiring only the standard second-order form of the MPD equations together with a Fourier series ansatz for the trajectory.  It is also the least efficient, as it requires computing Fourier coefficients for at least four quantities; see the first column of Table \ref{tab:lincompare}. The Hamilton-Jacobi method transforms the system into a set of first-order equations and reduces the Fourier decompositions to only two quantities, thereby improving efficiency (see second column of Table \ref{tab:lincompare}).  The fully analytic worldline-shift method of Ref.\ \cite{SkoupyWitzany2024_2} is the most efficient (see third column of Table \ref{tab:lincompare}).  No GW flux implementations using this approach have been presented yet.

\section{Computing spinning-body fluxes using the shifted-geodesic method}
\label{sec:shiftedgeo}

The speed of computation of the spinning-secondary corrections to GW amplitudes and fluxes is limited by the time needed to calculate the spinning-body trajectory presented in Sec.\ \ref{sec:ComputinggenericGWFluxes}. This motivates an efficient approximate model for the trajectory, which we call the \textit{shifted geodesic} approximation. In this approach, we approximate the spinning-body motion as a geodesic orbit, but with its frequencies and integrals of motion shifted to match those of a spinning body. This allows us to leverage existing infrastructure for computing adiabatic GW fluxes along geodesic orbits, while capturing the dominant secular effects due to the secondary's spin.

We propose that the shifted geodesic method offers a simple and computationally efficient alternative to the approaches discussed in Section \ref{sec:HamiltonJacobi}. It avoids the evaluation of oscillatory corrections to the trajectory, using only shifts to orbital constants. Since all methods require computing corrections to these constants (as shown in the bottom row of Table \ref{tab:lincompare}), the shifted geodesic method is particularly efficient because it only requires these constant expressions, which have closed-form analytic solutions and do not involve computing large numbers of Fourier coefficients. This makes it both as fast as the fully analytic formulation and conceptually straightforward to implement. In practice, we combine elements of the existing formalisms: the Hamilton-Jacobi framework is used to compute frequency corrections and shifts to the constants of motion, while the frequency-domain method is employed to obtain the $t$ and $\phi$ corrections, which are less expensive since they are governed by first-order differential equations. 

However, this efficiency comes at the cost of reduced accuracy \textit{which must be carefully assessed for its impact on EMRI waveform models}.  One of the main goals of this paper is to undertake a first-cut assessment of this kind. While the shifted geodesic method provides a rapid and convenient way to compute spinning-body orbits, we emphasize that it is {\it not} intended to replace more accurate approaches. We instead propose it as a tool to be used when computational speed and ease of implementation can be prioritized over full accuracy.  Our results demonstrate its effectiveness across much of the extreme mass-ratio parameter space, though further work is needed to assess how its reduced accuracy impacts EMRI parameter inference.

\subsection{Framework for shifted geodesic trajectories}
\label{sec:shiftedgeoframework}

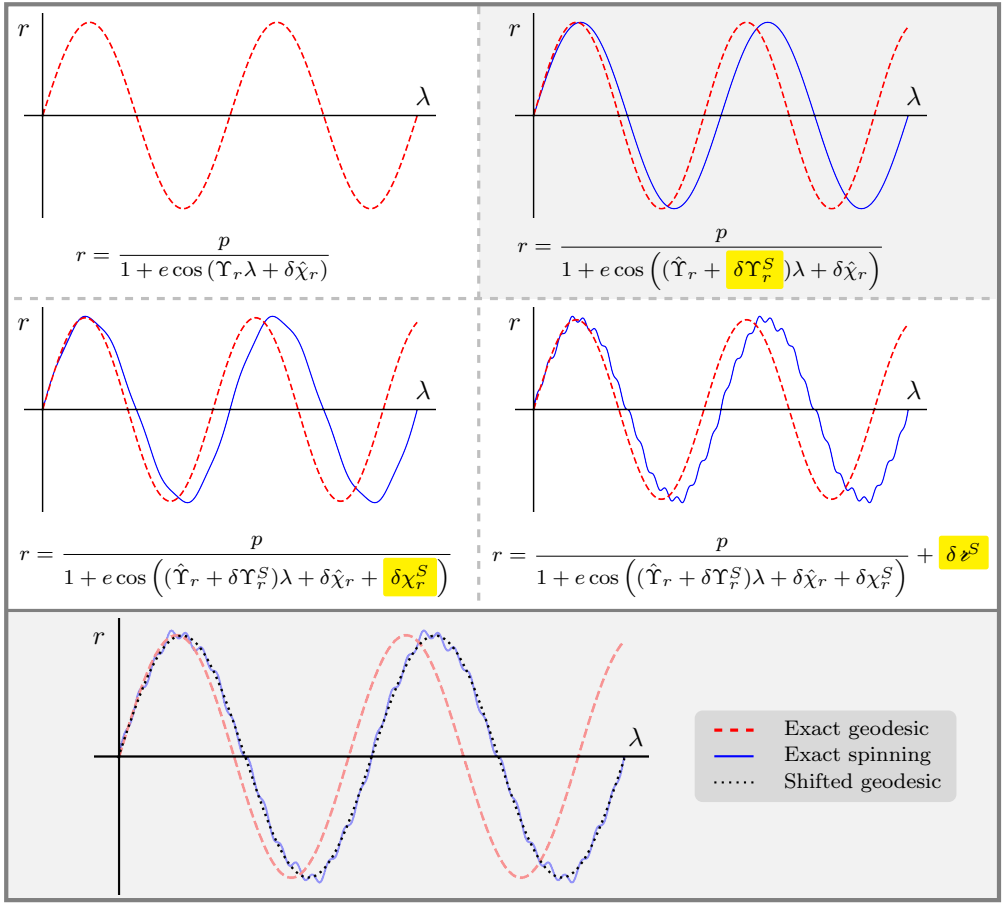
\begin{figure}[htbp]
\centering
\begin{tikzpicture}
  \begin{scope}[scale=0.87]

    \fill[gray!10] (4.9,-2.8) rectangle (12.8,1.69);

    \fill[gray!10] (-2.32,-12) rectangle (12.85,-7.57);

    \draw[dashed, lightgray, very thick] (4.9, 1.67) -- (4.9, -7.54);
    \draw[dashed, lightgray, very thick] (-2.2, -2.8) -- (12.75, -2.8);

    \draw[ultra thick,gray] (-2.32, -12) rectangle (12.85, 1.7);

    \node at (1.1,0) {\includegraphics[width=0.348\textwidth]{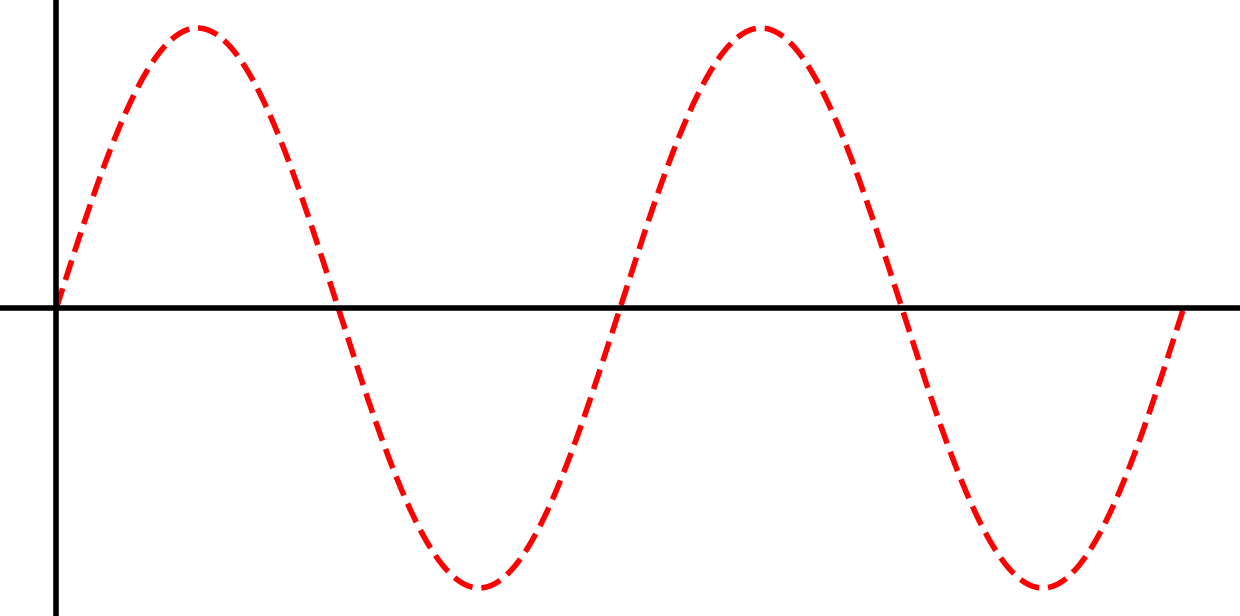}};
    \node at (8.6,0) {\includegraphics[width=0.348\textwidth]{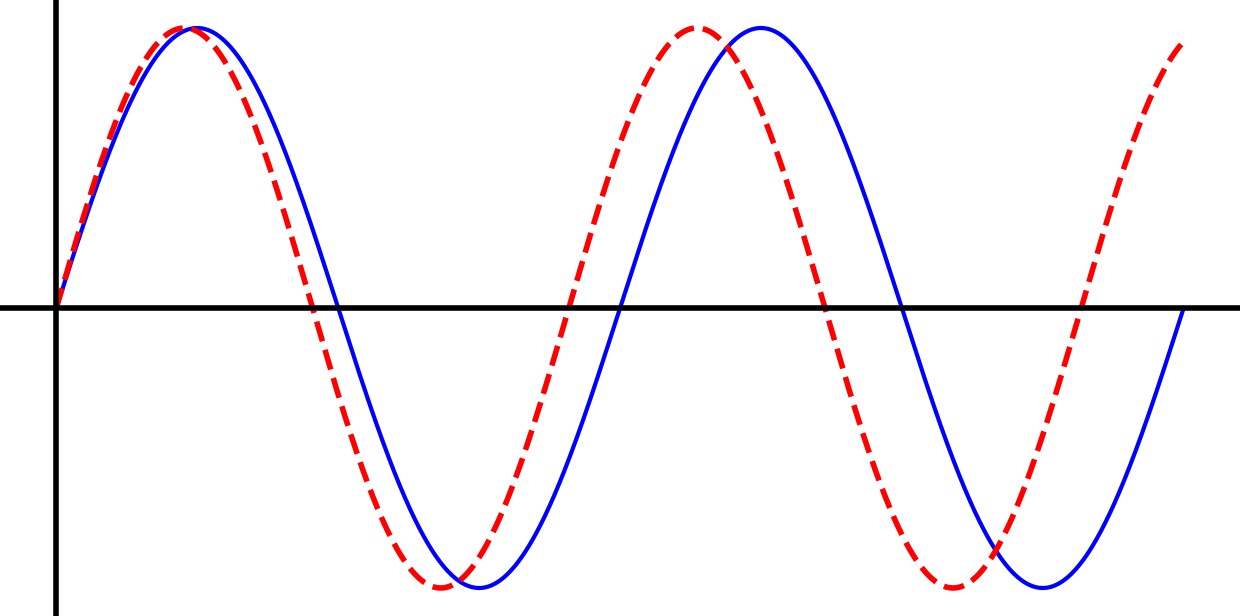}};
    \node at (1.1,-4.5) {\includegraphics[width=0.348\textwidth]{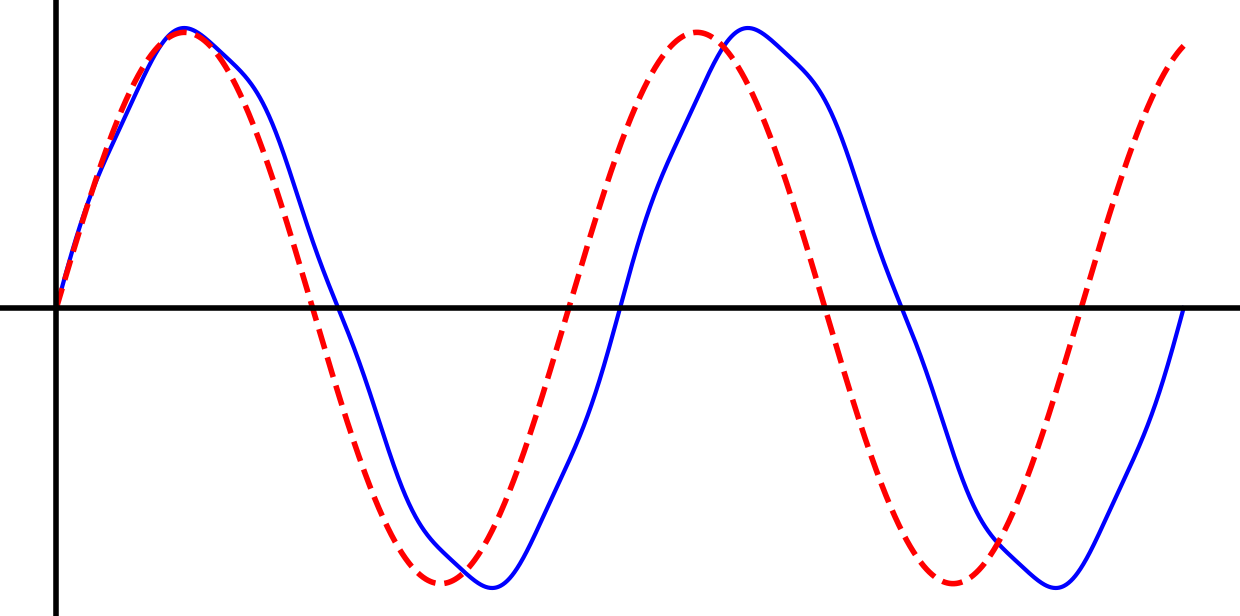}};
    \node at (8.6,-4.5) {\includegraphics[width=0.348\textwidth]{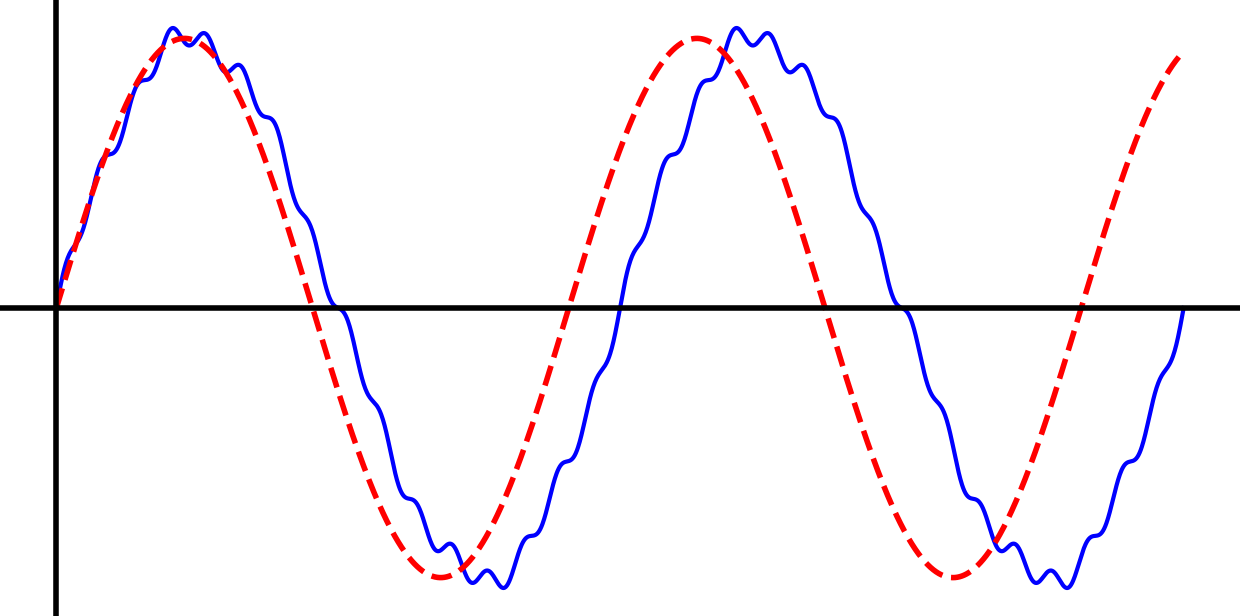}};

    \node at (3.26,-9.8) {\includegraphics[width=0.47\textwidth]{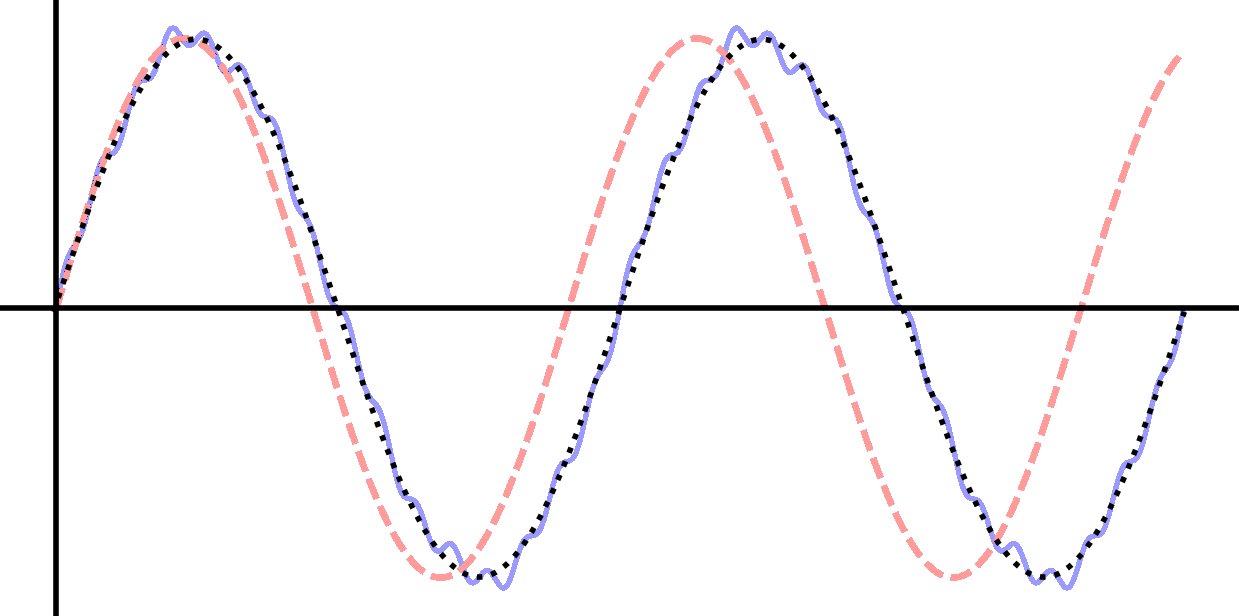}};

    \node at (-2.05,1.3)              {\footnotesize{$r$}};
    \node at (4.05,0.3)               {\footnotesize{$\lambda$}};
    \node at (5.45,1.3)               {\footnotesize{$r$}};
    \node at (11.55,0.3)              {\footnotesize{$\lambda$}};
    \node at (-2.05,-3.2)             {\footnotesize{$r$}};
    \node at (4.05,-4.2)              {\footnotesize{$\lambda$}};
    \node at (5.45,-3.2)             {\footnotesize{$r$}};
    \node at (11.55,-4.2)            {\footnotesize{$\lambda$}};
    \node at (-0.9,-8)           {\footnotesize{$r$}};
    \node at (7.3,-9.5)           {\footnotesize{$\lambda$}};

    \node at (0.7,-2.2) {\scriptsize{$r=\dfrac{p}{1+e\cos\left(\Upsilon_r\lambda+\delta\hat\chi_r\right)}$}};
    \node at (8.3,-2.2) {
      \scriptsize{
        $r = \dfrac{p}{1 + e\cos\left((\hat\Upsilon_r + \tikz[baseline] \node[fill=yellow, anchor=base, rounded corners=1pt] {$\delta\Upsilon_r^S$}; )\lambda + \delta\hat\chi_r\right)}$
      }
    };
    \node at (1.2,-6.9) {
      \scriptsize{
        $r=\dfrac{p}{1+e\cos\left((\hat\Upsilon_r+\delta\Upsilon_r^S)\lambda+\delta\hat\chi_r+\tikz[baseline] \node[fill=yellow, anchor=base, rounded corners=1pt] {$\delta\chi_r^S$};\right)}$
      }
    };
    \node at (8.9,-6.9) {
      \scriptsize{
        $r=\dfrac{p}{1+e\cos\left((\hat\Upsilon_r+\delta\Upsilon_r^S)\lambda+\delta\hat\chi_r+\delta\chi_r^S\right)}+
        \tikz[baseline]{\node[fill=yellow, anchor=base, rounded corners=1pt] {$\delta\mathcalligra{r}^S$};}$
      }
    };

    \draw[solid, gray, very thick] (-2.3, -7.57) -- (12.85, -7.57);

    \begin{scope}[shift={(8.5,-10.3)}]
      \fill[gray!30, rounded corners=3pt] (-0.3,1.2) rectangle (3.7,-0.2);
      \draw[red, dashed, very thick] (0,0.9) -- (0.6,0.9);
      \node[anchor=west] at (0.9,0.9) {\scriptsize Exact geodesic};
      \draw[blue, thick] (0,0.5) -- (0.6,0.5);
      \node[anchor=west] at (0.9,0.5) {\scriptsize Exact spinning};
      \draw[black, dotted, thick] (0,0.1) -- (0.6,0.1);
      \node[anchor=west] at (0.9,0.1) {\scriptsize Shifted geodesic};
    \end{scope}
    
  \end{scope}
\end{tikzpicture}
\caption{Schematic illustration of the progression from a reference geodesic trajectory to the full spinning-body trajectory for the radial coordinate. The trajectories shown are representative cartoons rather than physical trajectories; for realisic parameter choices, the spin-induced shifts too small to be visible over a small number of cycles. The top left panel shows the reference geodesic orbit (red dashed), with unshifted radial frequency $\Upsilon_r$. The top right panel (shaded gray) depicts the \emph{shifted geodesic} approximation (solid blue), where the radial frequency is modified by the secular shift $\delta \Upsilon_r^S$, capturing the dominant long-term effect of spin. The middle left panel includes both the secular frequency shift $\delta \Upsilon_r^S$ and the purely oscillatory correction $\delta \chi_r^S$, introducing periodic deviations around the shifted geodesic. The middle right panel shows the full spinning-body trajectory, incorporating all $\mathcal{O}(\sigma)$ corrections: the secular shift $\delta \Upsilon_r^S$, the oscillatory term $\delta \chi_r^S$ and the libration correction $\delta \mathcalligra{r}^S$, which accounts for coupling between radial and polar motion. In the bottom panel, the reference geodesic trajectory (red dashed) and the full spinning-body trajectory (solid blue) are shown, with the shifted geodesic approximation (black dotted) overlaid for direct comparison. This figure highlights how secular and oscillatory corrections successively modify the trajectory, with the shifted geodesic approximation capturing the key secular behavior relevant for gravitational-wave flux computations.}
\label{fig:shiftedgeodesictraj}
\end{figure}
As outlined in Sec.\ \ref{sec:lineartrajfreq}, the fixed turning-point framework assumes a reference geodesic, specified by the parameters $(p,e,x_I)$. The particle oscillates between radial and polar turning points that, on average, coincide with those of the reference geodesic. However, unlike in the geodesic case, the radial turning points now depend on $z$ and the polar turning points on $r$.  The resulting oscillations split naturally into two parts: $\delta\chi_r^S$ and $\delta\chi_\theta^S$, which describe purely radial and purely polar motion, and $\delta\mathcalligra{r}^S$ and $\delta\mathcalligra{z}^S$, which encode couplings between the radial and polar motion. 

The explicit parameterization of the spinning-body orbit is:
\begin{align}
r & =\frac{pM}{1+e\cos\left((\hat{\Upsilon}_r + \Upsilon_r^S)\lambda+\delta\hat{\chi}_r+\delta\chi_r^S\right)}+\delta\mathcalligra{r}_S\;,\label{eq:rparamgen}\\
\cos\theta & =\sin I\cos \left((\hat{\Upsilon}_\theta + \Upsilon_\theta^S)\lambda+\delta\hat{\chi}_{\theta}+\delta\chi_{\theta}^S\right)+\delta\mathcalligra{z}_S\;,\label{eq:thetaparamgen}
\end{align}
where $\hat \Upsilon_r + \Upsilon_r^S$ and $\hat \Upsilon_\theta + \Upsilon_\theta^S$ are the radial and polar frequencies respectively (divided into the geodesic contribution with hat accent, and the spin-curvature correction with $S$ superscript).  The oscillating secondary spin corrections to the true anomaly angles are given by
\begin{align}
\delta\chi_{r}^{S} =\sum_{n^s=-\infty}^{\infty}\delta\chi_{r,n^s}^{S}e^{-in^s w_r}\;,  \ \ \text{and} \ \ 
\delta\chi_{\theta}^{S} =\sum_{k^s=-\infty}^{\infty}\delta\chi_{\theta,k^s}^{S}e^{-ik^sw_{\theta}}\;.\label{eq:deltachigen}
\end{align}
The radial libration variation is
\begin{align}
\delta\mathcalligra{r}_S =\sum_{j=-1}^{1}\sum_{n^s,k^s=-\infty}^{\infty}\delta\mathcalligra{r}_{S,jn^sk^s}e^{-i(n^s w_r+k^s w_{\theta}+j w_p)}\;, \label{eq:deltachir2gen}
\end{align}
where $k^s$ and $j$ cannot both be zero; and the polar libration variation is
\begin{align}
\delta\mathcalligra{z}_S =\sum_{j=-1}^{1}\sum_{n^s,k^s=-\infty}^{\infty}\delta\mathcalligra{z}_{S,,jn^sk^s}e^{-i(n^s w_r+k^s w_{\theta}+j w_p)}\;,\label{eq:deltachitheta2gen}
\end{align}
where $n^s$ and $j$ cannot both be zero. The full spinning-body motion is therefore parameterized by:

\begin{enumerate}

\item secular shifts in orbital frequencies $\Upsilon_i^S$ and conserved quantities $\delta C^S_i$;

\item oscillatory corrections to the true anomaly angles $\delta\chi_r^S$ and $\delta\chi_\theta^S$;

\item oscillatory libration corrections arising from radial-polar coupling, $\delta\mathcalligra{r}_S$ and $\delta\mathcalligra{z}_S$;

\item oscillatory pieces of $t$ and $\phi$, i.e., $\Delta t$ and $\Delta \phi$.

\end{enumerate}

Figure~\ref{fig:shiftedgeodesictraj} schematically illustrates the structure of this decomposition using the radial coordinate as an example. The top left panel shows a reference geodesic orbit (red dashed), with unperturbed frequency $\hat{\Upsilon}_r$. The top right panel (shaded) introduces the dominant secular correction from spin: a shift in the radial frequency $\Upsilon_r^S$, resulting in a modified orbit (blue). The middle left panel adds the oscillatory correction $\delta\chi_r^S$, which induces periodic modulations around the shifted geodesic. The middle right panel shows the full $\mathcal{O}(\sigma)$ spinning-body trajectory, incorporating all corrections: the frequency shift $\Upsilon_r^S$, the true anomaly correction $\delta\chi_r^S$, and the libration term $\delta\mathcalligra{r}^S$ due to coupling between radial and polar motion. Finally, the bottom panel overlays all three curves (reference geodesic, full spinning-body trajectory and the shifted geodesic approximation) for direct comparison. 

The shifts in the radial and polar frequencies, $\Upsilon_i^S$ with $i = r,\theta$, introduce additional secular growth in the orbital phases and thereby modify the long-term structure of the orbit.  Secular effects associated with $\Upsilon_\phi$ and $\Gamma$ are consistently incorporated through the corrected integrals of motion and so do not require separate treatment at this order. The other corrections are strictly oscillatory at $\mathcal{O}(\sigma)$ and average to zero over many orbits. This motivates an approximation in which we retain only the secular frequency and integral-of-motion shifts and discard the oscillatory corrections; to calculate the secular shifts, we use the closed-form Hamilton-Jacobi expressions given in Ref.\ \cite{Piovano2024}. Specifically, we replace Eqs.~(\ref{eq:rparamgen}) and~(\ref{eq:thetaparamgen}) with
\begin{align}
r &= \frac{pM}{1 + e\cos\left((\hat{\Upsilon}_r + \Upsilon_r^S)\lambda + \delta\hat{\chi}_r\right)}\;, \label{eq:rparamgenSG} \\
\cos\theta &= \sin I\cos\left((\hat{\Upsilon}_\theta + \Upsilon_\theta^S)\lambda + \delta\hat{\chi}_\theta\right)\;.\label{eq:thetaparamgenSG}
\end{align}
These forms define the \textit{shifted geodesic approximation}. They retain the structure of geodesic motion while encoding the long-term impact of secondary spin via frequency shifts. Although this approximation discards oscillatory corrections in the radial and polar motion, we retain the oscillatory contributions to the temporal and azimuthal coordinates, $\Delta t$ and $\Delta \phi$, in our numerical implementation. As shown in ~\ref{sec:SGandLOtest}, these terms play an important role in maintaining the accuracy of the shifted-geodesic construction. The frequency-domain method is employed to compute $\Delta t$ and $\Delta \phi$, which are more inexpensive since they are determined by first-order differential equations. 

We also consider a ``leading-order'' (LO) spinning-body approximation that includes the four lowest harmonics of the oscillatory terms:
\begin{equation}
n^s_{\text{max}} = 4 \;, \qquad k^s_{\text{max}} = 4 \;.
\end{equation}
In practice, a minimal LO truncation, $n^s_{\max}=k^s_{\max}=1$, offers little improvement over the shifted-geodesic approximation.  We find that additional oscillatory harmonics are required for the LO construction to provide a clear accuracy gain.  A detailed justification of this choice is given in ~\ref{sec:SGandLOtest}, as well as Section \ref{sec:orbitalparams}. This approximation serves as an intermediate step between the shifted-geodesic approximation and the full spinning-body model.

\subsection{Evaluating fluxes in the shifted geodesic framework}
\label{sec:evalualtingSGfluxes}

We now turn our attention to the computation of gravitational-wave flux corrections arising from the spin of the secondary in the shifted-geodesic framework. The averaged energy and angular momentum fluxes from an orbiting spinning-body are given by\footnote{To avoid confusion between the mode indices associated with the spinning-body trajectory and those appearing in the Teukolsky decomposition, we adopt the notation $n^s$ and $k^s$ for the frequency mode indices in the spin-corrected trajectory. These should be distinguished from the $n$ and $k$ indices labeling the Fourier-harmonic modes of the Teukolsky function. Although the physical frequencies involved are related, the roles of the indices differ: $(n^s, k^s)$ parametrize the harmonic content of the perturbed motion, while $(n, k)$ label the radiative modes of the spacetime curvature.}
\begin{subequations}\label{eq:fluxes}
\begin{align}
    \left\langle\mathcal{F}^{E}\right\rangle&\equiv\left(\frac{dE}{dt} \right)^\infty+\left(\frac{dE}{dt} \right)^H = \sum_{lmnkj} \mathcal{F}^{E,\infty}_{lmnkj}+\sum_{lmnkj} \mathcal{F}^{E,H}_{lmnkj}=\sum_{lmnkj} \mathcal{F}^E_{lmnkj} \; ,\\[1mm]
    \left\langle\mathcal{F}^{J_z}\right\rangle&\equiv\left(\frac{dJ_z}{dt} \right)^\infty+\left(\frac{dJ_z}{dt} \right)^H = \sum_{lmnkj} \mathcal{F}^{J_z,\infty}_{lmnkj} + \sum_{lmnkj} \mathcal{F}^{J_z,H}_{lmnkj} = \sum_{lmnkj} \mathcal{F}^{J_z}_{lmnkj}\; ,
\end{align}
with
\begin{align}
    \mathcal{F}^{E}_{lmnkj} &= \frac{\abs{Z^\infty_{lmnkj}}^2 + \alpha_{lmnkj}\abs{Z^H_{lmnkj}}^2}{4\pi \omega_{mnkj}^2} \; , \qquad
    \mathcal{F}^{J_z}_{lmnkj} = \frac{m\qty(\abs{Z^\infty_{lmnkj}}^2 + \alpha_{lmnkj}\abs{Z^H_{lmnkj}}^2 )}{4\pi \omega_{mnkj}^3} \; .
\end{align}
\end{subequations}
See \ref{sec:computegenericflux} for the expressions for $\abs{Z^{\infty,H}_{lmnkj}}$ [Eq.\eqref{eq:Cpm_lmnkj}] and $\omega_{mnkj}$ [Eq.~\eqref{eq:Cpm_lmnkj2}].  The expression for $\alpha_{lmnkj}$ is given in Eq.\ (57) of Ref.\ \cite{Hughes2021}. Sums over the indices $(l,m,k,n)$ are taken with $l$ ranging from $2$ to $\infty$, $m$ from $-l$ to $l$, and $k,n$ from $-\infty$ to $\infty$. Observe that the superscripts $H$ and $\infty$ correspond to the asymptotic fluxes evaluated at the horizon and at infinity respectively. We truncate the above sums at $n_{\rm max}$, $k_{\rm max}$ and $l=10$ with sums taken from $-n_\text{max}$, $-k_\text{max}$ to $n_\text{max}$, $k_\text{max}$; determining reasonable values of $n_{\rm max}$ and $k_{\rm max}$ for the spinning-body flux corrections is discussed in Section~\ref{sec:modeindices}.  The relatively low truncation order in $l$ reflects the coarse grid adopted in early adiabatic generic waveform studies \cite{Hughes2021}, rather than a fundamental limitation of the formalism.  Recent advances enabled by the FastEMRIWaveforms (FEW) framework have demonstrated the importance of extending to substantially higher harmonic content for accurate generic-orbit waveforms \cite{Chua:2020stf,Katz:2021yft,Speri:2023jte,Chapman-Bird2025}. While a fully optimized generic $(l,m,n,k)$ mode grid for spinning-body fluxes is still under active development, the present choice provides a practical compromise for exploratory studies and will be systematically improved in future work.

Because terms proportional to the perpendicular component $\sigma_\perp$ oscillate with frequency $\Omega_s$, they only appear in the $j=\pm 1$ modes.  Their amplitudes are thus proportional to $\sigma_\perp$, so associated fluxes enter at quadratic order and can be neglected in a linear-order treatment of $\sigma$. Throughout this work, we thus restrict to $j=0$ when summing over $l$, $m$, $n$, and $k$. The fluxes $\langle\mathcal{F}^{E}\rangle$ and $\langle\mathcal{F}^{J_z}\rangle$ used here follow the formulation of Ref.\ \cite{Skoupy2023}. For completeness, we note that while an expression for the evolution of the spinning-secondary Carter-constant analogue has been derived \cite{Grant2024}, further work is needed to evaluate it across parameter space. Existing self-consistent inspiral computations for spinning bodies remain restricted to equatorial and spherical orbits \cite{Skoupy2022,Skoupy2025}.

Our pipeline builds on and extends the internal Mathematica implementation developed for \cite{Skoupy2023}, which we have adapted for the present analysis. The full code used in this work, along with documentation, will be made publicly available in a dedicated GitHub repository\footnote{The GitHub repository can be found here: \url{https://github.com/lisadrummond/SpinningSecondary-freqdom/}.} to facilitate reproducibility and cross-comparison with future implementations.  We retain the frequency-domain Teukolsky solver and (i) incorporate closed-form expressions for spin-induced shifts of the fundamental frequencies \cite{Witzany2019,Piovano2024} into the trajectory's geodesic structure via the shifted-geodesic replacement, and (ii) implement parallelized mode-sum machinery to assess convergence and compute fluxes for inspiral construction.  In this setup, the fluxes are computed by summing mode-summed Teukolsky amplitudes over $(l,m,n,k)$ at fixed $j=0$. While the production code used here is in Mathematica, a generic C\texttt{++} implementation with the same modular structure is in development to further improve performance. For accuracy, we retain the frequency-domain construction of the oscillatory corrections to the temporal and azimuthal coordinates, $\Delta t$ and $\Delta \phi$. Omitting these terms significantly degrades the accuracy of the shifted-geodesic construction and makes it difficult to obtain agreement with the exact evolution (see ~\ref{sec:SGandLOtest}).

\begin{table*}
\centering
\small
\caption{Typical speed-up from \textit{one evaluation} of the GW flux for a fixed $l$, $m$, $n$ and $k$. The orbit used here is $a=0.9$, $p=12$, $e=0.2$ and $x_I=\sqrt3/2$. Here calculations are performed in Mathematica on an Apple M1 chip with a $\sim3.2$ GHz clock speed, with wallclock timings evalauted using the function \texttt{AbsoluteTiming}. }
\label{tab:benchmark}
\begin{tabular}{ccccccc}
\toprule
& \multicolumn{5}{c}{\textbf{Computation method}} \\
\cmidrule(lr){2-7}
& \multicolumn{2}{c}{Exact (sec)} & LO (sec) & SG (sec) & \multicolumn{2}{c}{Relative gain} \\
\cmidrule(lr){2-3} \cmidrule(lr){4-4} \cmidrule(lr){5-5} \cmidrule(lr){6-7}
\textbf{Computation step} & $n^s_{\max}=16$ & $n^s_{\max}=8$ & $n^s_{\max}=4$ & - & LO & SG \\
\midrule
$r$-equations &  33.0546  & 2.61780  & 0.358028 & - & $92.32\times$   &  - \\
$\theta$-equations & 4.33018 & 0.80757 & 0.224426 & - & $19.29\times$   & -  \\
Norm-equations & 3.28321 & 0.59022  & 0.164353 & - & $5.56\times$  & -  \\
$\Delta t$ and $\Delta \phi$ & 38.4979 & 4.24740 & 2.02675 & 2.00249 & $18.99\times$ & $19.23\times$ \\
\textbf{Total Trajectory} & 90.8440 & 10.4805 & 2.84665 &  2.01826 & $31.91\times$ & $45.01\times$ \\
\textbf{Total GW flux} & 96.0118   & 15.4477 & 7.48689 &  6.86855& $12.82\times$ &$13.98\times$  \\
\bottomrule
\vspace{0.1em} 
\end{tabular}
\parbox{0.9\textwidth}{\footnotesize
\textit{Note.} When $k^s_{\max}$ is not explicitly listed, it is taken equal to $n^s_{\max}$. The term ``Relative gain" refers to the ratio of computational times, comparing the \textit{Exact} trajectory with $n^s_{\rm max} =k^s_{\rm max}= 16$, against the SG (shifted geodesic) and LO (leading order) approximations. Entries marked with ``-'' indicate steps that have no direct analogue in the shifted-geodesic method, since this approach does not explicitly compute the full $r(\lambda)$ and $\theta(\lambda)$ trajectories or perform a normalization calculation, but instead evaluates the frequencies and integrals of motion analytically. As a result, we do not quote a relative gain for these steps.
}
\end{table*}

In Table~\ref{tab:benchmark}, we compare runtimes for trajectory and flux computations for ``exact'' (using the full spinning-body-orbit framework\footnote{One should keep in mind that this framework for computing spinning-body orbits is itself an expansion.  ``Exact'' is thus (rather ironically) a somewhat inexact label; we use it because it defines the most practically exact standard against which to define approximations to spinning-body orbital motion.  The precise meaning of ``exact'' we use is that we have included enough terms in the expansion to ensure that further terms in the expansion negligibly change the orbit's properties.}, ``shifted-geodesic'' (SG), and ``leading-order'' (truncating the spinning-body-orbit calculation at $n^s_{\rm max} = 4$).  Comparison with the full method provides a benchmark for evaluating the accuracy of each approximation, as discussed in the next section. We note that two definitions of the exact trajectory appear in Table~\ref{tab:benchmark}. The choice $n^s_{\max}=k^s_{\max}=16$ corresponds to a trajectory accuracy comparable to that of the geodesic trajectory calculation used for the 0PA terms, especially at higher eccentricities. A second definition with $n^s_{\max}=k^s_{\max}=8$ is also listed; this truncation is used in Sec.~\ref{sec:results}, where the large number of repeated trajectory evaluations required to explore parameter space and compute fluxes makes the higher truncation computationally expensive. A central point of our analysis is that the 0PA-level of trajectory accuracy is not required for computing post-adiabatic fluxes. For the range of eccentricities and inclinations considered in Sec.~\ref{sec:results}, we find that the less expensive truncation $n^s_{\max}=k^s_{\max}=8$ provides sufficient accuracy for the relevant comparisons.

We emphasize that the timings reported in Table~\ref{tab:benchmark} correspond to a single choice of $(p,e,x_I)$ and to a single $(l,m,n,k)$ mode. Constructing a full inspiral requires repeating this calculation across a large number of such configurations, involving of order $\sim10^7$ distinct $(p,e,x_I,l,m,n,k)$ combinations, depending on the resolution choices adopted.  A factor of $\sim45$ speed-up will compound across the parallel workload, thus driving a corresponding significant reduction in total computational cost. While the trajectory corresponding to a given $(p,e,x_I)$ could, in principle, be reused for many modes, in practice each $(p,e,x_I,l,m,n,k)$ tuple is often computed independently within a massively parallel, high-throughput framework. In such an approach, each configuration is processed as a separate job, and the trajectory must be recomputed within every evaluation. It is thus crucial that the trajectory calculation itself is highly efficient.

The fluxes $\mathcal F^E$ and $\mathcal F^{L_z}$ in Eqs.\ (\ref{eq:fluxes}) can be linearized in secondary spin:
\begin{align}
\label{eq:fluxesspin}
    \mathcal{F}^{E}&\equiv\hat{\mathcal{F}}^{E}+\sigma \delta\mathcal{F}^{E} , \ \ \ \ \ \
    \mathcal{F}^{L_z}\equiv\hat{\mathcal{F}}^{L_z}+\sigma\delta\mathcal{F}^{L_z} \; .
\end{align}
The spin flux corrections $\delta \mathcal{F}^b$, with index $b$ denoting $(E,L_z)$, can be expanded as a Taylor series about the background geodesic solution $\hat{x}_i(\lambda)$. The time-averaging procedure, i.e., computing $\langle \mathcal F^{E, L_z} \rangle$, involves integrating $I^{\infty,H}_{lmnkj}(r(w_r, w_z), z(w_r, w_z))$ from Eq.\ (\ref{eq:Ipmlmomega}) over the fast orbital phases $w_r$ and $w_z$, which enter nonlinearly through the orbital position functions $r(w_r, w_z)$ and $z(w_r, w_z)$.\footnote{The integrand depends nonlinearly on the orbital functions but is linear with respect to secondary spin in the black hole perturbation theory sense.}  Trajectory corrections induced by secondary spin give rise to three distinct flux contributions, $\delta \mathcal{F}^b_{\text{osc}}$, $\delta \mathcal{F}^b_{\text{nonlin}}$, and $\delta \mathcal{F}^b_{\text{sec}}$:
\begin{enumerate}
    \item \textit{Purely oscillatory terms} $\delta \mathcal{F}^b_{\text{osc}}$: These vary periodically with the fundamental frequencies $\Upsilon_i$ and average to zero over an orbital period, leading to no secular growth.
    \item \textit{Nonlinear terms} $\delta \mathcal{F}^b_{\text{nonlin}}$: These terms originate from non-linear couplings between oscillatory components of the orbital motion. Because they do not necessarily average to zero over a phase cycle, they can yield non-negligible contributions to the phase-averaged fluxes.
    \item \textit{Secular frequency shift terms} $\delta \mathcal{F}^b_{\text{sec}}$: These arise due to spin-induced modifications of the orbital frequencies themselves. Unlike the purely oscillatory terms, they accumulate over time and drive the long-term evolution of the orbit.
\end{enumerate}
To derive expressions for the above terms $\delta \mathcal{F}^b_{\text{osc}}$, $\delta \mathcal{F}^b_{\text{nonlin}}$ and $\delta \mathcal{F}^b_{\text{sec}}$, we present a more explicit formulation of the above argument using near-identity transformations (NITs). When only the conservative MPD force is present, the spinning test-body system can be expressed as\footnote{For notational convenience, we choose $s_\parallel=1$, so that henceforth $\sigma = \varepsilon$.}
\begin{align}
\frac{d \mathcal{C}_{b}}{d\lambda} &= 0 \;, \\
\frac{d \zeta_i}{d\lambda} &= \hat\Upsilon_i(\vec{\mathcal{C}})+\varepsilon f_i^{(1)}(\vec{\mathcal{C}},\vec{\zeta}) \;,
\end{align}
where $\vec{\zeta} = ({\zeta_r,\zeta_z})$ denote the angle variables defined in Eq.\ (3.46) of Ref.\ \cite{Piovano2024}. We employ osculating elements $\mathcal{C}_b=(E,L_z,K)$ corresponding to the conserved quantities of a spinning-body orbit. This choice differs from the standard osculating-geodesic framework used in the NIT formulation of Refs.\ \cite{VanDeMeent:2018cgn, Lynch2021, McCart:2021upc,Lynch:2022zov, Lynch2024, Drummond2024,Lynch2024_2}, where the osculating elements are the geodesic constants of motion. The function $f_i^{(1)}(\vec{\mathcal{C}},\vec{\zeta})$ captures the oscillatory part of the conservative spinning-body trajectory (given by the linear-in-spin term multiplying $q$ on the right-hand side of Eq.\ (3.49) in Ref.\ \cite{Piovano2024}). The full equations including dissipative evolution then become
\begin{align}
\frac{d \mathcal{C}_b}{d\lambda} &= \varepsilon \hat{\mathcal{F}}^b (\vec{\mathcal{C}},\vec{\zeta})+\varepsilon^2  \delta \mathcal{F}^{b} (\vec{\mathcal{C}},\vec{\zeta})\;, \\
\frac{d \zeta_i}{d\lambda} &= \hat\Upsilon_i(\vec{\mathcal{C}})+\varepsilon f_i^{(1)}(\vec{\mathcal{C}},\vec{\zeta})\;,
\end{align}
where $\hat{\mathcal{F}}^b (\vec{\mathcal{C}},\vec{\zeta})$ represents the leading-order dissipative flux\footnote{Note that we do not include the conservative or oscillating dissipative corrections to the first-order gravitational self-force in this work.}, and $\delta\mathcal{F}^{b} (\vec{\mathcal{C}},\vec{\zeta})$ accounts for the spinning-secondary contribution to the flux.

Here, we reformulate the differential equations using NIT variables. As given in Ref. \cite{Piovano2024}, we define a near-identity transformation:
\begin{equation}
\zeta_i = w_i - \varepsilon \xi_i(\vec{\mathcal{C}},\vec{\zeta})+  \mathcal{O}(\varepsilon^2) \;,
\end{equation}
such that
\begin{align}
\frac{d w_i}{d\lambda} &= \hat\Upsilon_i(\vec{\mathcal{C}})+\varepsilon \delta\Upsilon^S_{i}(\vec{\mathcal{C}}) \;.
\end{align}
The NIT variable $\nit{\mathcal{C}}_{b}$ relates to $\mathcal{C}_{b}$ via
\begin{subequations}\label{eq:transformation}
\begin{equation}\label{eq:transformation1}
\nit{\mathcal{C}}_{b} = \mathcal{C}_{b} + \sp Y_b^{(1)}(\vec{\mathcal{C}},\vec{\zeta}) + \sp^2 Y_b^{(2)}(\vec{\mathcal{C}},\vec{\zeta})+  \mathcal{O}(\varepsilon^3)\;,
\end{equation}
\end{subequations}
where the transformation functions $Y_b$ are chosen to be smooth periodic functions of the orbital phases $\vec{\zeta}$. The equations of motion for the NIT variables $\nit{\mathcal{C}}_{b}$ and $w_i$ then take the form
\begin{subequations}\label{eq:transformed_EoM}
\begin{align}
\frac{d \nit{\mathcal{C}}_{b}}{d \lambda} &=\varepsilon \nit{\hat{\mathcal{F}}}^b(\vec{\nit{\mathcal{C}}}) + \varepsilon^2 \delta\nit{\mathcal{F}}^b(\vec{\nit{\mathcal{C}}}) +  \mathcal{O}(\varepsilon^3)\;, \\
\frac{d w_i}{d\lambda} &= \hat\Upsilon_i(\vec{\nit{\mathcal{C}}}) +\varepsilon \delta\Upsilon^S_i(\vec{\nit{\mathcal{C}}}) + \mathcal{O}(\varepsilon^2)\;.
\end{align}
\end{subequations}

These equations of motion are now independent of the orbital phases. For convenience, we define an averaged quantity $\avg{A} (\vec{P})$ as
	\begin{equation}\label{eq:average}
		\avg{A} (\vec{P}) = \frac{1}{(2\pi)^{i_{\text{max}}}} \idotsint_{\vec{Q}} A(\vec{P},\vec{Q}) d q_1 \dots d q_{i_{\text{max}}}\;,
	\end{equation}
	 and an oscillating quantity as
	\begin{equation}\label{eq:oscillating}
		\osc{A}(\vec{P},\vec{q}) = A(\vec{P},\vec{q}) - \avg{A}(\vec{P})  = \sum_{\vec{\kappa} \neq \vec{0}} A_{\vec{\kappa}}(\vec{P}) e^{i \vec{\kappa} \cdot \vec{Q}}\;. 
	\end{equation} 
We select the average values of the transformation terms such that $\avg{Y^{(1)}_b} = \avg{Y^{(2)}_b}= \avg{\xi_i} = 0$, leading to the transformed forcing functions \cite{Lynch2021,Lynch2024,Lynch2024_2,Drummond2024}:
\begin{subequations}
\begin{gather}
\nit{\hat{\mathcal{F}}}^b = \left<\hat{\mathcal{F}}^{b}\right>, \quad \delta\Upsilon^S_{i}= \left<f_{i}^{(1)}\right>\;, \tag{\theequation a-b}
\end{gather}
\begin{equation}
\delta\nit{\mathcal{F}}^b = \left<\delta\mathcal{F}^b \right> + \left<\frac{\partial \osc{Y}_b^{(1)}}{\partial w_i} \osc{f}_i^{(1)} \right> + \left<\frac{\partial \osc{Y}_b^{(1)}}{\partial \nit{\mathcal{C}}_k} \osc{\hat{\mathcal{F}}}^k \right>\;, \tag{\theequation c}
\end{equation}
\end{subequations}
where
\begin{equation}\label{eq:NIT_Y}
\osc{Y}_b^{(1)} \coloneqq \sum_{\vec{\kappa} \neq \vec{0}} \frac{i}{\vec{\kappa} \cdot \vec{\Upsilon}} \hat{\mathcal{F}}^b_{\vec{\kappa}} e^{i \vec{\kappa} \cdot \vec{\zeta}}\;.
\end{equation}
Here $\vec{\kappa}\in \mathbb{Z}^N$ denotes Fourier indices paired with the radial and polar frequency set $\vec{\Upsilon}=\{\Upsilon_r,\Upsilon_z\}$.  Note that the term $\left<\frac{\partial \osc{Y}_j^{(1)}}{\partial \nit{\mathcal{C}}_k} \osc{\hat{\mathcal{F}}}^k \right>$ corresponds to the oscillating dissipative part of the first order self-force; we exclude this in our analysis because we are only considering the secondary spin contribution to the 1PA terms.  

Finally, the transformed equations that describe the secondary-spin corrected motion are:
\begin{subequations}\label{eq:transformed_EoM2}
\begin{align}
\frac{d \nit{\mathcal{C}}_{b}}{d \lambda} &=\varepsilon \nit{\hat{\mathcal{F}}}^b(\vec{\nit{\mathcal{C}}}) + \varepsilon^2 \left<\delta \mathcal{F}^b \right> + \varepsilon^2  \left<\frac{\partial \osc{Y}_b^{(1)}}{\partial w_i} \osc{f}_i^{(1)} \right> +  \mathcal{O}(\varepsilon^3)\;, \\
\frac{d w_i}{d\lambda} &= \hat\Upsilon_i(\vec{\nit{\mathcal{C}}}) +\varepsilon \delta\Upsilon^S_i(\vec{\nit{\mathcal{C}}}) + \mathcal{O}(\varepsilon^2)\;.
\end{align}
\end{subequations}
When post-adiabatic effects such as spin corrections from the secondary body are included, we linearize Eq.~\ref{eq:fluxes} in the mass ratio $\varepsilon$, yielding 
\begin{align}
\label{eq:fluxesspin2}
    \left\langle\mathcal{F}^{E}\right\rangle&\equiv\left\langle\hat{\mathcal{F}}^{E}\right\rangle+\varepsilon\delta\tilde{\mathcal{F}}^{E}=\sum_{lmnk} \left(\mathcal{F}^E_{lmnk}+\varepsilon\delta\tilde{\mathcal{F}}^E_{lmnk}\right) ,\\[1mm]
    \left\langle\mathcal{F}^{L_z}\right\rangle&\equiv\left\langle\hat{\mathcal{F}}^{L_z}\right\rangle+\varepsilon\delta\tilde{\mathcal{F}}^{L_z}=\sum_{lmnk} \left(\mathcal{F}^{L_z}_{lmnk}+\varepsilon\delta\tilde{\mathcal{F}}^{L_z}_{lmnk}\right) \; .
\end{align}
In these expressions, the hat denotes the leading-order adiabatic contribution to the flux, and the $\delta$-terms represent post-adiabatic contributions. In this work, the only post-adiabatic contributions to the flux will be the linear-in-spin corrections due to the secondary's spin. Note that we have dropped the $j$-mode index corresponding to spin-precessional frequencies in the equations above and in our subsequent analysis. This is well-motivated for our analysis because the orthogonal spin components which lead to spin precession are expected to contribute only in terms beyond first post-adiabatic order \cite{Skoupy2023, Piovano2024, Mathews2025}; only $j = 0$ contributes at the order we are considering.

We now identify the key components of the spinning-secondary corrections to the fluxes. The secular contribution arises because the spin-induced frequency shifts $\delta\Upsilon_i^S(\vec{\mathcal{C}})$ modify the torus frequencies that enter the averaged fluxes.  Expanding the leading-order averaged flux $\langle\hat{\mathcal{F}}^b\rangle$ around the
background frequencies $\hat{\Upsilon}_i(\vec{\mathcal{C}})$ yields the spin-induced secular correction to the averaged flux:
\begin{equation}
\delta\mathcal{F}^b_{\mathrm{sec}}
=\frac{\partial\langle\hat{\mathcal{F}}^b\rangle}{\partial\Upsilon_i}
\,\delta\Upsilon_i^S(\vec{\mathcal{C}})\;.
\end{equation}
The nonlinear contribution arising from the interaction of oscillatory terms is given by
\begin{equation}
\delta \mathcal{F}^b_{\text{nonlin}} = \left<\frac{\partial \osc{Y}_b^{(1)}}{\partial w_i} \osc{f}_i^{(1)} \right>\;,
\end{equation}
while the purely oscillatory term, $\delta \mathcal{F}^b_{\text{osc}}$, averages to zero and does not contribute. The vanishing of $\delta \mathcal{F}^b_{\text{osc}}$ provides a partial justification for the shifted geodesic approximation, which neglects this term from the outset. 

To validate this approximation, we must also show that the omission of $\delta \mathcal{F}^b_{\text{nonlin}}$ does not significantly impact waveform accuracy. In the next section, we demonstrate that across the parameter space explored, the dominant correction arises from the secular term $\delta \mathcal{F}^b_{\text{sec}}$, while the nonlinear term $\delta \mathcal{F}^b_{\text{nonlin}}$ remains subdominant. Nonetheless, the precise domain of validity depends on the relative magnitudes of these contributions, which we quantify through a numerical investigation in the following section.

\section{Validity of approximation: Fluxes computed with ``exact'' trajectories versus with shifted geodesics}
\label{sec:results}
In this section, we assess the validity of the shifted-geodesic approximation by quantifying both its accuracy and its domain of applicability across parameter space. Our focus is on evaluating how well this approximation captures the key features of the ``exact'' spinning-body trajectory\footnote{We remind the reader that ``exact'' is our shorthand for an expansion taken to high enough order that including further terms has a negligible impact on our model for orbital motion.}, particularly across a range of orbital parameters $(p,e,x_I)$ and mode indices $(n, k)$.  The analysis is motivated by two key observations. First, post-adiabatic corrections need not be captured with high precision, since they contribute subdominantly to the overall evolution. Second, among the various components of the spinning-secondary trajectory, the dominant contributions arise from shifts in the fundamental frequencies and conserved quantities; subleading oscillatory corrections typically have a minor (in some cases, negligible) impact.

\subsection{Roadmap for this section: What is compared and why}
\label{sec:roadmap}
 
Throughout this section, we quantify the accuracy of an approximate trajectory model by comparing its gravitational-wave fluxes to those computed from the full spinning-body trajectory.  The comparisons in this section involve several different trajectories/models:
\begin{enumerate}
\item \textbf{Exact:} the full spinning-body trajectory of Eqs.~(\ref{eq:rparamgen})--(\ref{eq:thetaparamgen}), evaluated with a sufficiently large harmonic truncation $(n^s_{\max},k^s_{\max})$ so that the resulting fluxes are converged to our target tolerance. Unless otherwise stated, we take $(n^s_{\max},k^s_{\max})=(8,8)$ as the reference ``exact'' choice. This choice provides a balance between numerical cost and accuracy; it is sufficiently converged for the range of $(n^s,k^s)$ modes that contribute to the post-adiabatic terms considered here (see Ref.\ \cite{Skoupy2023}), while avoiding the significantly higher computational expense associated with truncating the sums at larger values of $(n^s,k^s)$.  
\item \textbf{Shifted geodesic (SG):} the approximation that \emph{retains} the secular frequency shifts $\Upsilon_i^S$ (with $i=r,\theta$) and the shifts in the integrals of motion $\delta\mathcal{C}_j$, but \emph{discards} oscillatory corrections $\delta\chi_i^S$, $\delta\mathcalligra{r}_S$, and $\delta\mathcalligra{z}_S$.
\item \textbf{Leading-order oscillatory (LO):} an approximation that includes the four lowest oscillatory harmonics, $n^s_{\max}=4$ and $k^s_{\max}=4$ (in addition to the SG components), motivated by the  study in Sec.~\ref{sec:orbitalparams}.
\item \textbf{Two diagnostic variations ($\Upsilon_i^S=0$ and $\delta\mathcal{C}_j=0$):} (1) the exact trajectory with only frequency shifts omitted ($\Upsilon_i^S=0$), and (2) exact trajectory with integral-of-motion shifts omitted ($\delta\mathcal{C}_j=0$). These are not intended as standalone models, but as controlled  tests that isolate which physical ingredients dominate the flux corrections.
\end{enumerate}

We now outline the sequence of figures presented throughout Section \ref{sec:results} and the rationale for the comparisons of fluxes that we perform. The analysis addresses two complementary questions and proceeds in two stages. 

\begin{figure}[htbp]
\centering
\hspace{-0.8cm}
\includegraphics[width=0.85\columnwidth]{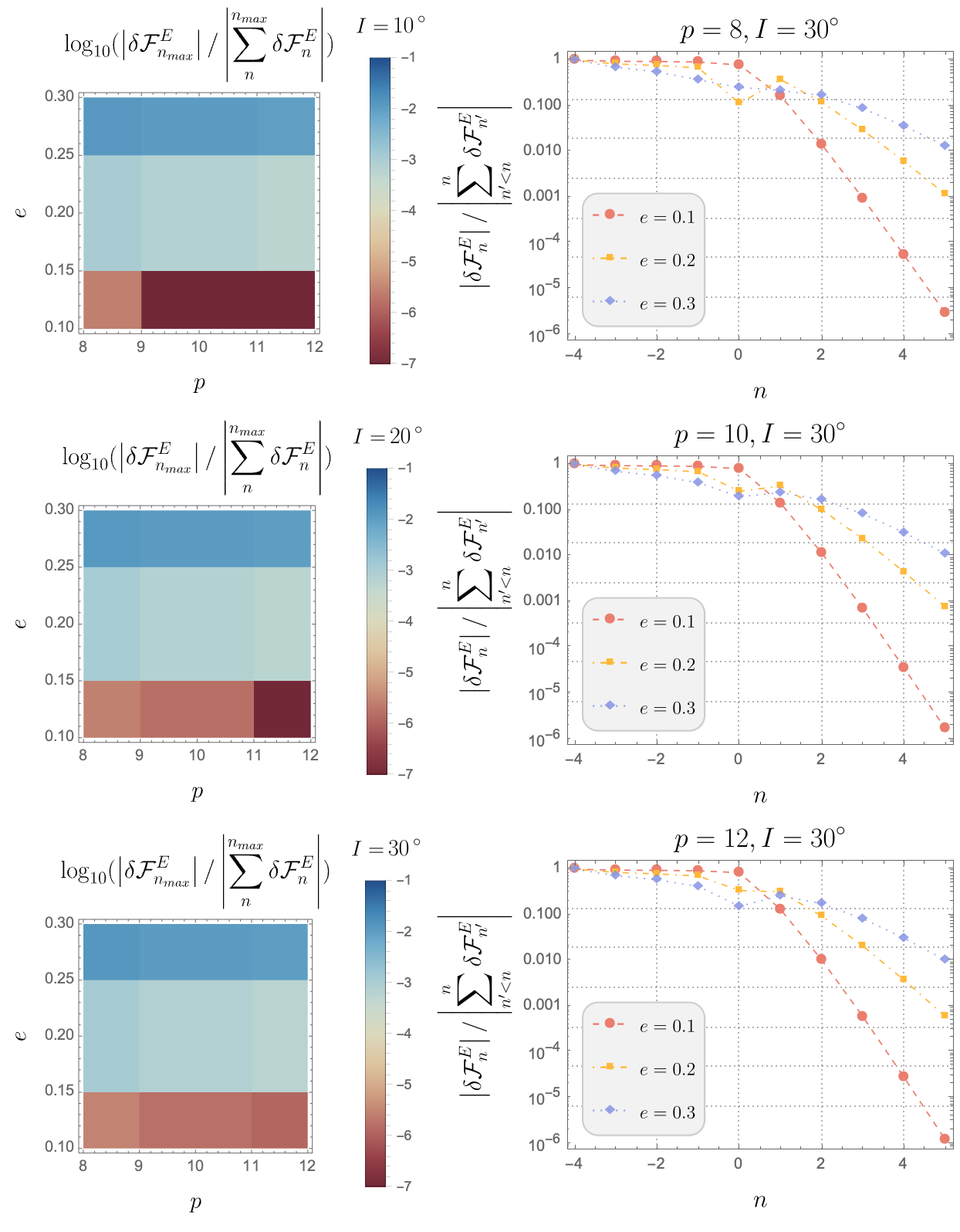}
\caption{Convergence of the exact spin-induced energy flux as a function of $n$, shown as the ratio of the contribution at $n_{\max}$ to the partial sum over all modes with $n \leq n_{\max}$. The left column shows $\left| \delta \mathcal{F}_{n_{\max}} \right| \Big/\left|  \sum_{n \leq n_{\max}} \delta \mathcal{F}_n \right|$, summed over $l$, $m$ and $k$, as a function of $p$ and $e$, with the rows corresponding to different values of inclination I. The right column illustrates how the choice of $n_{\max}$ affects the convergence criterion. We plot $\left| \delta \mathcal{F}_n \right| \Big/ \left|\sum_{n' \leq n}  \delta \mathcal{F}_{n'} \right|$, such that the rightmost point on each curve corresponds to $\left| \delta \mathcal{F}_{n_{\max}} \right| \Big/ \left| \sum_{n \leq n_{\max}}  \delta \mathcal{F}_n \right|$. Here the rows correspond to different values of semilatus rectum $p$ and the three curves in each panel correspond to different eccentricities $e$. Parameters: $a=0.9$, $n_{max}=5$, $k_{max}=4$, $n^s_{\textrm{max}}=8$, $k^s_{\textrm{max}}=8$.}
\label{fig:convcriteria}
\end{figure}

\begin{figure}[htbp]
\hspace{-0.6cm}
\includegraphics[width=1.07\columnwidth]{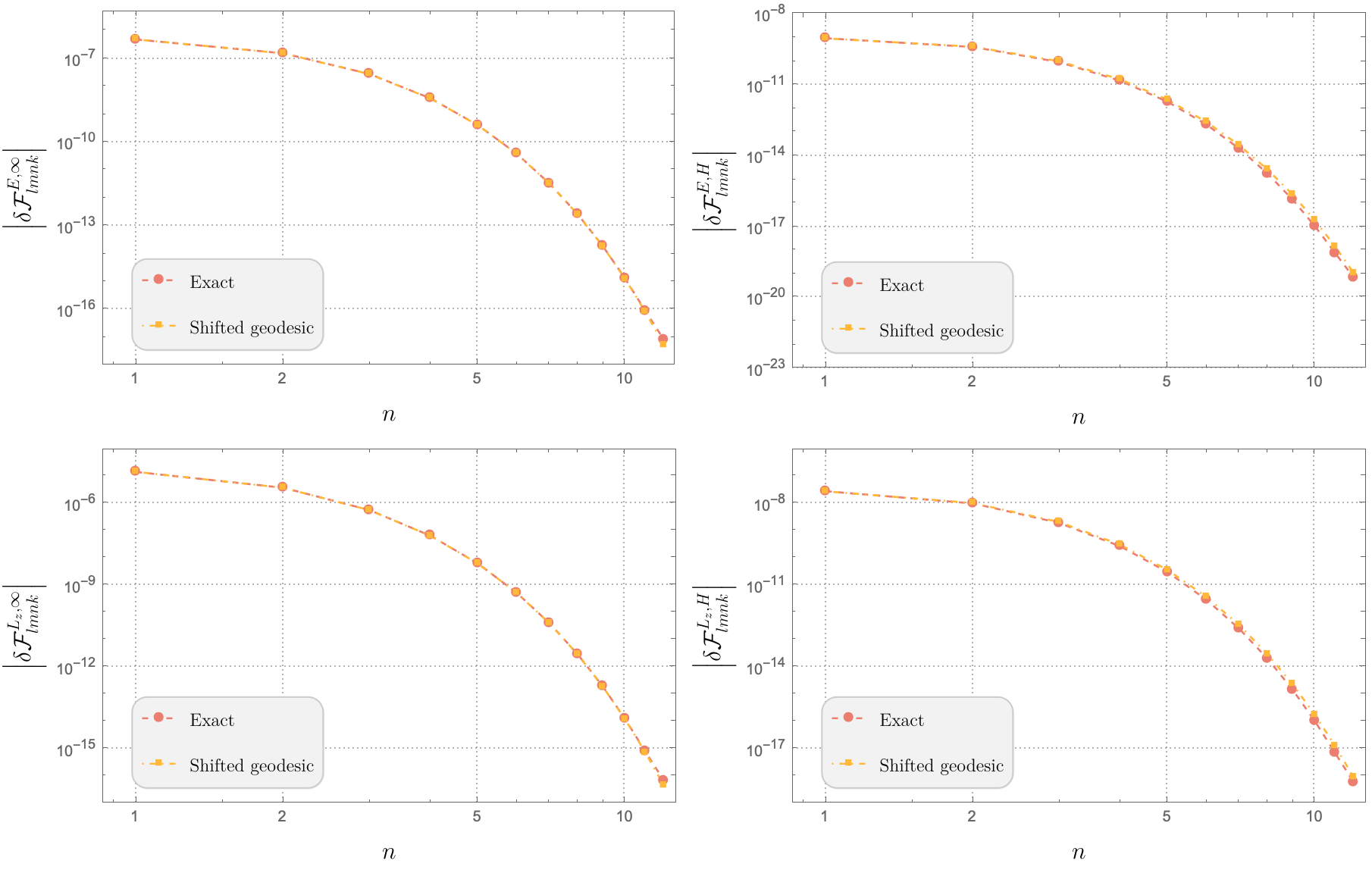}
\caption{Corrections to the gravitational wave energy flux due to the spin of the secondary as a function of $ n $. The top left panel shows the correction to the outgoing energy flux at infinity due to the spin of the secondary, $\mathcal{F}^{E,\infty}_{lmnk}$, while the top right panel shows the correction to the ingoing energy flux at the horizon due to the spin of the secondary, $\mathcal{F}^{E,H}_{lmnk}$. The bottom left and bottom right panels depict the corresponding corrections to the angular momentum fluxes, $\mathcal{F}^{L,\infty}_{lmnk}$ and $\mathcal{F}^{L,H}_{lmnk}$, respectively. The red circles and dashed curves represent fluxes computed with the exact trajectory, while the yellow squares and dash-dotted curves correspond to fluxes computed with the shifted geodesic approximation. The shifted geodesic approximation generally provides a close match to the exact results, though discrepancies are visible at higher $ n $, particularly in the outgoing flux components. Parameters: $p=12$, $a=0.9$, $e=0.2$, $x_I=\sqrt{3}/2$, $l=2$, $m=2$, $k=0$. For the exact trajectory, as highlighted earlier, we use $n^s_\textrm{max}=8$, $k^s_\textrm{max}=8$.}
\label{fig:fluxcomparen}
\end{figure}

\begin{figure}[htbp]
\hspace{-0.8cm}
\includegraphics[width=1.07\columnwidth]{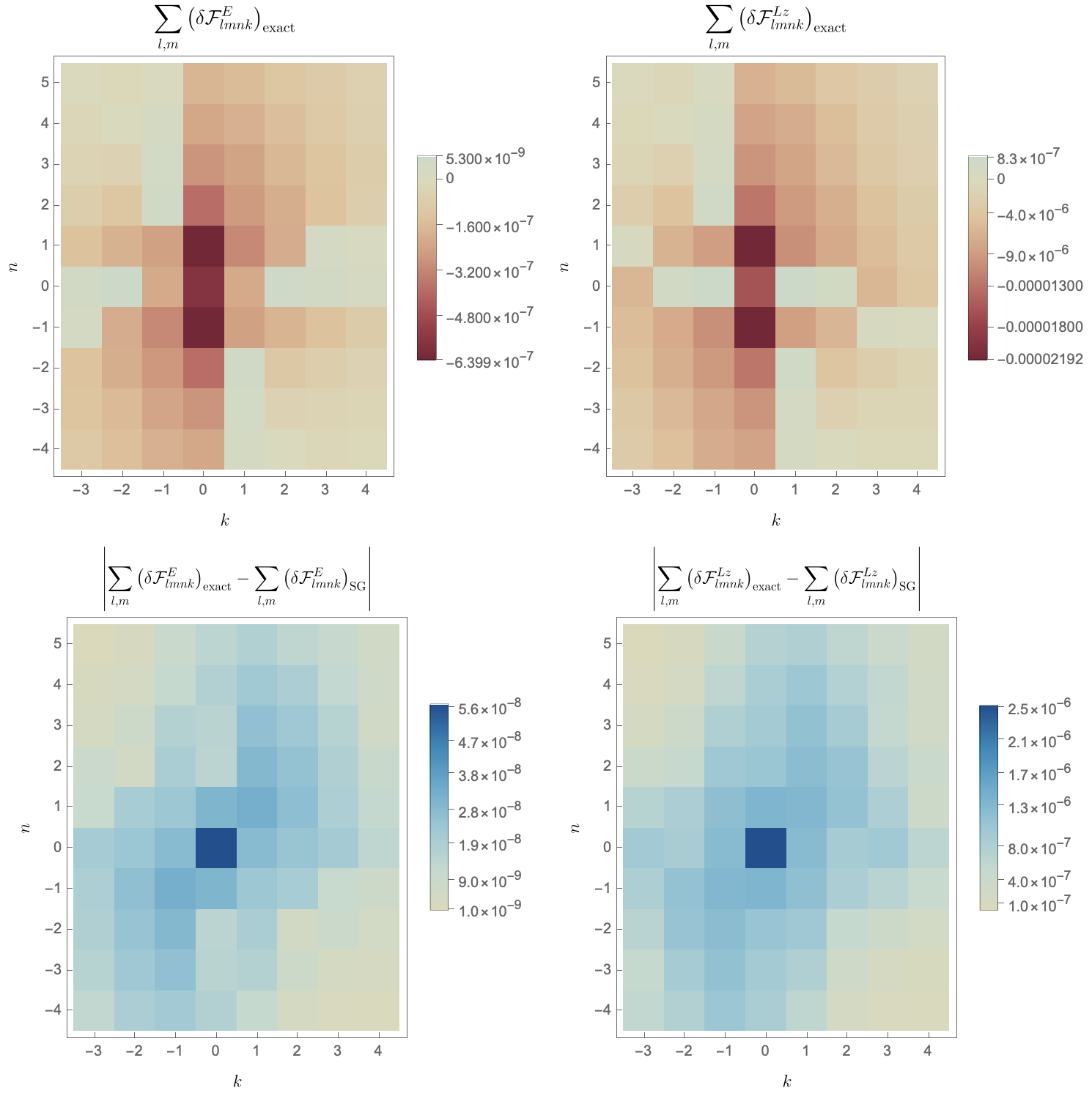}
\caption{Absolute difference between the exact gravitational-wave flux corrections due to the secondary's spin and those computed using the shifted geodesic approximation, shown as a function of $n$ and $k$. The top panels show the exact spin-induced corrections to the energy flux $E$ (left) and angular momentum flux $L_z$ (right), summed over $l$ and $m$. The bottom panels display the corresponding differences between the exact and shifted geodesic fluxes for $E$ (left) and $L_z$ (right).
 Parameters: $p=12$, $a=0.9$, $e=0.2$, $x_I=\sqrt{3}/2$. For the exact trajectory, $n^s_{\textrm{max}}=8$, $k^s_{\textrm{max}}=8$. }
\label{fig:fullsum}
\end{figure}

In Sec.~\ref{sec:modeindices}, we examine convergence of the spinning-body fluxes with respect to harmonic mode indices $n$ and $k$, focusing on the sensitivity of the fluxes to truncation choices and on where the shifted-geodesic (SG) approximation departs from the exact result at the level of individual harmonics. We summarise the figures that appear in Sec.~\ref{sec:modeindices} below:

\begin{enumerate}
\item \textbf{Mode truncation order for spinning-body fluxes (Fig.~\ref{fig:convcriteria}).}
We first determine a suitable truncation of the spinning-body mode sum by examining the convergence of the exact spinning-body energy flux with respect to the radial harmonic index $n$. Figure~\ref{fig:convcriteria} establishes the minimum $n_{\max}$ required to ensure that the flux computation itself is converged across the relevant range of $(p,e,x_I)$. This choice defines the baseline used in all subsequent comparisons.

\item \textbf{SG vs.\ exact flux comparison across $n$ (Fig.~\ref{fig:fluxcomparen}).}
Fig.~\ref{fig:fluxcomparen} compares exact and shifted-geodesic (SG) spinning-body flux corrections at fixed $(l,m,k)$ as functions of the harmonic index $n$ for representative orbital parameters. This figure highlights at which modes the SG approximation deviates the most from the exact method.

\item \textbf{SG errors in $(n,k)$-space (Fig.~\ref{fig:fullsum}).}
Figure~\ref{fig:fullsum} extends the previous comparison to the full $(n,k)$-plane while summing over $(l,m)$, showing both the exact flux content and the absolute differences between the exact and SG results. This makes explicit which regions of mode space dominate the SG error.
\end{enumerate}

Having quantified the accuracy of the approximations in mode space, in Sec.~\ref{sec:orbitalparams} we investigate how this accuracy varies across the orbital parameter space. We do so using comparisons designed to isolate which secondary-spin effects are the most important and to determine the region of parameter space over which the approximate models satisfy the adiabatic accuracy criterion. In particular, we compute the spinning-body fluxes under the various approximations while varying $p$ and $e$. We present the following figures in Sec.~\ref{sec:orbitalparams}:

\begin{enumerate}
\setcounter{enumi}{3}

\item \textbf{Diagnostic comparison (exact, SG, $\Upsilon_i^S=0$ and $\delta\mathcal{C}_j=0$) versus orbital parameters (Figs.~\ref{fig:fluxcomparep} and~\ref{fig:fluxcompareecc}).}
We examine how the agreement between SG and the exact result depends on the orbital parameters. Figures~\ref{fig:fluxcomparep} and~\ref{fig:fluxcompareecc} compare exact fluxes to SG, fluxes without frequency shifts in the trajectory ($\delta\Upsilon_i^S=0$) and fluxes without  integral-of-motion shifts in the trajectory  ($\delta\mathcal{C}_j=0$), as functions of $p$ and $e$. These comparisons show which components of the SG approximation are most important for capturing the dominant spinning-secondary effects.

\item \textbf{Motivating the leading-order (LO) model (Fig.~\ref{fig:comparepvalues_leadingharmonics}).}
Figure~\ref{fig:comparepvalues_leadingharmonics} quantifies the improvement obtained by including a small number of oscillatory harmonics beyond SG. By comparing absolute errors relative to the exact result, this figure shows that a low-harmonic truncation yields a substantial accuracy gain at modest additional cost, motivating the leading-order (LO) extension.

\item \textbf{Accuracy of SG and LO across orbital parameter space (Figs.~\ref{fig:parameterspaceE} and~\ref{fig:parameterspaceLz}).}
Finally, Figs.~\ref{fig:parameterspaceE} and~\ref{fig:parameterspaceLz} summarize the accuracy of the SG and LO approximations across $(p,e,I)$ by presenting fractional differences relative to the exact fluxes. These maps delineate the regime of validity of each approximation.
\end{enumerate}

We find that shifts in both the fundamental frequencies and the constants of motion must be included in order to capture the dominant effects of secondary spin in the gravitational-wave fluxes.  In contrast, the oscillatory corrections typically contribute subdominantly and can often be truncated without significantly degrading accuracy.  In some regions of parameter space, particularly as the orbit approaches the LSO, the accuracy of the shifted-geodesic approximation degrades.

Motivated by this behavior, we assess the leading-order (LO) spinning-body approximation that supplements the shifted-geodesic model with the four lowest oscillatory harmonics in the trajectory, $n^s_{\max}=4$ and $k^s_{\max}=4$.  This extension improves agreement with the exact fluxes while retaining much of the computational efficiency of the SG approximation, as shown in Sec.\ ~\ref{sec:orbitalparams}.

Throughout, we regard an approximation as sufficiently accurate for practical adiabatic inspiral calculations when the resulting flux differences fall below the tolerance $\epsilon_{\rm ad}=10^{-7}$ \cite{Khalvati:2025znb}; this threshold also sets the target accuracy for post-adiabatic terms \cite{Burke2024}, as discussed in Sec.~\ref{sec:modeindices}.

\subsection{Dependence on mode indices}
\label{sec:modeindices}
 In this section, we examine the convergence of the secondary-spin flux contributions as a function of mode indices $n$ and $k$. This is important as truncating the mode sum too early can lead to biases in parameter estimation of EMRI systems \cite{Khalvati:2025znb}. 

Adiabatic fluxes are typically computed by truncating the mode sum when individual contributions fall below some threshold $\epsilon_{\rm ad}$ of the total. We define the truncation tolerance at each post-adiabatic order by requiring that the first neglected mode satisfies
\begin{equation}
|\mathcal{F}_{n_{\max}+1}^{E}| \le \epsilon_{\textrm{ad}} \left|\sum_{n=-n_{\max}}^{n_{\max}} \mathcal{F}_n^{E}\right|.
\end{equation}
The choice of $\epsilon_{\rm ad}$ determines values for $(n_{\text{max}}, k_{\text{max}})$ and allows us to assess the convergence of the fluxes computed from spinning body orbits. The most thoroughly tested convergence criteria for EMRI flux generation are those developed for the FastEMRIWaveforms (FEW) models \cite{Chua:2020stf,Katz:2021yft,Speri:2023jte,Chapman-Bird2025}, which thus far have been applied primarily to equatorial orbits. Following these analyses, we use $\epsilon_{\rm ad} = 10^{-7}$, a value that has been used in recent studies of adiabatic EMRI waveform models \cite{Khalvati:2025znb}. The present work addresses fully generic orbits, where the convergence behavior of the mode-sum in the $0$PA limit has not been explored to the same depth.  To our knowledge, Ref.\ \cite{Hughes2021} provides the most up-to-date published study of $0$PA generic EMRI inspirals.  For guidance, we base some elements of our analysis on Ref.\ \cite{Hughes2021}, particularly the truncation choice for the ($l,m$)-mode summation. 

As shown in \ref{sec:convcriteriapostad}, controlling the total truncation error in a post-adiabatic expansion implies that the convergence tolerance for the $i$th-order flux correction may be relaxed relative to the leading-order adiabatic tolerance, with 
\begin{equation}
\label{eq:convcriteria1}
\epsilon_i \lesssim \varepsilon^{-i}\epsilon_{\rm ad}.
\end{equation}
Since the tolerance for the leading-order adiabatic fluxes is set to $\epsilon_{\textrm{ad}} \sim 10^{-7}$, Eq.~\eqref{eq:convcriteria1} implies that a looser threshold of $\epsilon_i \sim 10^{-2}$ is sufficient for the post-adiabatic spinning-secondary correction when the mass ratio is $\varepsilon \sim 10^{-5}$. This is in line with the recommendation for the fractional accuracy of the adiabatic fluxes given in Ref.~\cite{Khalvati:2025znb} and the recommendation for the fractional accuracy of $\sim 10^{-2}$ for the post-adiabatic corrections given in \cite{Burke:2021xrg}.  At less extreme mass ratios than $\varepsilon = 10^{-5}$, a more stringent criterion (i.e., a smaller value of $\epsilon_i$) may need to be used (though one should bear in mind that linear perturbation theory is likely to be quite a bit less reliable as we push to less extreme mass ratios).

We begin by examining flux convergence with the aim of identifying suitable choices for $n_{\max}$ and $k_{\max}$ when evaluating spinning-secondary corrections. To guide our choice of truncation thresholds for the post-adiabatic corrections, we examine the convergence behavior of the exact spinning-body fluxes in detail. Figure~\ref{fig:convcriteria} visualizes this convergence with respect to the $n$-mode summation. Specifically, the left column of the figure shows the ratio $ \left| \delta \mathcal{F}_{n_{\max}} \right| \Big/ \left|\sum_{n \leq n_{\max}}  \delta \mathcal{F}_n \right| $, summed over $l$, $m$, and $k$, plotted across the two-dimensional space of orbital parameters $(p, e)$, with rows corresponding to different inclination angles $I$. This quantity represents the relative contribution of the final retained mode to the total spin-induced flux and provides a direct measure of convergence. Smaller values of the convergence ratio indicate that the flux has effectively converged by the time the sum reaches $n_{\max}$. From the results shown in Fig.~\ref{fig:convcriteria}, we find that adopting $n_{\max} = 5$ and $k_{\max} = 4$ yields a fractional error that ranges from $\sim 10^{-7}$ at low eccentricity to $\sim 10^{-2}$ at higher eccentricity. According to the criteria Eq.\ \eqref{eq:convcriteria1}, These values therefore provide a reasonable choice for computing the spin-induced flux corrections over these parameter ranges.

In Figures~\ref{fig:fluxcomparen} and \ref{fig:fullsum}, we examine how the flux corrections depend on the Teukolsky indices $n$ and $k$. Figure~\ref{fig:fluxcomparen} shows that both the exact and shifted-geodesic approximations follow a similar overall trend in $n$, with flux contributions decaying at higher harmonics as expected.  Discrepancies between the two methods are more pronounced in the outgoing fluxes to infinity (left panels), where the shifted-geodesic approximation tends to overestimate contributions at higher $ n $. In contrast, the ingoing fluxes at the horizon (top right and bottom right panels) exhibit agreement across all harmonics. The largest deviations between the shifted-geodesic and exact trajectories occur at high $n$, where flux contributions are typically smaller and less significant in the overall sum. This reflects the fact that the ``exact'' motion contains high-frequency structure not present in the shifted geodesic approximation due to the neglect of various oscillatory terms.

Figure~\ref{fig:fullsum} compares the exact spinning-body gravitational-wave flux corrections to those computed using the shifted geodesic approximation as a function of mode indices $n$ and $k$. The top panels show the exact energy and angular momentum flux corrections (summed over $l$ and $m$), while the bottom panels display the absolute differences between the exact and approximate results, displaying the accuracy of the shifted geodesic approach across mode space. 

\subsection{Dependence on orbital parameters}
\label{sec:orbitalparams}
\begin{figure}[htbp]
\hspace{-0.6cm}
\includegraphics[width=1.07\columnwidth]{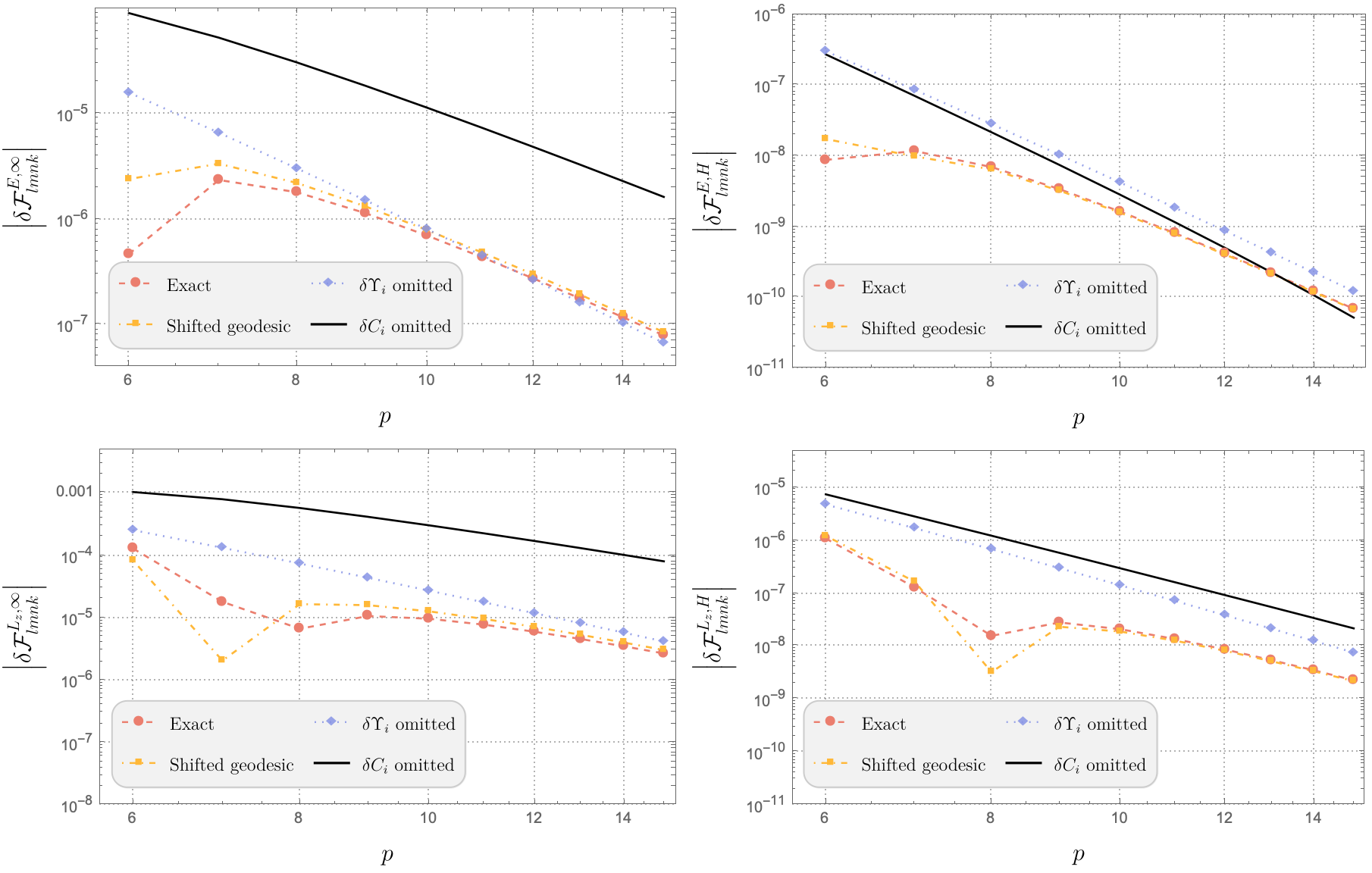}
\caption{Corrections to the gravitational wave fluxes due to the spin of the secondary as a function of semilatus rectum, $p$. The four panels correspond to the same flux components as Figure~\ref{fig:fluxcomparen}, with the top left showing the outgoing energy flux to infinity, the top right showing the ingoing energy flux at the horizon, and the bottom panels representing the corresponding angular momentum fluxes.  In each panel, the red dashed curve with circular markers represents the exact trajectory, while the orange dot-dashed curve with square markers corresponds to the shifted geodesic approximation. The blue dotted curve represents the case where frequency shifts due to secondary spin, $\delta \Upsilon^S_i$, are omitted, and the solid black curve corresponds to the case where the corrections to the constants of motion due to secondary spin, $\delta \mathcal{C}_i$, are omitted. The results demonstrate that omitting $\delta C_i$ leads to the largest deviations, while the shifted geodesic approximation provides the best agreement with the exact trajectory. Parameters for all curves: $a=0.9M$, $e=0.2$, $x_I=\sqrt{3}/2$, $l=2$, $m=2$, $n=0$, $k=0$. For the exact trajectory, $n^s_{\textrm{max}}=8$, $k^s_{\textrm{max}}=8$.}
\label{fig:fluxcomparep}
\end{figure}

\begin{figure}[htbp]
\hspace{-0.6cm}
\includegraphics[width=1.07\columnwidth]{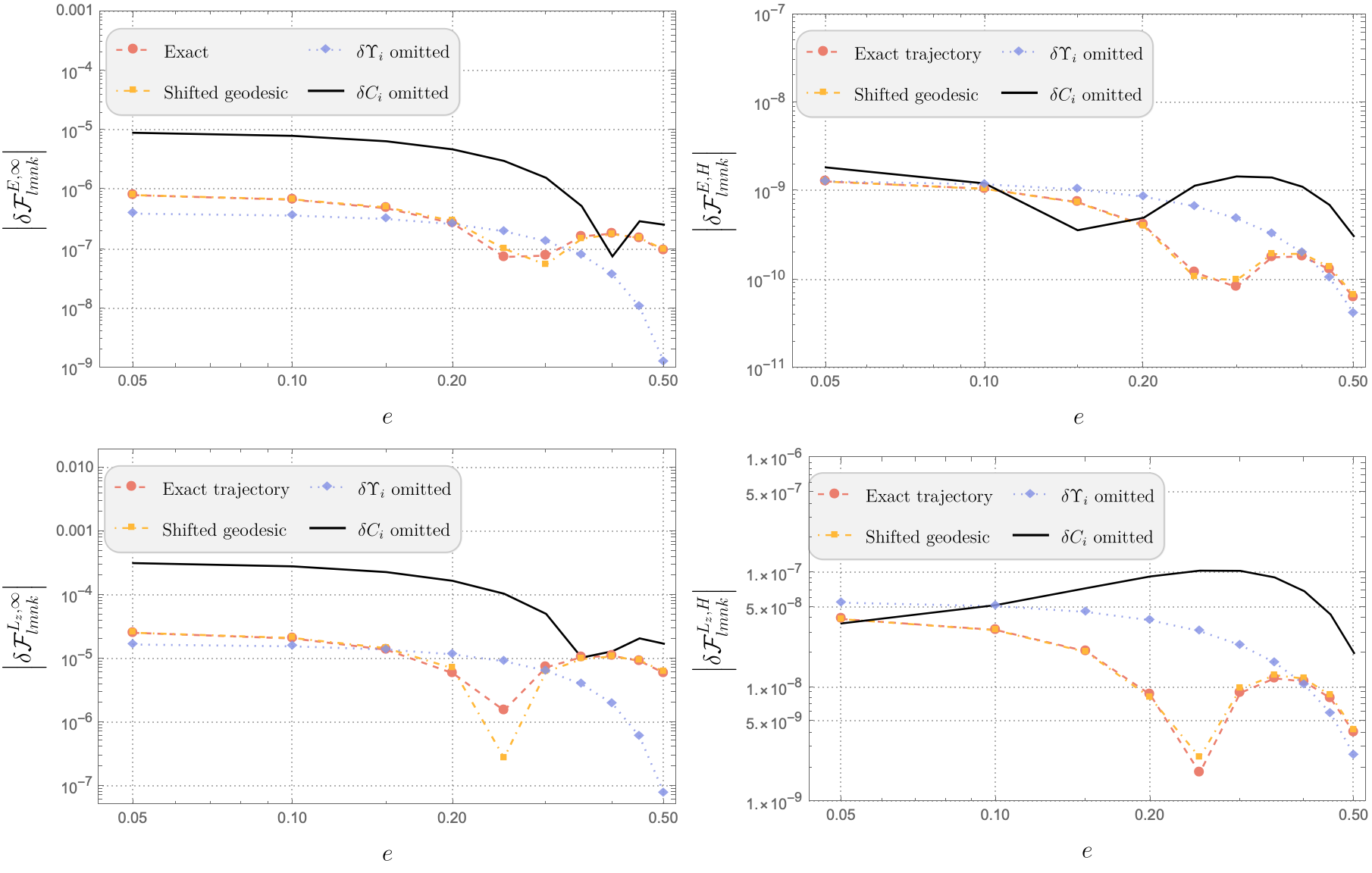}
\caption{Corrections to the gravitational wave energy flux due to the spin of the secondary as a function of eccentricity $e$. The four panels display the same quantities as in Fig.~\ref{fig:fluxcomparep}, following the same color and line style conventions. Similar to the $p$-dependence, we observe that neglecting the constant of motion shifts $\delta \mathcal{C}_i$ results in the largest discrepancies, 
whereas the shifted geodesic approximation provides the closest match to the exact trajectory.  Parameters for all curves: $a=0.9M$, $p=12$, $x_I=\sqrt{3}/2$, $l=2$, $m=2$, $n=0$, $k=0$. For the exact trajectory, $n^s_{\textrm {max}}=8$, $k^s_{\textrm {max}}=8$.}
\label{fig:fluxcompareecc}
\end{figure}

\begin{figure}[htbp]
\hspace{-0.6cm}
\includegraphics[width=1.07\columnwidth]{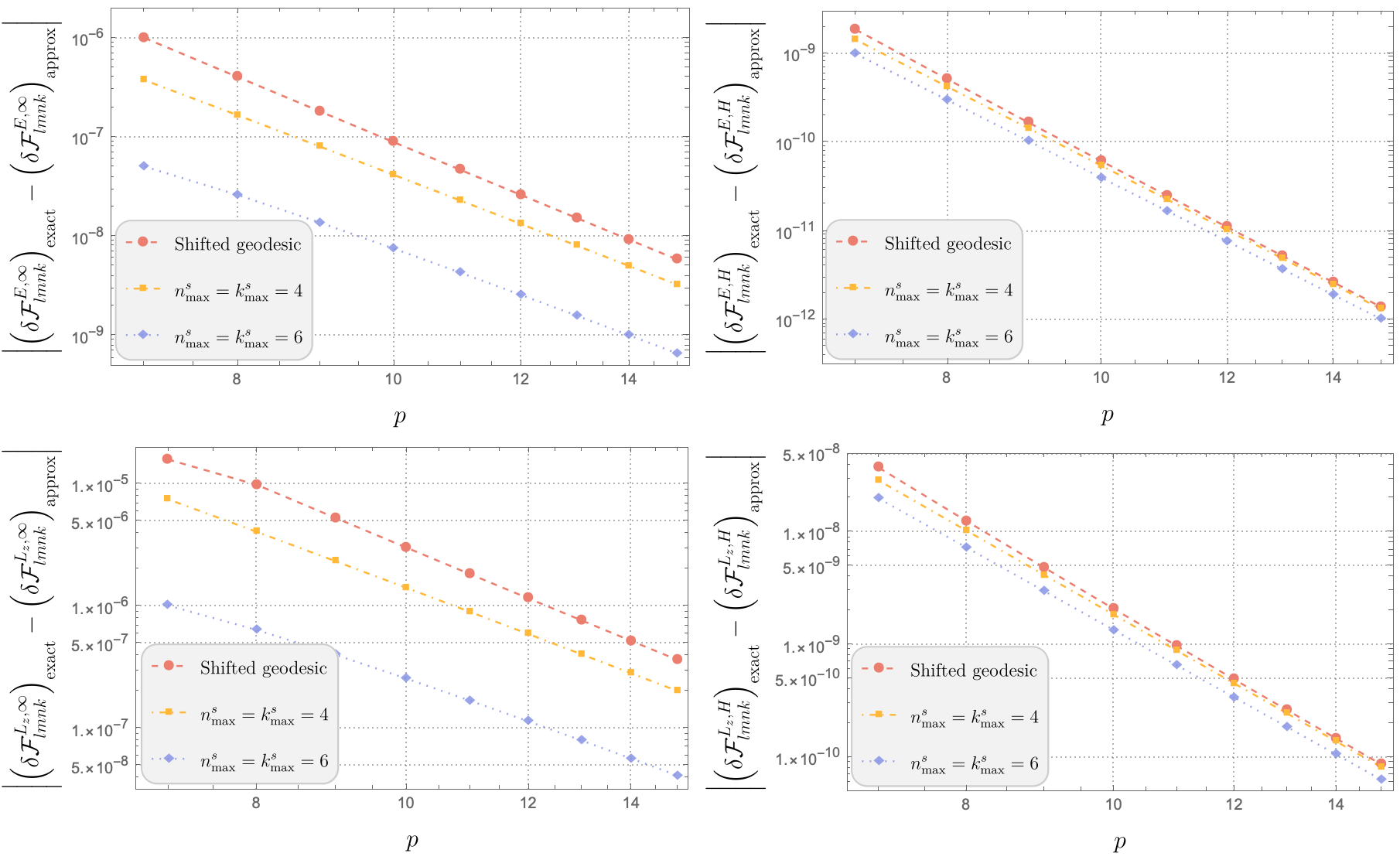}
\caption{Absolute difference between the exact gravitational wave flux corrections due to the secondary's spin and those computed using trajectories with varying numbers of harmonics, as a function of $p$. The $x$-axis represents the semilatus rectum, $p$, and the $y$-axis represents the absolute difference in fluxes. The four panels correspond to the same flux components as Figures \ref{fig:fluxcomparen}, ~\ref{fig:fluxcomparep} and \ref{fig:fluxcompareecc} with the top left showing the outgoing energy flux to infinity, the top right showing the ingoing energy flux at the horizon, and the bottom panels representing the corresponding angular momentum fluxes. The red dashed line corresponds to the fluxes computed with the shifted geodesic trajectory, the orange dash-dotted line includes the first four leading harmonics, and the blue dotted line includes two additional harmonics in the trajectory. Parameters for all curves: $a=0.9M$, $e=0.2$, $x_I=\sqrt{3}/2$, $l=2$, $m=2$, $n=0$, $k=0$. For the exact trajectory, $n^s_{\textrm{max}}=8$, $k^s_{\textrm{max}}=8$.}
\label{fig:comparepvalues_leadingharmonics}
\end{figure}

\begin{figure}[htbp]
\centering
\includegraphics[width=0.7\columnwidth]{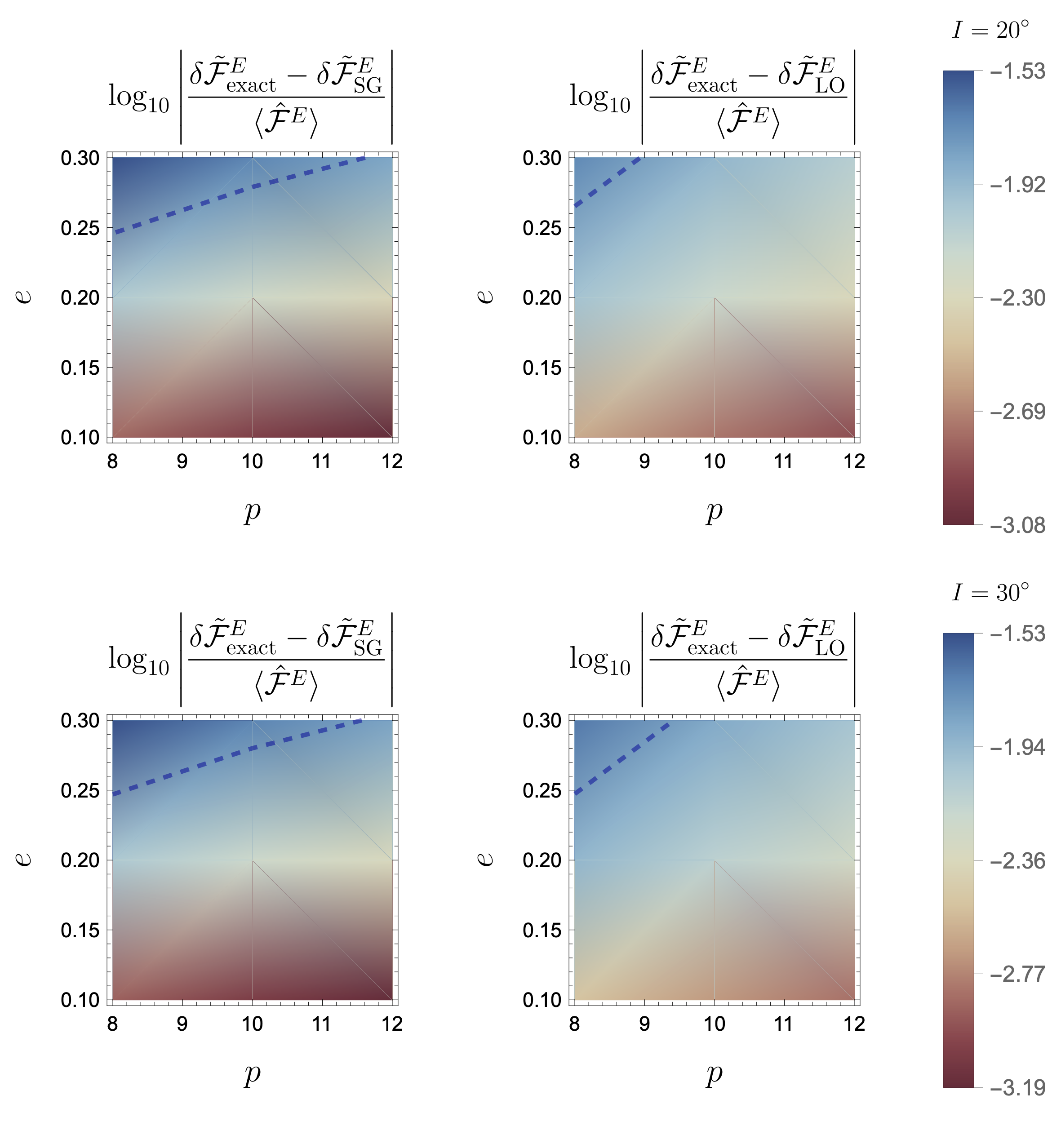}
\caption{Fractional differences in the secondary spin contributions to the total energy flux $\delta \tilde{\mathcal{F}}^E$ across a range of semi-latus rectums $p$, eccentricities $e$, and inclinations $I$. Each flux is computed using three methods: (1) the full ``exact'' result including all secondary spin-dependent corrections, (2) the ``shifted geodesic'' (SG) approximation, and (3) the ``leading order'' (LO) correction to SG. Left panels show the fractional difference between SG and exact fluxes; right panels show the difference between LO and exact. Rows correspond to $I = 20^\circ$ and $30^\circ$. The color scale indicates $\log_{10}\left|\delta \tilde{\mathcal{F}}^E_\textrm{exact}-\delta \tilde{\mathcal{F}}^E_\textrm{approx}\right|/\left|\mathcal{F}^E_{\mathrm{exact}}\right|$, with the dashed dark blue contour marking a threshold of $10^{-2}$, the level below which the shifted-geodesic contributions are sufficiently accurate for mass ratios $\varepsilon \lesssim 10^{-5}$. The leading order (LO) correction maintains high accuracy throughout the explored parameter space, with fractional differences mostly remaining below the $10^{-2}$ threshold marked by the dashed dark blue contour. Parameters: $a=0.9M$, $n_\textrm{max}=6$, $k_\textrm{max}=4$. For the exact trajectories, $n^s_{\textrm{max}}=8$, $k^s_{\textrm{max}}=8$. }
\label{fig:parameterspaceE}
\end{figure}

\begin{figure}[htbp]
\centering
\includegraphics[width=0.7\columnwidth]{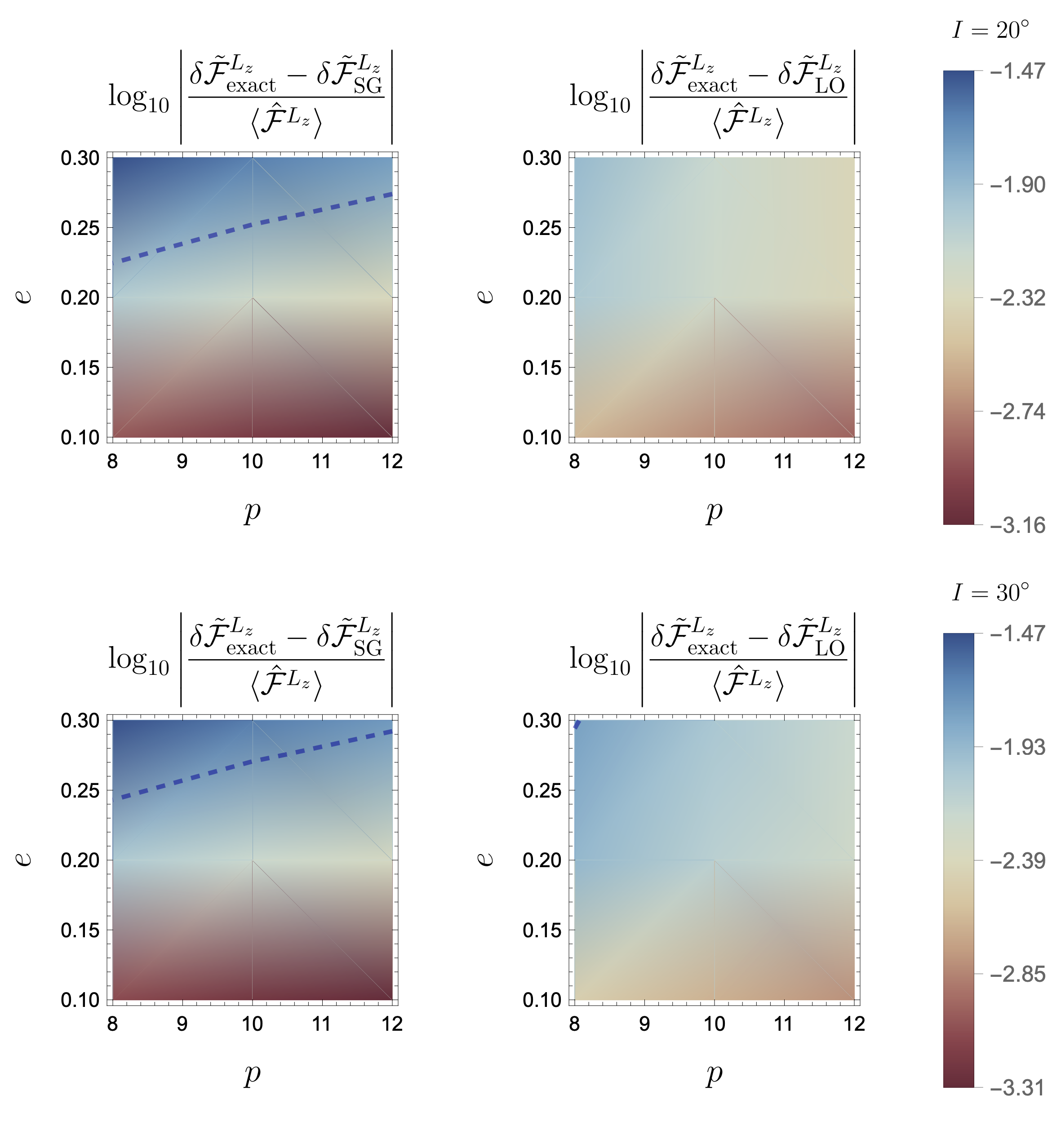}
\caption{Same as Fig.~\ref{fig:parameterspaceE}, but for the $z$-component of the angular momentum flux $\delta \tilde{\mathcal{F}}^{L_z}$. The overall behavior mirrors the energy flux results: the shifted geodesic (SG) approximation yields percent-level accuracy at low eccentricity and inclination, but breaks down at higher values. The leading-order (LO) result lies entirely below this threshold and therefore there is no dashed blue contour plotted on the right panels.  Parameters: $a=0.9M$, $n_\textrm{max}=6$, $k_\textrm{max}=4$. For the exact trajectories, $n^s_{\textrm{max}}=8$, $k^s_{\textrm{max}}=8$. }
\label{fig:parameterspaceLz}
\end{figure}

In this section, we investigate the accuracy of different choices of approximation across orbital parameter space. We begin by evaluating the accuracy of the shifted-geodesic approximation across a range of semilatus rectum $p$ and eccentricity $e$ values. Figures~\ref{fig:fluxcomparep}, ~\ref{fig:fluxcompareecc}, and ~\ref{fig:comparepvalues_leadingharmonics} compare different approximations for spin-induced corrections to the gravitational wave energy and angular momentum fluxes. Each figure contains four panels, corresponding to the outgoing and ingoing energy fluxes ($\mathcal{F}^{E,\infty}$ and $\mathcal{F}^{E,H}$) and the outgoing and ingoing angular momentum fluxes ($\mathcal{F}^{L_z,\infty}$ and $\mathcal{F}^{L_z,H}$).  

Figures~\ref{fig:fluxcomparep} and~\ref{fig:fluxcompareecc} compare three approximations. The first is the \textit{shifted-geodesic approximation}, which neglects oscillatory shifts in the trajectory due to secondary spin ($\delta\chi_r^S$, $\delta\chi_z^S$, $\delta\mathcalligra{r}^S$, and $\delta\mathcalligra{z}^S$). The other two approximations involve omitting either the shifts in constants of motion ($\delta \mathcal{C}_i$) or the shifts in frequencies ($\delta \Upsilon_i$). In the figures, the exact trajectory is shown with red dashed curves, the shifted-geodesic approximation with orange dot-dashed curves, the approximation omitting frequency shifts with blue dotted curves, and the approximation omitting integral-of-motion shifts with solid black lines. 

In both figures, the largest discrepancies occur when $\delta C_i$ is neglected (solid black curves), indicating that shifts in the constants of motion play a dominant role in determining the correct fluxes. The frequency shifts ($\delta \Upsilon_i$) also introduce significant errors, particularly for the horizon fluxes ($\mathcal{F}^{L_z,H}$ and $\mathcal{F}^{E,H}$). Overall, the shifted-geodesic approximation -- where oscillatory corrections are omitted -- provides the best agreement with the exact trajectory compared to approximations that neglect either frequency shifts ($\delta \Upsilon_i$) or integral-of-motion of shifts ($\delta C_i$). These results suggest that the dominant effects of secondary spin on the trajectory arise from adjustments to the constants of motion and frequency shifts, rather than from oscillatory corrections. For practical modeling, the shifted-geodesic approximation thus captures the most important effects of the spinning-body trajectory.

Figure~\ref{fig:comparepvalues_leadingharmonics} further investigates the role of oscillatory corrections as a function of $p$. Here, we compare the shifted-geodesic approximation (which includes no oscillatory corrections) to cases where 4 or 6 harmonics are included in the secondary-spin correction to the trajectory. The figure shows the absolute difference between the exact fluxes and those computed using different numbers of harmonics in the orbital trajectory correction. As expected, incorporating more harmonics systematically improves accuracy across all flux components. Including the first four leading harmonics (orange dash-dotted line) can reduce the error in the flux by $\sim$ half an order of magnitude for the infinity fluxes, while incorporating two additional harmonics (blue dotted line) suppresses flux differences by more than an order of magnitude. 

To assess the accuracy of the shifted-geodesic (SG) and leading-order (LO; defined by $n^s_{max} = k^s_{max} = 4$), approximations, we compute the secondary spin contributions to the gravitational wave energy flux $\langle\delta \mathcal{F}^E\rangle$ and $z$-component of angular momentum flux $\langle\delta \mathcal{F}^{L_z}\rangle$ across a grid of semi-latus rectum $p$, eccentricity $e$, and inclination angle $I$. The fluxes are obtained via mode summation over $(l, m, k, n)$ using Teukolsky amplitudes, and compared against the full ``exact'' fluxes incorporating all spin-dependent terms. The fractional differences between the SG and exact fluxes, as well as between the LO and exact fluxes, are plotted in Figs.~\ref{fig:parameterspaceE} and~\ref{fig:parameterspaceLz} for two inclination angles: $I =20^\circ$, and $30^\circ$. The data is linearly interpolated in $(p, e)$ space to generate smooth color maps of the log-scaled fractional errors. In both the energy and angular momentum fluxes, the SG approximation performs well in the low-eccentricity, low-inclination regime, but exhibits growing deviations at higher values of $e$ and $I$. Moreover, the error increases at smaller values of the semi-latus rectum, indicating reduced accuracy of the SG approximation in the strong-field regime near the black hole. 

As discussed in the previous section, in the context of adiabatic waveform models, the $n$-mode summation is truncated when the fractional contribution of additional modes falls below a fixed threshold, chosen to be $\sim10^{-7}$ relative to the cumulative sum. When evaluating secondary spin corrections, the typical amplitude of these post-adiabatic terms is several orders of magnitude smaller than the leading-order geodesic flux, proportional to the mass-ratio. If the secondary-spin contributions are suppressed by a mass-ratio of $\varepsilon\sim10^{-5}$, it is sufficient to have fractional errors in the secondary spin contributions that are less than the $10^{-2}$ level.  A dark blue contour line marks the threshold where the fractional error reaches $10^{-2}$, corresponding to the level at which post-adiabatic corrections are typically negligible for extreme mass ratios $\mu/M \lesssim 10^{-5}$. Importantly, the LO correction significantly reduces the error across the entire parameter space, with almost all regions remaining below the $10^{-2}$ threshold. These results show that the SG approximation remains accurate in a subset of the tested parameter space, with the LO extension captures spin-induced corrections more accurately across the entire domain. Therefore, the LO approach in particular supports accurate and computationally efficient modeling of spinning-secondary fluxes in EMRIs.

These results demonstrate that both the SG and LO approximations may have particularly good performance in the weak-field regime ($p \gtrsim 10$), where IMRIs spend a substantial fraction of their inspiral, with higher eccentricity and higher inclination configurations still modeled accurately. In this regime, the SG or LO approximations may be particularly effective for computing fluxes and exploring a large portion of parameter space efficiently. Deep in the strong field, one can switch to a more accurate scheme and map the flux corrections between different spin gauges if necessary.

\section{Approximate inspiral and dephasing estimate}
\label{sec:inspiral}
In this section, we quantify the dephasing from the shifted geodesic (SG) and leading-order (LO) approximations by evolving a realistic inspiral, representative of a canonical EMRI. Using the orbit-averaged equations from Sec.~\ref{sec:evalualtingSGfluxes}, we evolve the same initial data and flux interpolants under two models: a reference evolution that retains the leading adiabatic fluxes and their $\varepsilon$-order spin corrections, and an SG/LO evolution in which the motion is approximated by shifted constants of motion and frequencies. We then compare the resulting orbital phases $w_i$ (and, by extension, the accumulated GW phase) to isolate the dephasing attributable to the SG approximation. The aim here is purely diagnostic (i.e., to assess the accuracy of the approximation) not to present a production-grade generic inspiral model; accordingly, we omit ingredients that should not materially affect this comparison but would be required in a full pipeline.

\begin{figure}[htbp]
\centering
\resizebox{0.9\textwidth}{!}{%
\begin{tikzpicture}[node distance=0.5cm and 0.8cm]

\fill[gray!30!, fill opacity=0.5] (-2, -2) rectangle (15.5, 1.5);
\draw[gray!50!, line width=1.5pt] (-2, -2) rectangle (15.5, 1.5);

\begin{scope}[shift={(-0.1cm, 0.6cm)}]
\node (kerr) [box] {\footnotesize{Kerr geodesics and } \\ \footnotesize{parallel transport}};
\node (spin) [box, right=of kerr] {\footnotesize{Spin-curvature} \\\footnotesize{force  $f^\alpha_S$}};
\node (dipole) [box, below=of spin] {\footnotesize{Dipole term} \\ \footnotesize{in $T_{\mu\nu}$}};
\node (kinematics) [box, right=of spin] {\footnotesize{Spin-perturbed constants} \\ \footnotesize{$(\delta\Upsilon_r^S,\delta\Upsilon_z^S,\delta\Upsilon_\phi^S,\delta E^S,\delta L_z^S,\delta K^S)$}};
\node (frequencies) [box, below=of kinematics] {\footnotesize{Spin-perturbed orbital} \\ \footnotesize{ kinematics $(\delta \chi_r^S,\delta \chi_z^S,\delta \mathcalligra{r}^S,\mathcalligra{z}^S$)}};
\node (fluxes) [box, right=of kinematics] {\footnotesize{Corrections to} \\\ \footnotesize{GW fluxes $\delta\mathcal{F}^b$}};
\node (inspiral) [highlight, right=of frequencies] {\footnotesize{Inspiral with} \\ \footnotesize{secondary spin}};
\node (full) [thickbox, below=of kerr] {\footnotesize{\textbf{Full spinning-}} \\ \footnotesize{\textbf{body calculation}}};

\draw [arrow] (kerr) -- (spin);
\draw [arrow] (kerr) -- (dipole);
\draw [arrow] (spin) -- (kinematics);
\draw [arrow] (dipole) -- (fluxes);
\draw [arrow] (kinematics) -- (frequencies);
\draw [arrow] (kinematics) -- (fluxes);
\draw [arrow] (frequencies) -- (fluxes);
\draw [arrow] (kinematics) -- (inspiral);
\draw [arrow] (fluxes) -- (inspiral);
\end{scope}

\fill[gray!30!, fill opacity=1] (-2, -5.5) rectangle (15.5, -2);
\draw[gray!50!, line width=1.5pt] (-2, -5.5) rectangle (15.5, -2);
\begin{scope}[shift={(-0.1cm, -3cm)}]
\node (kerr) [box] {\footnotesize{Kerr geodesics and } \\ \footnotesize{parallel transport}};
\node (spin) [box, right=of kerr] {\footnotesize{Spin-curvature} \\\footnotesize{force $f^\alpha_S$}};
\node (dipole) [box, below=of spin] {\footnotesize{Dipole term} \\ \footnotesize{in $T_{\mu\nu}$}};
\node (frequencies) [box, right=of spin] {\footnotesize{Closed-form corrections} \\ \footnotesize{$(\delta\Upsilon_r^S,\delta\Upsilon_z^S,\delta\Upsilon_\phi^S,\delta E^S,\delta L_z^S,\delta K^S)$}};
\node (fluxes) [box, below=of frequencies] {\footnotesize{Corrections to} \\\ \footnotesize{GW fluxes $\delta\mathcal{F}^b$}};
\node (inspiral) [highlight, right=of fluxes] {\footnotesize{Inspiral with} \\ \footnotesize{secondary spin}};
\node (full) [thickbox, below=of kerr] {\footnotesize{\textbf{Shifted-geodesic }} \\ \footnotesize{\textbf{calculation}}};

\draw [arrow] (kerr) -- (spin);
\draw [arrow] (kerr) -- (dipole);
\draw [arrow] (spin) -- (frequencies);
\draw [arrow] (dipole) -- (fluxes);
\draw [arrow] (frequencies) -- (fluxes);
\draw [arrow] (frequencies) -- (inspiral);
\draw [arrow] (fluxes) -- (inspiral);
\end{scope}

\end{tikzpicture}}
\caption{Schematic flowchart illustrating the structure of the complete spinning-body calculation. At leading order, the dynamics are governed by geodesic motion and parallel transport in Kerr spacetime, which inform the computation of the spin-curvature force and dipole contributions to the stress-energy tensor. These effects perturb the orbital kinematics; the influence of these corrections determines the inspiral evolution of a spinning secondary. The lower panel illustrates the simplified structure of the \textit{shifted-geodesic} (SG) approximation. Instead of solving for the fully perturbed trajectory, we use closed-form Hamilton-Jacobi corrections to the constants of motion and fundamental frequencies. These corrections feed the flux computation and, in turn, drive an approximate inspiral that captures spin effects at reduced cost compared to the full spinning-body calculation. 
}
\label{fig:flowchart}
\end{figure}
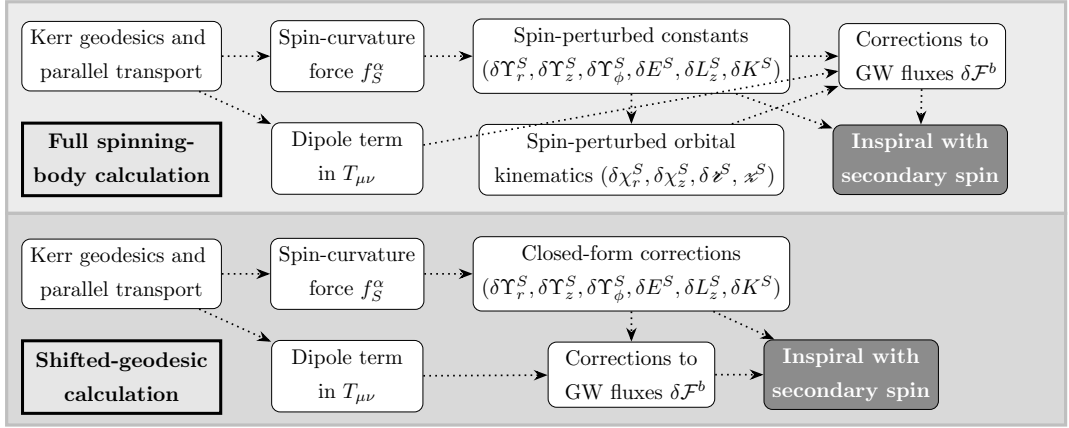

As discussed in Section \ref{sec:evalualtingSGfluxes}, after an averaging (NIT) transformation we obtain:
\begin{subequations}\label{eq:transformed_EoM3}
\begin{align}
\frac{d \nit{\mathcal{C}}_{b}}{d \lambda} &=\varepsilon \nit{\hat{\mathcal{F}}}^b(\vec{\mathcal{C}}) + \varepsilon^2 \delta \nit{\mathcal{F}}^b(\vec{\mathcal{C}}) +  \mathcal{O}(\varepsilon^3)\;, \\
\frac{d w_i}{d\lambda} &= \hat\Upsilon_i(\vec{\mathcal{C}}) +\varepsilon \delta\Upsilon^S_i(\vec{\mathcal{C}}) + \mathcal{O}(\varepsilon^2)\;.
\end{align}
\end{subequations}
These fluxes govern the secular evolution of $\vec{\mathcal{C}}=(E,J_z,K)$ and, via the Jacobian $\mathbf{J}=\partial \vec{\mathcal{C}}/\partial\vec{P}$, yield evolution equations for the orbital elements $\vec{\mathcal{P}}=(p,e,x_I)$ (see: e.g., Refs.~\cite{Skoupy2022,Hughes2021}). Specifically,
\begin{align}
\dot{\mathcal{P}}_a &= \sum_{b\in\{E,J_z,Q\}} (J^{-1})_{ab}\,\dot{\mathcal{C}}_b\;,
\qquad
J_{ba} = \frac{\partial \mathcal{C}_b}{\partial \mathcal{P}_a}\;.
\label{eq:Jacobian-def}
\end{align}
In \ref{ref:explicitev}, we write out the explicit equations for the evolution of orbital parameters. This leads to a coupled system for the orbital elements:
\begin{align}
\frac{d\nit{\mathcal{P}}_a}{d\lambda} &=  \varepsilon\nit{\hat{F}}^a(\vec{\mathcal{P}})+\varepsilon^2\delta \nit{F}^a(\vec{\mathcal{P}}) +  \mathcal{O}(\varepsilon^3)\;, \\
\frac{d w_i}{d\lambda} &= \hat\Upsilon_i(\vec{\mathcal{P}})+\varepsilon\delta\Upsilon^S_i(\vec{\mathcal{P}}) +  \mathcal{O}(\varepsilon^2)\;,
\end{align}
where $\nit{\hat{F}}^a$ are the leading-order adiabatic fluxes, and $\delta \nit{F}^a$ the first-order secondary-spin corrections to the fluxes; both are now expressed as functions of $\vec{\mathcal{P}}$ rather than the original variables $\vec{\mathcal{C}}$.

We now outline the framework for modeling the inspiral of a spinning body. The specific steps in this calculation are as follows:
\begin{enumerate}
\item Generate the reference geodesic and the linear-in-spin perturbed (spinning-body) trajectory for a selected grid of orbital elements $\vec{\mathcal{P}}=(p,e,x_I)$.
\item Solve the Teukolsky equation along this worldline to obtain mode amplitudes and the orbit-averaged fluxes $\nit{\hat{\mathcal{F}}}^b(\vec{\mathcal{P}})$ and $\delta\nit{\mathcal{F}}^b(\vec{\mathcal{P}})$, treating secondary-spin effects to first order in $\varepsilon$.
\item Build smooth interpolants of these fluxes across $(p,e,x_I)$-space.
\item Map fluxes $\vec{\dot{\mathcal{C}}}$ to orbital element evolution $\vec{\dot{\mathcal{P}}}$ via the Jacobian $J=\partial\vec{\mathcal{C}}/\partial\vec{\mathcal{P}}$ and integrate the averaged equations of motion to evolve $(\vec{\mathcal{P}},\vec{w})$ along the inspiral.
\end{enumerate}
Lastly, we compute the accumulated phase over the inspiral, via
\begin{align}
\Phi_i(t)=\int_0^t\Omega_i(p(t'),e(t'),x_I(t'))\textrm{d}t'\;,
\end{align}
where $\Omega_i\in(\Omega_r,\Omega_z,\Omega_\phi)$. Figure~\ref{fig:flowchart} shows where the shifted-geodesic (SG) approximation framework differs from the full spinning-body calculation. 

In order to generate the necessary data to compute an inspiral, we evaluate fluxes on a grid in eccentricity $e$, inclination parameter $x_I$ and semi-latus rectum $p$. The structure of the flux data grid used here is derived from, but not identical to, the dataset of Ref.~\cite{Hughes2021}. That grid was designed to span a wide region of parameter space and to extend close to the last stable orbit, enabling the study of generic inspirals across a broad range of orbital configurations. This grid is almost certainly not sufficient for precision LISA science studies, but it remains, at present, the only example of flux data covering enough of the generic inspiral parameter space to enable exploratory studies of this type. 

\begin{figure}[htbp]
\hspace{-0.6cm}
\includegraphics[width=1.07\columnwidth]{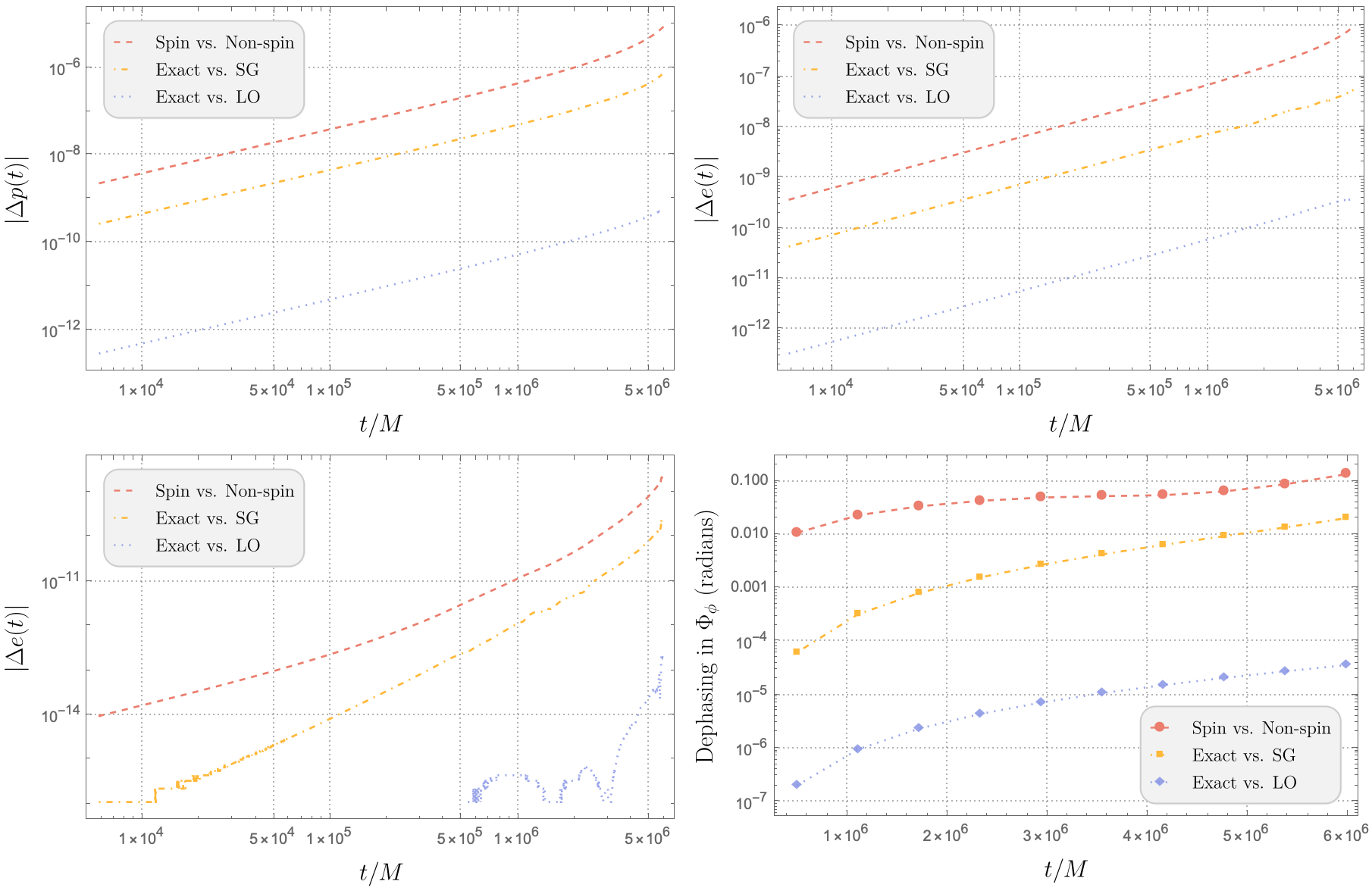}
\caption{Absolute differences in orbital parameters and accumulated gravitational wave phase between spinning-body trajectories and their geodesic counterparts (red dashed), between spinning-body exact results and leading-order (LO) approximations (orange dash-dotted) and between spinning-body exact results and shifted-geodesic (SG) approximations (blue dotted), as functions of coordinate time $t/M$. The four panels correspond to: (top left) semilatus rectum $p$, (top right) eccentricity $e$, (bottom left) inclination parameter $x_I$, and (bottom right) accumulated phase $\Phi_{\rm \phi}$. The $x$-axis shows the elapsed coordinate time in units of $M$, and the $y$-axis shows the absolute difference in the corresponding quantity. Parameters for all curves: $a=0.7M$, $p_\text{init}=8$, $e_\text{init}=0.2$, $x_\text{I,init}=0.95$, $\varepsilon=10^{-5}$.}
\label{fig:dephasing}
\end{figure}

To capture rapid variation in flux quantities with smaller values of $p$, this grid rescales orbit separation $p$ to the variable
\begin{align} 
u \equiv \frac{1}{\sqrt{\,p - \beta}\,}\;,\qquad \beta \equiv 0.9p_{\rm LSO}\;. 
\end{align}
The grid structure we use, as described in Ref.\ \cite{Hughes2021}, consists of a 40-point mesh laid down between $p_{\min}=p_{\mathrm{LSO}}+0.02$ and $p_{\max}=p_{\min} + 10$, uniformly spaced in $u$:
\begin{align}
u_{\min} &= \frac{1}{\sqrt{p_{\min}-\beta}}\;, &
u_{\max} &= \frac{1}{\sqrt{p_{\max}-\beta}}\;, &
\Delta u &= \frac{u_{\min}-u_{\max}}{39}.
\end{align}
(Note that $u_{\rm max} < u_{\rm min}$; the labels ``max'' and ``min'' indicate that these values of $u$ correspond to the maximum and minimum values of $p$.)  The $b$th point of that mesh is
\begin{align}
u_b = u_{\max} + b\,\Delta u\;, \qquad
p_b = \beta + \frac{1}{u_b^2}\;, \qquad b=0,\dots,39.
\end{align}
In the present work, however, we use only a subset of this grid corresponding to $b=0,\dots,10$, covering the range $p\simeq13.63$ down to $p\simeq5.64$. The inspirals we analyze therefore terminate well before the last stable orbit at $p_{\rm LSO}\sim4$. This earlier cutoff is because we restrict the evolution to the region where the post-adiabatic flux corrections are numerically well-behaved and reliably converged. Although the underlying flux grid extends to $p_{\rm LSO}+0.02$, the inspirals shown in Fig.~\ref{fig:dephasing} are stopped at $p\simeq5.64$ in order to remain within this regime. The final points shown in Fig.~\ref{fig:dephasing} therefore correspond to this earlier cutoff rather than to the edge of the flux grid.

For the remaining orbital parameters, we evaluate fluxes on the sets:
\begin{align}
e \in \{0.25,\;0.15,\;0.05\}\;, 
\qquad x_I \in \{0.95,\;0.949,\;0.948\}\;. 
\end{align}
These values are chosen to bracket the inspiral trajectory in $(p,e,x_I)$. Unlike the grid of Ref.~\cite{Hughes2021}, which was constructed to cover a large fraction of parameter space for generic inspirals; the present calculation focuses on a single representative inspiral. Therefore, we require flux data only in the narrow region of parameter space actually traversed by that trajectory. The ranges in $e$ and $x_I$ are selected based on the portion of parameter space covered by the adiabatic inspiral; the spacings are chosen to be at least as fine as those used in Ref.~\cite{Hughes2021} (in the case of $x_I$, the spacing is much finer in this work). This ensures interpolation accuracy along the inspiral track while avoiding the need to compute fluxes across the full parameter space. The resulting grid contains $99$ parameter tuples $(p,e,x_I)$.

We initialize the inspiral at $(p_0, e_0, x_{I0}) = (8,\;0.2,\;0.95)$ with a mass-ratio of $\varepsilon=10^{-5}$. Figure \ref{fig:dephasing} shows the absolute differences in orbital parameters and the accumulated phase between different trajectory models as a function of coordinate time $t/M$ over the inspiral. Three sets of comparisons are shown: (i) spinning versus non-spinning trajectories (red dashed curves), which illustrate the cumulative effect of the secondary's spin relative to a non-spinning evolution, (ii) the exact spinning trajectory versus the leading-order harmonics (LO) approximation (orange dash-dotted curves) and (iii) the exact spinning trajectory versus the shifted-geodesic (SG) approximation (blue dotted curves). The four panels correspond to the semilatus rectum $p$ (top right), eccentricity $e$ (top left), inclination $x_I$ (bottom left), and accumulated dephasing in \ $\Phi_{\phi}$ (bottom right). 

Across all panels, the differences between spinning and non-spinning trajectories grow secularly over the inspiral, while the comparison between the exact and LO approximations reveals a systematic accumulation of error, though at deviations many orders of magnitude smaller than those induced by the secondary's spin. In the bottom-left panel, the deviation between the exact and LO solutions for the evolution of $x_I(t)$ appears negligible, approaching the level of the numerical noise floor.

The bottom-right panel illustrates the dephasing of the GW signal, a quantity of direct relevance for data analysis. This dephasing refers specifically to the $(l,m,n,k) = (2,2,0,0)$ mode, which is typically dominant for lower eccentricity, lower inclination inspirals such as the one considered here, though this need not be the case for generic EMRIs. Over a simulated inspiral of $1.6\times10^7M$ (about 1 year for a primary with $M = 10^6 M_\odot$), the phase difference between spinning and non-spinning models accumulates to $0.14$ radians. In contrast, the dephasing introduced by using the approximate methods in place of the exact spinning trajectory is much smaller: $3.65\times10^{-5}$ radians for the LO approximation and $2.06\times10^{-2}$ radians for the SG approximation.

These results can be interpreted in the context of a commonly used heuristic for waveform distinguishability: an accumulated phase difference of roughly one radian over the lifetime of the inspiral \cite{Kocsis2011}. By this measure, errors from LO truncations appear negligible on the timescales shown, whereas neglecting the spin altogether would likely lead to mismatches that exceed the heuristic threshold. It is important to emphasize, however, that the one-radian criterion is only a rule of thumb. The true detectability of waveform differences depends sensitively on parameter correlations and the detector sensitivity curve. A full assessment of waveform accuracy will therefore require detailed Bayesian parameter estimation \cite{Burke2024,Khalvati:2025znb} or Fisher-matrix analyses.

It is important to emphasize that the inspiral calculation presented here is intended \textit{solely as a diagnostic comparison to assess the effectiveness of the approximation scheme}, rather than as a fully generic inspiral computation. Several key ingredients are either simplified or omitted. First, only the $0$PA evolution of the Carter constant has been included, so secondary-spin contributions to the Carter-constant flux are absent. Second, the calculation is restricted in eccentricity and inclination: the grid was designed to capture the behavior of a single representative inspiral, rather than spanning the full $(p,e,x_I)$ parameter space required for a generic analysis. Finally, the inspiral sampling is relatively sparse. While the resolution is at least comparable to that used in the generic adiabatic flux study by Hughes et al. \cite{Hughes2021}, it is not dense enough to quantify interpolation errors, which should be estimated in future work.

These limitations mean that the results should be interpreted qualitatively, as an assessment of how well the LO and SG approximations capture secular trends and GW phase evolution, rather than quantitatively as predictions for generic EMRI waveforms with secondary spin. A complete analysis would require consistently incorporating the secondary's spin into the evolution of the Carter constant, broader coverage in orbital parameter space, and a denser sampling strategy to rigorously control interpolation errors.

\section{Conclusion}
\label{sec:conclusion}
 
The results presented in this paper show the shifted-geodesic approximation provides an effective and computationally simple way to capture the impact of secondary spin on the computation of gravitational-wave fluxes.  Conceptually, the method offers a clean mapping between geodesics and spinning-body orbits: each spinning trajectory can be viewed as a geodesic with shifted frequencies and conserved quantities.  Practically, it provides a fast and straightforward framework which combines well-understood elements: infrastructure for computing adiabatic fluxes together with closed-form frequency and conserved quantity corrections.  As discussed below, in regions of the parameter space where the SG approximation is less accurate, a great detail of precision can be regained by using a leading-order approximation (LO) which is still less computationally expensive than a ``full'' spinning-body solution, such as has been used in previous analyses of spinning-body radiation reaction.
 
The accumulated GW dephasing introduced by this approximation is small: over an inspiral of $\sim1$ year, the discrepancy between exact and SG trajectories is $\sim10^{-2}$ radians.  Although detailed comparison studies are needed to fully assess the impact of systematic errors due to this shift, these values suggest that the shifted-geodesic approximation captures the leading secular effects of spin with sufficient fidelity for many purposes. In particular, it appears that the SG approximation successfully isolates the dominant secular effects of spin while offering an efficient framework that is straightforward to implement. This approach may perform particularly well for intermediate-mass-ratio inspirals (IMRIs), which typically spend more time in band in the weak-field regime than EMRIs. In this regime, the shifted-geodesic approximation can capture a broad portion of the parameter space, including higher eccentricities and inclinations, by accurately describing the early inspiral of the spinning secondary.  As inspirals approach the last stable orbit, it should be possible to adjust methods to use a less approximate model for flux computation. Definitively determining whether the shifted-geodesic approximation is sufficient to avoid biasing EMRI parameter estimation with LISA will require follow-up systematic parameter-estimation studies such as Refs.~\cite{Burke2024} and \cite{Khalvati:2025znb}. A natural next step is to incorporate shifted-geodesic fluxes into practical waveform frameworks such as FastEMRIWaveforms (FEW) and compare them against existing flux grids~\cite{Skoupy2025}, to determine whether the approximation can be used reliably without introducing significant biases, and whether it offers a computationally efficient alternative within certain regions of parameter space.

It is important to emphasize that the shifted-geodesic method is not intended to replace a full inspiral calculation. A complete treatment remains vital for reasons such as assessing the validation of approximations and accurately modeling the strong-field evolution of intermediate-mass-ratio inspirals.  Indeed, it is plausible that newly developed analytic methods \cite{SkoupyWitzany2024_2} may enable computation using a ``full'' description of the orbital motion with great efficiency.

Several caveats to our present analysis should also be borne in mind.  First, the inspiral calculation here was designed only to enable an example comparison of the techniques we present with a ``full'' spinning-body inspiral analysis.  It is not a complete generic evolution, as it includes only the $0$PA evolution of the Carter constant.  It is also worth bearing in mind that the grid which provides our $0$PA inspiral data is restricted in eccentricity and inclination. These omissions mean that the results should be viewed as illustrative of the approximation's performance rather than definitive accuracy statements.  Second, the performance of the SG scheme is likely to degrade near the separatrix, where spin corrections to the trajectories may diverge, depending on the chosen spin gauge. It has been shown that, for nearly equatorial orbits, such divergences occur in all but one spin gauge~\cite{Piovano:2025aro}.  Similar behavior has been observed for spherical and generic orbits in the fixed turning point gauge~\cite{Skoupy2025}.  These divergences stem from neglecting the spin-induced separatrix shift, implying that near the LSO the approximation and its associated errors may diverge; this could potentially be mitigated by incorporating the correct separatrix shifts.

As an inspiral evolves through the deep strong field, it is likely that the SG approximation will not be sufficient for many purposes.  In such a regime, we find that retaining several oscillatory Fourier harmonics of the spinning-body motion significantly improves agreement with full calculations.  When a small number of such harmonics are included, the resulting hybrid description can in many cases exceed the accuracy requirements anticipated for LISA data analysis.  While exploration of the full parameter space remains ongoing, these results suggest that a combined treatment --- using the SG approximation's shifted frequencies and integrals of motion, augmenting in the strong field with a modest number of oscillatory corrections --- can provide a robust and efficient pathway for modeling secondary spin effects in EMRI waveforms.

\ack
This work makes use of the Black Hole Perturbation Toolkit \cite{BHPToolkit}, specifically the \textit{KerrGeodesics}, \textit{Teukolsky} and \textit{SpinWeightedSpheroidalHarmonics} packages. The \textit{Spinning-Body-Hamilton-Jacobi} package was also used \cite{repoHJproject}. The computations presented here were conducted in the Resnick High Performance Computing Center, a facility supported by Resnick Sustainability Institute at the California Institute of Technology. L.V.D.\ is supported by the Sherman Fairchild Postdoctoral Fellowship at the California Institute of Technology.  S.A.H.'s work on this problem has been supported by National Science Foundation Grants No.\ PHY-2110384 and PHY-2409644.  V.S.\ is supported by the Czech Science Foundation grant 26-23696S. G.A.P.\ acknowledges the support of the Win4Project grant ETLOG of the Walloon Region for the Einstein Telescope.
\vspace{2em}

\bibliographystyle{IEEEtran}
\bibliography{References}

\appendix{}

 \section{Motion of a spinning test body}
\label{sec:motionspin}
 Finite-size effects arise because real bodies are not point-like; their extent interacts with spacetime curvature, generating forces relative to free-fall trajectories. For a black hole, the dominant finite-size effect is its spin angular momentum, which precesses as it moves through spacetime and couples to curvature, producing a time-varying spin-curvature force. Geodesic orbits describe the motion of a pointlike body freely falling in spacetime, governed by the parallel transport equation for its 4-momentum:
\begin{equation}
    \frac{Dp^\mu}{d\tau} = 0\;,
    \label{eq:geod_general}
\end{equation}
where $D/d\tau \equiv u^\alpha\nabla_\alpha$ denotes a covariant derivative along the trajectory and $p^\alpha$ is the 4-momentum. 

If the body has internal structure, such as spin, this structure couples to the spacetime, modifying its trajectory. The equation of motion becomes \cite{Mathisson2010G_2, Papapetrou1951, Dixon1970}:
\begin{equation}
    \frac{Dp^\mu}{d\tau} = -\frac{1}{2}{R^\mu}_{\nu\lambda\sigma}u^\nu S^{\lambda\sigma}\;.
    \label{eq:spinforce}
\end{equation}
where the spin-curvature force is the term on the right-hand side of the equation, depending on the spacetime's Riemann tensor ${R^\mu}_{\nu\lambda\sigma}$, the body's spin tensor $S^{\lambda\sigma}$ and its 4-velocity $u^\alpha = dx^\alpha/d\tau$. This force is entirely conservative. The small body's spin tensor evolves via the spin precession equation:
\begin{equation}
    \frac{DS^{\mu\nu}}{d\tau} = p^\mu u^\nu - u^\mu p^\nu\;.
    \label{eq:spinprec}
\end{equation}
Observe that Eq.(\ref{eq:spinforce}) together with Eq.\ (\ref{eq:spinprec}) are called the Mathisson-Papapetrou-Dixon (MPD) equations \cite{Mathisson2010,Papapetrou1951,Dixon1970}. To fully specify the motion, a spin supplementary condition (SSC) is is imposed, selecting a unique worldline. In this work we use the Tulczejew SSC \cite{Tulczyjew1959} $p_\mu S^{\mu\nu} = 0$. 

The relationship between the spin vector and spin tensor for the Tulczejew SSC is given by \cite{Kyrian2007}  
\begin{equation}
S^{\mu} = -\frac{1}{2} {\epsilon^{\mu\nu}}_{\alpha\beta} u_{\nu} S^{\alpha\beta},
\label{eq:spinvec}
\end{equation}
where
\begin{equation}
\epsilon_{\alpha\beta\gamma\delta} = \sqrt{-g}[\alpha\beta\gamma\delta]\;,
\end{equation}
$\sqrt{-g}$ is the metric determinant and $[\alpha\beta\gamma\delta]$ is the totally antisymmetric symbol. The evolution of the spin vector in the linear in spin limit can be described using a tetrad-based approach, with spin components expressed as:
\begin{equation} S_\alpha = \mu^2\bigl(s_\perp\cos\phi_s,e_{1\alpha} + s_\perp\sin\phi_s,e_{2\alpha} + s_\parallel,e_{3\alpha}\bigr)\;, \label{eq:Smisalign1} \end{equation}
where $s = \sqrt{s_\perp^2 + s_\parallel^2}$ and $\phi_s$ determines the spin orientation. This formulation expresses $S_\alpha$ in terms of the parallel and perpendicular spin components of the small-body's dimensionless spin parameter $s$. Secondary spin precession occurs only when $s_\perp \neq 0$. We use the closed-form solution for a parallel-transported vector from \cite{vandeMeent2020}, based on a tetrad originally developed in \cite{Marck1983}. This tetrad consists of four legs ${e_{0\alpha}, e_{1\alpha}, e_{2\alpha}, e_{3\alpha}}$, where $e_{0\alpha}$ is the four-velocity $u_\alpha$ of the orbiting body. Legs $1$ and $2$ are related to auxiliary legs $\tilde{e}_{1\alpha}$ and $\tilde{e}_{2\alpha}$ via a precession phase rotation:
\begin{align} 
e_{1\alpha} &= \cos\psi_p,\tilde{e}_{1\alpha} + \sin\psi_p\tilde{e}_{2\alpha}\;,\label{eq:tetradleg1} \\ \ e_{2\alpha} &= -\sin\psi_p\tilde{e}_{1\alpha} + \cos\psi_p\tilde{e}_{2\alpha}\;.\label{eq:tetradleg2} 
\end{align}
Expressions for $\tilde{e}_{1\alpha}$, $\tilde{e}_{2\alpha}$, and $e_{3\alpha}$ are given in Eqs.\ (48), (50), and (51) of \cite{vandeMeent2020} and the $\psi_p$ is precession phase of the secondary spin vector.

\section{Details for the Hamilton-Jacobi method for computing generic spinning-body trajectories}
\label{sec:altmethoddetails}
In Refs.\ \cite{Witzany2019_2,Piovano2024}, the corrections to the radial and polar trajectories are parameterized in terms of shifts to the turning points $r_{1s}$, $r_{2s}$, $z_{1s\parallel}$, $z_{1s\parallel}$, and the anomaly angles $\chi_r \in [0, 2\pi)$ and $\chi_z \in [0, 2\pi)$, along with the spin precession angle $\psi_{\rm p}$.  Note that $\tilde{s}^{CD}$ is defined by
\begin{align}
\tilde{s}^{0D}=0, \ \ \ \ \tilde{s}^{12}=0, \ \ \ \ \tilde{s}^{23}=\sqrt{s^2-s_\parallel^2}\sin\psi_p, \ \ \ \ \tilde{s}^{31}=\sqrt{s^2-s_\parallel^2}\cos\psi_p,
\end{align}
 and the expressions for $w^{\prime}_y$ are in Eqs.\ (26)--(27) of Ref.\ \cite{Witzany2019_2}. The condition for vanishing 4-velocity provides the turning point criteria:
\begin{align}
    (w^{\prime}_y)^2 - e_{0y} e^{\kappa}_{C;y} e_{D\kappa}s^{CD} = 0, \label{eq:turningpointcondition}
\end{align}
where $y = r, \theta$. This enables the corrections to the turning points of the trajectories due to secondary spin to be expressed analytically relative to reference geodesic turning points $r_g$ and $z_g$.  The radial and polar spin-induced corrections to the trajectory are then given as:
\begin{align}
    \delta r(\lambda) &= r_{\rm s}(\chi_r(\lambda), \chi_z(\lambda), \psi_{\rm p}(\lambda)), \label{eq:rad_shift_turning_points} \ \ \ \
    \delta z(\lambda) = z_{\rm s}(\chi_r(\lambda), \chi_z(\lambda), \psi_{\rm p}(\lambda)),
\end{align}
where
\begin{align}
    r_{\rm s}(\chi_r, \chi_z, \psi_{\rm p}) &= \frac{r_{1 \rm s}(z_{\rm g}, \psi_{\rm p}) + r_{2 \rm s}(z_{\rm g}, \psi_{\rm p})}{2} 
    + \frac{r_{1 \rm s}(z_{\rm g}, \psi_{\rm p}) - r_{2 \rm s}(z_{\rm g}, \psi_{\rm p})}{2} \sin{\chi_r}, \\
    z_{\rm s}(\chi_r, \chi_z, \psi_{\rm p}) &= z_{1{\rm s}\perp}(r_{\rm g}) + z_{1{\rm s}\parallel}(r_{\rm g}, \psi_{\rm p})\sin{\chi_z}.
\end{align}

In Ref.\ \cite{Piovano2024}, Piovano et al.\ employ a near-identity transformation to evaluate the trajectories explicitly, substituting $\vec{\zeta} = \wmean - q\vec \xi(\wmean)$ into $\chi_r(\zeta_r)$ and $\chi_z(\zeta_z)$. Using a near-identity transformation, the trajectories $r(\wmean)$ and $z(\wmean)$ can be expressed as:
\begin{align}
    r(\wmean) &= r_{\rm g}(w_r) + \varepsilon \delta r(\wmean), \ \ \ \ z(\wmean) = z_{\rm g}(w_z) + \varepsilon \delta z(\wmean),
\end{align}
where the spin corrections are
\begin{align}
    \delta r(\wmean) &= r_{\rm s}(\wmean) - \frac{\partial r_{\rm g}}{\partial \chi_r} \frac{\partial \chi_r}{\partial w_r} \xi_r (\wmean),  \ \ \ \
    \delta z(\wmean) = z_{\rm s}(\wmean) - \frac{\partial z_{\rm g}}{\partial \chi_z} \frac{\partial \chi_z}{\partial w_z} \xi_z(\wmean). \label{eq:explicitHJ}
\end{align}
The formal solution for $\vec \xi = (\xi_r, \xi_z, \xi_{\rm p})$ is:
\begin{align}
    \xi_y &= \sum_{n=-\infty}^{\infty} \sum_{k=-\infty}^{\infty} \sum_{j=-1}^{1} \frac{i e^{-i \vec \kappa \cdot \wmean}}{\vec \kappa \cdot \vec \Upsilon_{\rm g}}\left(\frac{\delta Y_y}{Y_{y \rm g}}\right)_{\!nkj}, \ \ \ \     
    \xi_{\rm p} = 0, \label{eq:xirz}\\
    \left(\frac{\delta Y_y}{Y_{y \rm g}}\right)_{\!nkj} &= \frac{1}{(2\pi)^3} \int_{(0,2\pi]^3} \dd^3 \mathsf{w} \frac{\delta Y_y(\wmean)}{Y_{y \rm g}(w_y)} e^{i \vec \kappa \cdot \wmean}, \label{eq:deltaYoverY}
\end{align}
where $\vec\Upsilon_{\rm g} = (\hat\Upsilon_{r}, \hat\Upsilon_{z}, \Upsilon_{\rm p})$, $y = r, z$ and $\vec \kappa = (n,k, j)$ with $j \in \{-1, 0, 1\}$. See Ref.\ \cite{Piovano2024} for further details.

\section{Computing generic spinning-body GW fluxes}
\label{sec:computegenericflux}

 We follow the notation in Ref.\ \cite{Skoupy2023}; see Refs.\ \cite{Hughes2021,Piovano2024} and many other works for additional background on the Teukolsky formalism. The relevant perturbations are encoded in the Newman-Penrose scalar
\begin{equation}\label{eq:NPscalar}
    \psi_4 = -C_{\alpha\beta\gamma\delta} n^\alpha \mbar^\beta n^\gamma \mbar^\delta \, ,
\end{equation}
which encodes the outgoing gravitational radiation observable at null infinity in asymptotically flat spacetimes. The scalar $\psi_4$ satisfies the Teukolsky equation,
\begin{equation} \label{eq:teuk}
{}_{-2}\mathcal{O} , \Psi(t,r,\theta,\phi) = 4\pi \Sigma T\;,
\end{equation}
where $\Psi \equiv \zeta^4 \psi_4$ is the rescaled Weyl scalar, ${}_{-2}\mathcal{O}$ is a second-order differential operator specific to spin weight $-2$, and $T$ is the source term constructed from the stress-energy tensor $T^{\mu\nu}$. In the frequency domain, the Teukolsky equation separates into a radial function $R_{lm}(r,\omega)$ and an angular function $S_{lm}(\theta,\omega)$, where $S_{lm}$ is the spin-weighted spheroidal harmonic. This separation is enabled by expressing $\psi_4$ in the form
\begin{equation}
\psi_4 = \frac{1}{(r - ia\cos\theta)^4} \sum_{lm} \int d\omega \, R_{lm}(r,\omega) S_{lm}(\theta,\omega) e^{i(m\phi - \omega t)}\;.
\end{equation}
which factorizes the angular and radial dependence of the perturbation.

The radial equation reads
\begin{equation}
    \mathcal{D}_{lm\omega} R_{lm}(r,\omega) = \mathcal{T}_{lm\omega}\,,
\end{equation}
where $\mathcal{D}_{lm\omega}$ is a second-order differential operator in $r$, and the source term $\mathcal{T}_{lm\omega}$ is given by
\begin{equation}\label{eq:source_term}
    \mathcal{T}_{lm\omega} = \int \dd t \dd \theta \dd \phi \Delta^2 \sum_{ab} \mathcal{T}_{ab} e^{i\omega t -i m \phi}\;,
\end{equation}
with the indices $ab = nn, n\bar{m}, \bar{m}\bar{m}$ corresponding to different projections of the stress-energy tensor onto the Newman-Penrose tetrad \cite{Newman1962}. The quantity $\mathcal{T}_{ab}$ is written as
\begin{equation}\label{eq:source_term2}
    \mathcal{T}_{ab} = \sum_{i=0}^{I_{ab}} \pdv[i]{r}\qty( f^{(i)}_{ab} \sqrt{-g} T_{ab} ) \, ,
\end{equation}
with $I_{nn} = 0$, $I_{n\bar{m}} = 1$, $I_{\bar{m}\bar{m}} = 2$. Here, the projections of the stress-energy tensor are given by
\begin{subequations}
\begin{equation}\label{eq:Tab}
    \sqrt{-g} T_{ab} = \int \dd \tau \qty( (A^{\rm m}_{ab} + A^{\rm d}_{ab}) \delta^4 - \partial_\rho \qty( B^\rho_{ab} \delta^4 ) ) \, ,
\end{equation}
where
\begin{align}
    A^{\rm m}_{ab} &= P_{(a} v_{b)} \; , \label{eq:Amab}\\
    A^{\rm d}_{ab} &= S^{c d} v_{(b} \gamma_{a)dc} + S^{c}{}_{(a} \gamma_{b)dc} v^d \; , \label{eq:Adab}\\
    B^\rho_{ab} &= S^{\rho}{}_{(a} v_{b)}\;, \label{eq:Brhoab}
\end{align}
\end{subequations}
and the spin coefficients are defined by
\begin{equation}
    \gamma_{adc} = \lambda_{a\mu;\rho} \lambda^\mu_d \lambda^\rho_c \; ,
\end{equation}
which capture the influence of the background curvature on the tetrad fields.

The amplitudes at infinity and at the horizon, $Z^{\infty,H}_{lm\omega}$, are determined using the Green function formalism:
\begin{equation}\label{eq:Cpm}
    Z^{\infty,H}_{lm\omega} = \frac{1}{W} \int_{r_+}^{\infty} \frac{R^{H,\infty}_{lm} \mathcal{T}_{lm\omega}}{\Delta^2} \dd r \, ,
\end{equation}
where $R^{H,\infty}_{lm}(r)$ are the homogeneous solutions satisfying the appropriate boundary conditions at the horizon and infinity, and $W$ is the invariant Wronskian. After substituting Eqs.~\eqref{eq:source_term}, \eqref{eq:source_term2}, and \eqref{eq:Tab} into Eq.~\eqref{eq:Cpm} and performing the integration over the delta functions, the amplitudes become
\begin{equation}
    Z^{\infty,H}_{lm\omega} = \int_{-\infty}^{\infty} \frac{\rmd \tau}{\Sigma} e^{i\omega t(\tau) - i m \phi(\tau)} I^{\infty,H}_{lm\omega}(r(\tau), z(\tau), u_a(\tau), S_{ab}(\tau)) \, ,
\end{equation}
with
\begin{align}
\label{eq:Ipmlmomega}
    I^{\infty,H}_{lm\omega} &= \frac{\Sigma}{W}  \sum_{ab} \sum_{i=0}^{I_{ab}} (-1)^i \Biggl[ \biggl( \Big( A^{\rm m}_{ab} + A^{\rm d}_{ab} + i \big(\omega B^t_{ab} - m B^\phi_{ab}\big) \Big) f^{(i)}_{ab} \nonumber \\ &\quad + B^r_{ab} \pdv{f^{(i)}_{ab}}{r} + B^z_{ab} \pdv{f^{(i)}_{ab}}{z} \biggr) \dv[i]{R^{H,\infty}_{lm\omega}}{r} + B^r_{ab} f^{(i)}_{ab} \dv[i+1]{R^{H,\infty}_{lm\omega}}{r} \Biggr] \, .
\end{align}

It can be shown that these amplitudes can be written as a sum over discrete frequencies,
\begin{equation}
\label{eq:Cpm_lmnkj2}
    Z^{\infty,H}_{lm\omega} = \sum_{m,n,k,j} Z^{\infty,H}_{lmnkj} \delta(\omega - \omega_{mnkj}) \quad \text{with} \quad \omega_{mnkj} = m \Omega_\phi + n \Omega_r + k \Omega_z + j \Omega_s \; .
\end{equation}
Here, $m$ is the azimuthal harmonic index, and $n$, $k$, and $j$ denote the radial, polar, and spin harmonic indices, respectively. Note that $\Omega_i$ corresponds to frequencies conjugate to coordinate-time $t$. The partial amplitudes are then given by
\begin{multline}\label{eq:Cpm_lmnkj}
    Z^{\infty,H}_{lmnkj} = \frac{1}{(2\pi)^2 \Gamma} \int_0^{2\pi} \dd w_r \int_0^{2\pi} \dd w_z \int_0^{2\pi} \dd w_p \; I^{\infty,H}_{lmnkj}(w_r, w_z, w_p) \\ \times \exp\Bigl(i \omega_{mnkj} \Delta t(w_r, w_z, w_p) - i m \Delta \phi(w_r, w_z, w_p) + i n w_r + i k w_z + i j w_p\Bigr) \, ,
\end{multline}
where 
\begin{equation}
I^\pm_{lmnkj}(w_r, w_z, w_p) = I^\pm_{lm\omega_{mnkj}}(r(w_r, w_z, w_p),z(w_r, w_z, w_p),u_a(w_r, w_z, w_p),S_{ab}(w_r, w_z, w_p))\;.
\end{equation}

\section{Accuracy tests for Leading-Order and Shifted-Geodesic approximations}
\label{sec:SGandLOtest}

\begin{figure}[htbp]
\hspace{-0.6cm}
\includegraphics[width=1.07\columnwidth]{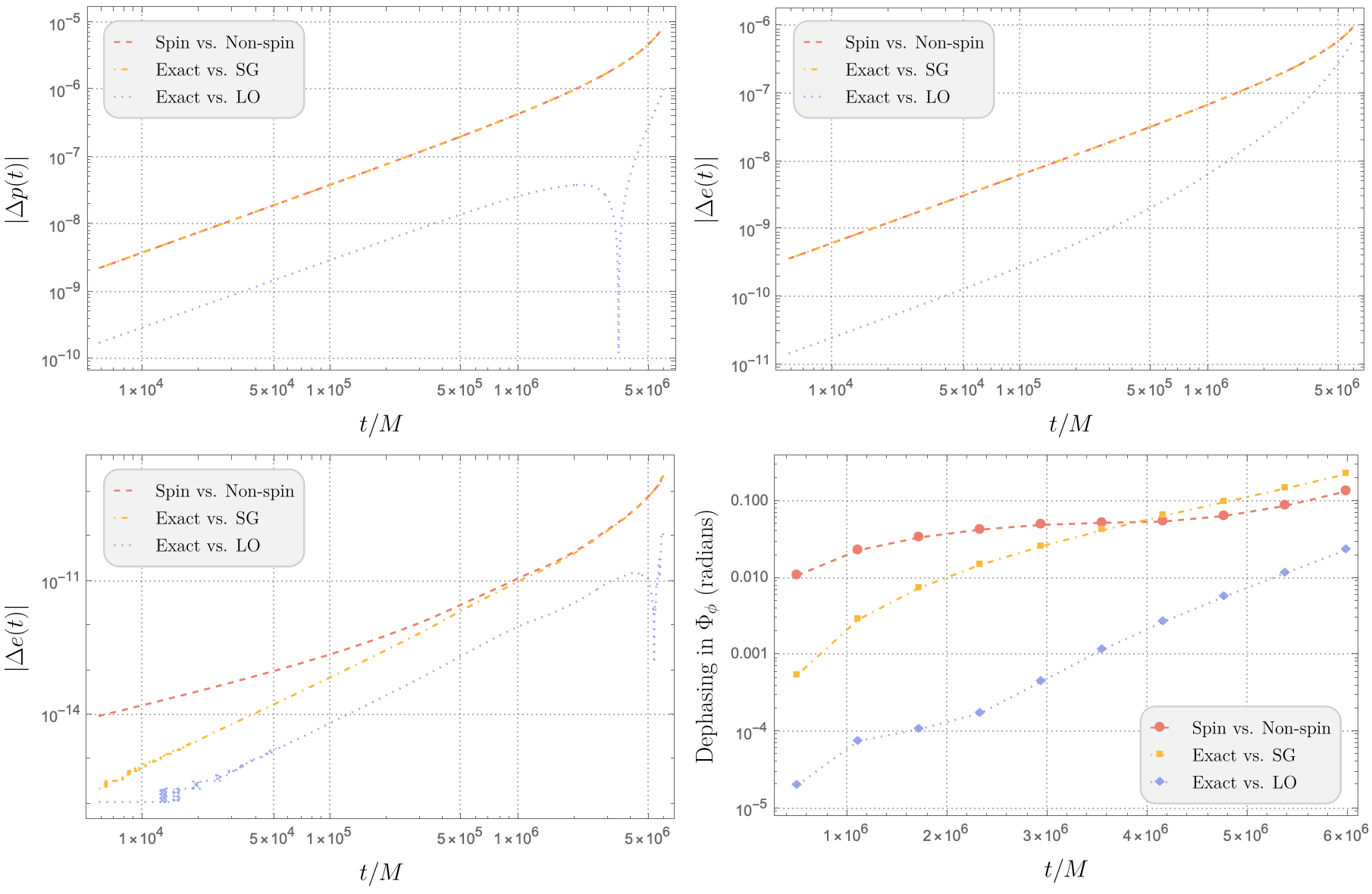}
\caption{Absolute differences in orbital parameters and accumulated gravitational-wave phase between spinning-body trajectories and two approximate models, shown as functions of coordinate time $t$. Red dashed curves show differences between the exact spinning and non-spinning trajectories. Orange dash-dotted curves show differences between the exact spinning-body evolution and a simplified shifted-geodesic (SG) approximation in which oscillatory corrections to the temporal and azimuthal coordinates are omitted. Blue dotted curves show differences between the exact spinning-body evolution and a leading-order (LO) spinning-body approximation with $n^s_{\max}=k^s_{\max}=1$. The four panels correspond to: (top left) semilatus rectum $p$, (top right) eccentricity $e$, (bottom left) inclination parameter $x_I$, and (bottom right) accumulated azimuthal phase $\Phi_\phi$. Parameters for all curves are $a=0.7M$, $p_{\rm init}=10$, $e_{\rm init}=0.2$, $x_{\rm I,init}=0.95$, and $\varepsilon=10^{-5}$.
}
\label{fig:dephasing2}
\end{figure}
Figure~\ref{fig:dephasing2} compares the exact spinning-body evolution to further simplified versions of the shifted-geodesic (SG) and leading-order (LO) approximations, included here to motivate the modeling choices adopted in the main text. These comparisons show how additional truncations reduce the accuracy of the inspiral evolution. In particular, we examine (i) a leading-order (LO) spinning-body approximation in which only the lowest oscillatory harmonics are retained, $n^s_{\max}=k^s_{\max}=1$, and (ii) a shifted-geodesic (SG) approximation in which the oscillatory corrections to the temporal and azimuthal coordinates, $\Delta t$ and $\Delta\phi$, are omitted. The figure shows absolute differences in the orbital elements $p$, $e$, and $x_I$, as well as the accumulated azimuthal phase, as functions of coordinate time.

The orange dash-dotted curves show the performance of the shifted-geodesic approximation when oscillatory corrections to the temporal and azimuthal coordinates are omitted. Although these corrections are oscillatory, neglecting them leads to a significant loss of accuracy in both the orbital elements and the accumulated phase. In particular, errors in $\Phi_\phi$ grow monotonically over the inspiral, indicating that the oscillatory contributions to $t$ and $\phi$ play an essential practical role in maintaining phase accuracy. 

The blue dotted curves demonstrate that truncating the oscillatory corrections at $n^s_{\max}=k^s_{\max}=1$ does not yield a systematic improvement over the SG approximation adopted in the main text. While this LO approximation can track the exact evolution reasonably well at early times, the errors grow over the inspiral and are comparable to those obtained from the SG model in some quantities. This is evident in the accumulated azimuthal phase (bottom right panel), where the LO dephasing grows to $\mathcal{O}(10^{-2})$ radians by the end of the evolution. We therefore find that truncating the LO approximation at $n^s_{\max}=k^s_{\max}=1$ does not yield a systematic improvement over the shifted-geodesic model adopted in the main text (i.e., compare the blue curve in \ref{fig:dephasing2} to the orange curve in Figure \ref{fig:dephasing}; the dephasing is comparable), and that higher oscillatory harmonics are required for the LO approximation to offer a clear accuracy gain.

Taken together, Fig.~\ref{fig:dephasing2} shows that neither a purely secular shifted-geodesic construction nor a minimal LO truncation is sufficient for accurate long-term evolution, motivating the combined treatment adopted in the main text.

\section{Convergence criteria for computing post-adiabatic order fluxes}
\label{sec:convcriteriapostad}

This appendix describes the convergence criteria used to truncate the mode sums in the computation of energy and angular-momentum fluxes and explains how these choices control the total truncation error in a post-adiabatic expansion. The goal is to justify the use of looser truncation thresholds for subleading post-adiabatic contributions while maintaining consistency with the adiabatic accuracy tolerance adopted throughout the paper.

To compute energy or angular momentum fluxes, we perform a mode decomposition of the form: 
\begin{equation}\mathcal{F}^E = \mathcal{F}^{E,(0)} + \varepsilon \, \mathcal{F}^{E,(1)} + \cdots,
\end{equation}
where $\mathcal{F}^{E,(0)}\equiv\hat{\mathcal{F}}^{E}$ is the leading-order adiabatic flux,
$\mathcal{F}^{E,(1)}\equiv\delta \mathcal{F}^{E}$ is the spinning-secondary correction and $\varepsilon \ll 1$ is a formal expansion parameter (i.e., the mass ratio). Each term $\mathcal{F}^{E,(i)}$ is computed as a mode sum over $l,m,n,k$; for clarity of presentation, we focus on the sum over $n$:
\begin{equation}
\mathcal{F}^{E,(i)} = \sum_{n=-\infty}^{\infty} \mathcal{F}_n^{E,(i)}\;,
\end{equation}
where $ \mathcal{F}_n^{E,(i)}$ denotes the $n$-th mode contribution at $i$-th order in $\varepsilon$. To truncate the sum at mode $n_{max}$, a common convergence criterion is:
\begin{equation}
|\mathcal{F}_{n_{max}+1}^{E,(i)}| \leq \epsilon_i \left| \sum_{n=-n_{max}}^{n_{max}} \mathcal{F}_n^{E,(i)} \right|\;,
\end{equation}
ensuring that the next mode's contribution is negligible compared to the partial sum.  The $i$-th convergence parameter $\epsilon_i\ll 1$; note that these are distinct from the mass-ratio parameter $\varepsilon$.

Let us now estimate the total fractional error in the flux $\mathcal{F}^{E}$ due to truncation. Each partial sum has a truncation error of order $\epsilon_i |\mathcal{F}^{E,(i)}|$, and contributes to $\mathcal{F}^{E}$ with a suppression factor $(\varepsilon)^i$. The total relative truncation error in the full flux is then approximately:
\begin{equation}
\frac{\mathcal{F}^{E}_{\rm error}}{\mathcal{F}^{E}} \sim \frac{\epsilon_0 |\mathcal{F}^{E,(0)}| + \varepsilon \epsilon_1 |\mathcal{F}^{E,(1)}| + \cdots}{|\mathcal{F}^{E,(0)}| + \varepsilon |\mathcal{F}^{E,(1)}|+ \cdots} \;.
\end{equation}
Let us take $\epsilon_0$ to be $\epsilon_{\rm ad}$.  Assuming all $|\mathcal{F}^{E,(i)}|$ are of comparable magnitude, the denominator can be approximated by $\mathcal{F}^{E,(0)}$, and we obtain:
\begin{equation}
\frac{\mathcal{F}^{E}_{\rm error}}{\mathcal{F}^{E}} \sim \epsilon_0 + \varepsilon (\epsilon_1 - \epsilon_0) + \cdots
\end{equation}
Assuming that $\epsilon_0 \ll \epsilon_1$, and requiring that the contribution of the first order contribution to the convergence accuracy be comparable to the contribution from the leading order piece leads to the criterion $\epsilon_1 \lesssim \varepsilon^{-1}\epsilon_0$, or $\epsilon_1 \lesssim \varepsilon^{-1}\epsilon_{\rm ad}$.  By induction, it is not difficult to see that the condition we impose at any order in the expansion can be written
\begin{equation}
\label{eq:tol}
\epsilon_i \lesssim \varepsilon^{-i} \epsilon_{\rm ad}\;.
\end{equation}
The result (\ref{eq:tol}) justifies the looser convergence criteria used for subleading post-adiabatic terms while maintaining control over the overall truncation error.  We will use this criterion to assess the accuracy of the approximation schemes described in this work. 

\section{Explicit evolution equations for orbital elements}
\label{ref:explicitev}
We work with orbital parameters $ \vec{\mathcal{P}}=(p,e,x_I) $ with  conserved quantities $ \vec{\mathcal{C}}=(E,L_z, Q) $. We use the Jacobian to relate the evolution of $\vec{\mathcal{P}}$ and $\vec{\mathcal{C}}$:
\begin{align}
\dot{\mathcal{P}}_a &= \sum_{b\in\{E,L_z,Q\}} (J^{-1})_{ab}\,\dot{\mathcal{C}}_b\;,
\qquad
J_{ba} = \frac{\partial \mathcal{C}_b}{\partial \mathcal{P}_a}\;.
\label{eq:Jacobian-def2}
\end{align}
\setcounter{footnote}{0}
We linearize in $\varepsilon$ as\footnote{Again, we put $\sigma=\varepsilon$ for notational convenience, without loss of generality.}
\begin{align}
 \mathcal{C}_b=&\hat{\mathcal{C}}_b+\varepsilon\,\delta\mathcal{C}_b\;, \qquad
\dot{\mathcal{C}}_b=\dot{\hat{\mathcal{C}}}_b+\varepsilon\,\delta\dot{\mathcal{C}}_b\;.\\
 \mathcal{P}_a=&\hat{\mathcal{P}}_a+\varepsilon\,\delta\mathcal{P}_a\;, \qquad
\dot{\mathcal{P}}_a=\dot{\hat{\mathcal{P}}}_a+\varepsilon\,\delta\dot{\mathcal{P}}_a\;.
\end{align}
Note that $\dot{\hat{\mathcal{C}}}_b\equiv\hat{\mathcal{F}}^b$ and $\delta\dot{\mathcal{C}}_b\equiv\delta\mathcal{F}^b$. Differentiating $ \dot{\vec{\mathcal{P}}}=\mathbf{J}^{-1}\dot{\vec{\mathcal{C}}} $ with respect to $ \varepsilon$ at $ \varepsilon=0 $ gives:

\begin{align}
\hat{\dot{\mathcal P}}_a
&= (\hat{J}^{-1})_{a b}\,\hat{\dot{\mathcal C}}_b\;,
\qquad
\hat{J}_{b a} \equiv
\left.\frac{\partial \mathcal C_b}{\partial \mathcal P_a}\right|_{\varepsilon=0}\;,  \label{eq:Pdot-hat} \\
\delta\dot{\mathcal P}_a &\equiv\partial_\varepsilon\dot{\mathcal{P}}_a\big|_{\varepsilon = 0}
= (\hat{J}^{-1})_{a b}\,\delta\dot{\mathcal C}_b
- (\hat{J}^{-1})_{a b}\,
\left(\partial_\varepsilon J_{b c}\right)_{\varepsilon=0}\,
\hat{\dot{\mathcal P}}_c\;.
\label{eq:Pdot-delta}
\end{align}

Now, we write out the explicit component-by-component formulas below:
\begin{align}
E&=\hat E+\varepsilon\ \delta E^S\;, &
L_z&=\hat L_z+\varepsilon\,\delta\hat L_z^S\;, &
Q&=\hat Q+\varepsilon\,\delta Q^S\;,
\\
\dot{\hat E}&=\dot{\hat E}+\varepsilon\,\delta\dot{E}^S\;, &
\dot{\hat L}_z&=\dot{\hat L}_z+\varepsilon\,\delta\dot{J}^S_z\;, &
\dot{Q}&=\dot{\hat Q}+\varepsilon\,\delta\dot{Q}^S\;.
\end{align}
All partial derivatives below are evaluated at $ \varepsilon=0 $.
\begin{align}
E_{,p}&\equiv\frac{\partial \hat E}{\partial p}\;,\quad
E_{,e}\equiv\frac{\partial \hat E}{\partial e}\;,\quad
E_{,x}\equiv\frac{\partial \hat E}{\partial x_I}\;,\qquad \\
L_{,p}&\equiv\frac{\partial \hat L_z}{\partial p}\;,\quad
L_{,e}\equiv\frac{\partial \hat L_z}{\partial e}\;,\quad 
L_{,x}\equiv\frac{\partial \hat L_z}{\partial x_I}\;,\qquad \\
Q_{,p}&\equiv\frac{\partial \hat Q}{\partial p}\;,\quad
Q_{,e}\equiv\frac{\partial \hat Q}{\partial e}\;,\quad
Q_{,x}\equiv\frac{\partial \hat Q}{\partial x_I}\;,
\end{align}
\begin{align}
\delta E_{,p}&\equiv\frac{\partial (\delta E^S)}{\partial p}\;, \quad
\delta E_{,e}\equiv\frac{\partial (\delta E^S)}{\partial e}\;, \quad
\delta E_{,x}\equiv\frac{\partial (\delta E^S)}{\partial x_I}\;,\qquad \\
\delta L_{,p}&\equiv\frac{\partial (\delta L_z^S)}{\partial p}\;,\quad
\delta L_{,e}\equiv\frac{\partial (\delta  L_z^S)}{\partial e}\;,\quad 
\delta L_{,x}\equiv\frac{\partial (\delta L_z^S)}{\partial x_I}\;,\qquad \\
\delta Q_{,p}&\equiv\frac{\partial (\delta Q^S)}{\partial p}\;,\quad 
\delta Q_{,e}\equiv\frac{\partial (\delta Q^S)}{\partial e}\;,\quad 
\delta Q_{,x}\equiv\frac{\partial (\delta Q^S)}{\partial x_I}\;.
\end{align}

The geodesic Jacobian and its determinant are
\begin{align}
J&=
\begin{pmatrix}
E_{,p} & E_{,e} & E_{,x}\\
L_{,p} & L_{,e} & L_{,x}\\
Q_{,p} & Q_{,e} & Q_{,x}
\end{pmatrix}\;,
\\
D&=\det J
=E_{,p}(L_{,e} Q_{,x}-L_{,x} Q_{,e})-E_{,e}(L_{,p} Q_{,x}-L_{,x} Q_{,p})+E_{,x}(L_{,p} Q_{,e}-L_{,e} Q_{,p})\;.
\end{align}

The 0PA-order orbital element evolution equations are:
\begin{align}
\dot{\hat{p}}&=\frac{(L_{,e} Q_{,x}-L_{,x} Q_{,e})\,\dot{\hat E}
+(E_{,x} Q_{,e}-E_{,e} Q_{,x})\,\dot{\hat L}_z
+(E_{,e} J_{,x}-E_{,x} J_{,e})\,\dot{\hat Q}}{D}\;,
\\
\dot{\hat{e}}&=\frac{(L_{,x} Q_{,p}-L_{,p} Q_{,x})\,\dot{\hat E}
+(E_{,p} Q_{,x}-E_{,x} Q_p)\,\dot{\hat L}_z
+(E_{,x} L_{,p}-E_{,p} L_{,x})\,\dot{\hat Q}}{D}\;,
\\
\dot{\hat{x}}_I&=\frac{(L_{,p} Q_{,e}-L_{,e} Q_{,p})\,\dot{\hat E}
+(E_{,e} Q_{,p}-E_{,p} Q_{,e})\,\dot{\hat L}_z
+(E_{,p} L_{,e}-E_{,e} L_{,p})\,\dot{\hat Q}}{D}\;.
\end{align}

It is convenient to collect terms in the following way:
\begin{align}
S_E&\equiv \delta E_{,p}\,\dot{\hat{p}}+\delta E_{,e}\,\dot{\hat{e}}+\delta E_{,x}\,\dot{\hat{x}}_I\;,\\
S_L&\equiv \delta L_{,p}\,\dot{\hat{p}}+\delta L_{,e}\,\dot{\hat{e}}+\delta L_{,x}\,\dot{\hat{x}}_I\;,\\
S_Q&\equiv \delta Q_{,p}\,\dot{\hat{p}}+\delta Q_{,e}\,\dot{\hat{e}}+\delta Q_{,x}\,\dot{\hat{x}}_I\;.
\end{align}

The linear-in-$\varepsilon$ corrections to the orbital evolution equations are then
\begin{align}
\delta \dot{p}\equiv \left.\frac{\partial \dot p}{\partial \varepsilon}\right|_{\varepsilon=0}
&=\frac{(L_{,e} Q_{,x}-L_{,x} Q_{,e})\,\delta\dot{E}^S
+(E_{,x} Q_{,e}-E_{,e} Q_{,x})\,\delta\dot{L}_z^S
+(E_{,e} L_{,x}-E_{,x} L_{,e})\,\delta\dot{Q}^S}{D}\nonumber
\\&-\frac{(L_{,e} Q_{,x}-L_{,x} Q_{,e})\,S_E
+(E_{,x} Q_{,e}-E_{,e} Q_{,x})\,S_L
+(E_{,e} L_{,x}-E_{,x} L_{,e})\,S_Q}{D}\;,
\end{align}
\begin{align}
\delta \dot{e}\equiv \left.\frac{\partial \dot e}{\partial \varepsilon}\right|_{\varepsilon=0}
&=\frac{(L_{,x} Q_{,p}-L_{,p} Q_{,x})\,\delta\dot{E}^S
+(E_{,p} Q_{,x}-E_{,x} Q_{,p})\,\delta\dot{L}_z^S
+(E_{,x} L_{,p}-E_{,p} L_{,x})\,\delta\dot{Q}^S}{D}\nonumber
\\&-\frac{(L_{,x} Q_{,p}-L_{,p} Q_{,x})\,S_E
+(E_{,p} Q_{,x}-E_{,x} Q_{,p})\,S_L
+(E_{,x} L_{,p}-E_{,p} L_{,x})\,S_Q}{D}\;,
\end{align}
\begin{align}
\delta \dot{x_I}\equiv \left.\frac{\partial \dot x}{\partial \varepsilon}\right|_{\varepsilon=0}
&=\frac{(L_{,p} Q_{,e}-L_{,e} Q_{,p})\,\delta\dot{E}^S
+(E_{,e} Q_{,p}-E_{,p} Q_{,e})\,\delta\dot{L}_z^S
+(E_{,p} L_{,e}-E_{,e} L_{,p})\,\delta\dot{Q}^S}{D}\nonumber
\\&-\frac{(L_{,p} Q_{,e}-L_{,e} Q_{,p})\,S_E
+(E_{,e} Q_{,p}-E_{,p} Q_{,e})\,S_L
+(E_{,p} L_{,e}-E_{,e} L_{,p})\,S_Q}{D}\;.
\end{align}

\end{document}